\documentclass[12pt,preprint]{aastex}
\usepackage{psfig,amsfonts,amsmath,graphicx,natbib,apjfonts,lscape,nicefrac,
url,longtable,datetime,threeparttable,threeparttablex}
\usepackage[breaklinks]{hyperref}

\usepackage{siunitx,booktabs}

\hypersetup{colorlinks=true, urlcolor=blue, citecolor=blue}

\newcommand{\Swift}{\textit{Swift}}
\newcommand{\Spitzer}{\textit{Spitzer}}
\newcommand{\Chandra}{\textit{Chandra}}
\newcommand{\nuc}{$\nu_{\rm c}$}
\newcommand{\numax}{$\nu_{\rm m}$}
\newcommand{\nua}{$\nu_{\rm a}$}
\newcommand{\E}{$E_{\rm K, iso}$}
\newcommand{\Efte}{$E_{\rm K, iso, 52}$}
\newcommand{\epse}{$\epsilon_{\rm e}$}
\newcommand{\epsb}{$\epsilon_{\rm B}$}
\newcommand{\dens}{$n_{0}$}

\newcommand{\tjet}{$t_{\rm jet}$}

\newcommand{\AV}{$A_{\rm V}$}

\newcommand{\emcee}{\textsc{emcee}}

\def\cfa{1}
\def\leicester{2}
\def\warwick{3}
\def\caltech{4}
\def\ucsc{5}
\def\mpibonn{6}
\def\ING{7}

\begin{document} 

\title{\textsc{GRB~120521C at $z\sim6$ and the Properties of High-redshift GRBs}}

\author{
Tanmoy Laskar\altaffilmark{\cfa},
Edo Berger\altaffilmark{\cfa},
Nial Tanvir\altaffilmark{\leicester},
B.~Ashley Zauderer\altaffilmark{\cfa},
Raffaella Margutti\altaffilmark{\cfa},
Andrew Levan\altaffilmark{\warwick},
Daniel Perley\altaffilmark{\caltech},
Wen-fai Fong\altaffilmark{\cfa},
Klaas Wiersema\altaffilmark{\leicester},
Antonino Cucchiara\altaffilmark{\ucsc},
Karl Menten\altaffilmark{\mpibonn}, and
Marie Hrudkova\altaffilmark{\ING}
}

\altaffiltext{\cfa}{Harvard-Smithsonian Center for Astrophysics, 60
Garden Street, Cambridge, MA 02138}
\altaffiltext{\leicester}{Department of Physics and Astronomy, University of Leicester, University 
Road, Leicester LE1~7RH, United Kingdom}
\altaffiltext{\warwick}{Department of Physics, University of Warwick, Coventry CV4~7AL, United 
Kingdom}
\altaffiltext{\caltech}{Department of Astronomy, California Institute of Technology,
MC 249-17, 1200 East California Blvd, Pasadena CA 91125, USA}
\altaffiltext{\ucsc}{Department of Astronomy and Astrophysics, UCO/Lick Observatory, University of 
California, 1156 High Street, Santa Cruz, CA 95064, USA}
\altaffiltext{\ING}{Isaac Newton Group of Telescopes, Apartado de Correos 321, E-387 00 Santa Cruz 
de la Palma, Canary Islands, Spain}
\altaffiltext{\mpibonn}{Max-Planck-Institut f{\" u}r Radioastronomie, Auf dem H{\" u}gel 69, 
D-53121 Bonn, Germany}

\shorttitle{GRB\,120521C at $z\sim 6$}
\shortauthors{Laskar et al.}

\begin{abstract}
We present optical, near-infrared, and radio observations of the afterglow of GRB 120521C. By 
modeling the multi-wavelength dataset, we derive a photometric redshift of $z\approx 6.0$, which we 
confirm with a low signal-to-noise ratio spectrum of the afterglow.  We find that a model with a 
constant-density environment provides a good fit to the afterglow data, with an inferred density of 
$n\lesssim 0.05$ cm$^{-3}$.  The radio observations reveal the presence of a jet break at $t_{\rm 
jet}\approx 7$ d, corresponding to a jet opening angle of $\theta_{\rm jet}\approx 3^\circ$.  The 
beaming-corrected 
$\gamma$-ray and kinetic energies are $E_\gamma\approx E_K\approx 3\times 10^{50}$ erg.  We quantify 
the uncertainties in our results using a detailed Markov Chain Monte Carlo analysis, which allows us 
to uncover degeneracies between the physical parameters of the explosion. To compare GRB\,120521C to 
other high-redshift bursts in a uniform manner we re-fit all available afterglow data for the two 
other bursts at $z\gtrsim 6$ with radio detections (GRBs 050904 and 090423).  We find a jet break at 
$t_{\rm jet}\approx 15$ d for GRB 090423, in contrast to previous work.  Based on these three 
events, we find that GRBs at $z\gtrsim 6$ appear to explode in constant-density environments, and 
exhibit a wide range of energies and densities that span the range inferred for lower redshift 
bursts.  On the other hand, we find a hint for narrower jets in the $z\gtrsim 6$ bursts, potentially 
indicating a larger true event rate at these redshifts.  Overall, our results indicate that long 
GRBs share a common progenitor population at least to $z\sim 8$.
\end{abstract}


%Draft version \date{\today} \date{\currenttime} EDT

\section{Introduction}
Long duration $\gamma$-ray bursts (GRBs) are known to be associated with the violent deaths of
massive stars \citep[eg][]{wb06}. In conjunction with the large luminosities of their afterglows, 
they can therefore serve as powerful probes of the high-redshift Universe \citep{ioc07},
providing clues to the formation environments of the first stars, the ionization and metal 
enrichment history of the Universe, and the properties of galaxies that are otherwise too faint to 
study through direct imaging and spectroscopy \citep{tkk+06,tlf+12,cbf+13}. Furthermore, modeling 
of multi-wavelength afterglow data allows us to constrain the densities and structure 
of massive star environments on parsec scales, as well as the energies of the explosions and the 
degree of ejecta collimation.

To use GRBs as effective probes of star-formation in the re-ionization era ($z\gtrsim6$; 
\citealt{fns+02,fsb+06}), it is important to understand whether there is any evolution in the 
properties of their progenitors with redshift. This is best achieved by studying the afterglows 
of the highest-redshift events to determine their explosion energy, circumburst density and degree 
of collimation, and by comparing these properties with those of their lower-redshift counterparts. 
In the long term, such studies have the potential to uncover the contribution of Population III 
stars, which have been speculated to be highly energetic ($E_{\rm iso}\sim10^{52}$--$10^{57}$\,erg) 
with relatively long durations \citep[$T_{\rm 90}\sim1000$\,s; 
e.g.][]{fwh01,byh03,hfw+03,mr10,si11,tsm11,wbg+12}. 

At present, there are only three GRBs with spectroscopically-confirmed redshifts of $z\gtrsim6$: 
GRB 050904 at $z=6.29$ \citep{tac+05,hnr+06,kka+06}, GRB~080913 at $z=6.70$
\citep{gkf+09}, and GRB~090423 at $z=8.23$ \citep{sdvc+09,tfl+09}. In addition, GRB~090429B has an 
inferred photometric redshift of $z\sim9.4$ \citep{clf+11}. To fully determine the physical 
properties of a GRB and its environment requires multi-wavelength observations spanning the radio 
through to the X-rays; only two of the $z\gtrsim6$ events have radio detections: GRB~050904 
\citep{fck+06,gfm07} and GRB~090423 \citep{tfl+09,cff+10}. 

Previous studies of GRB~050904 have found a high circumburst density ($n\sim10^2$--$10^3\,{\rm 
cm}^{-3}$; \citealt{fck+06,gfm07}), a high isotropic-equivalent $\gamma$-ray energy ($E_{\gamma,\rm 
iso}\approx10^{54}$\,erg; \citealt{cmc+06}), a large isotropic-equivalent kinetic energy 
($E_{\rm K,iso}\approx$~few~$\times10^{53}$\,erg; \citealt{fck+06,gfm07}), and no evidence for host 
extinction ($A_{\rm V}\lesssim0.1$\,mag; \citealt{gfm07,zwm+10}, although see also 
\citealt{smfd07,sgm11}). A jet break at $t_{\rm jet}\approx3$\,d \citep{tac+05} indicates 
a beaming-corrected $\gamma$-ray energy of $8\times10^{51}$\,erg and and kinetic energy of $E_{\rm 
K}\approx2\times10^{51}$\,erg, the latter being one of the largest known \citep{gfm07}. 
GRB~090423 has an inferred density of $n\lesssim1\,{\rm cm}^{-3}$ \citep{cff+10},
large isotropic-equivalent $\gamma$-ray energy ($E_{\gamma}\gtrsim10^{53}$\,erg) and kinetic 
energy ($E_{\rm K, iso}\gtrsim3\times10^{53}$\,erg), and no host extinction ($A_{\rm 
V}\lesssim0.1$\,mag; \citealt{tfl+09}). No jet break was seen for this event, resulting in a claim 
of $E_{\rm K}\gtrsim7\times10^{51}$\,erg, even larger than for GRB~050904. 

Whereas individual studies of these two GRBs have been undertaken, they employed different
implementations of afterglow synchrotron models and their results cannot be compared directly. Here 
we report multi-wavelength observations of GRB~120521C and deduce a photometric redshift of 
$z\approx6$, making this the third high-redshift GRB with multi-wavelength data from radio to 
X-rays. The availability of well-sampled light curves spanning several orders of magnitude in 
frequency and time allow us to perform broad-band afterglow modeling, and thereby to determine the 
energetics of the explosion, the density profile of the circumburst environment, the microphysical 
parameters of the relativistic shocks, and the collimation of the ejecta. We additionally 
re-analyze all available afterglow data for GRBs 050904 and 090423, enabling us to compare the 
three high-redshift GRBs in a uniform manner. Finally, we compare the properties of the 
high-redshift GRBs to those of bursts at $z\sim1$ to investigate whether high-redshift GRBs exhibit 
evidence for an evolution in the progenitor population or favor different environments than their 
lower-redshift counterparts. 
We present our observations and analysis for GRB~120521C in Section \ref{text:data_analysis} and 
determine a photometric redshift for this event in Section \ref{text:photoz}. We describe the 
theoretical model employed and our multi-wavelength modeling software in Section 
\ref{text:modelling} and present our broadband afterglow model for GRB~120521C in Section 
\ref{text:results}. We apply our modeling code to re-derive the properties of GRBs 050904 and 
090423 in Section \ref{text:othergrbs} and compare the results to those obtained for GRB~120521C 
and to lower-redshift events in Section \ref{text:comparisons}. We present our conclusions in 
Section \ref{text:conclusions}. We use the standard cosmological parameters, $\Omega_{\rm m} = 
0.27$, $\Omega_{\rm \Lambda} = 0.73$ and $H_{\rm 0} = 71\,{\rm km}\,{\rm s}^{-1}\,{\rm Mpc}^{-1}$. 
All magnitudes are in the AB system, unless stated otherwise.

\section{GRB Properties and Observations}
\label{text:data_analysis}
GRB~120521C was discovered with the \Swift\ Burst Alert Telescope
\citep[BAT;][]{bbc+05} on 2012 May 21 at 23:22:07\,UT \citep{gcn13318}. The burst duration was
$T_{90} = (26.7\pm0.4)$\,s, with a fluence of $F_{\gamma} = (1.1 \pm 0.1)
\times10^{-6}$\,erg\,cm$^{-2}$ \citep[15--150\,keV;][]{gcn13333}. The \Swift\ X-ray Telescope
\citep[XRT;][]{bhn+05} began observing the field 69\,s after the BAT trigger, leading to the 
detection of an X-ray afterglow at coordinates RA(J2000)  = 14$^{\rm h}$ 17$^{\rm m}$ 
08.73$^{\rm s}$, Dec(J2000) = +42$^{\circ}$ 08\arcmin 41.0\arcsec, with an
uncertainty radius of 1.6\arcsec (90\% containment)
\footnote{\url{http://www.swift.ac.uk/xrt_positions/522656}}.
XRT continued observing the afterglow for 1.5 days in photon counting (PC) mode, with the
last detection at about 0.5 days.

\subsection{X-rays}
\label{text:data_analysis:XRT}
We analyzed the XRT data using the latest version of the HEASOFT package (v6.11) and corresponding
calibration files. We utilized standard filtering and screening criteria, and generated a count-rate
light curve following the prescriptions by \citet{mgg+10}. The data were re-binned with
the requirement of a minimum signal-to-noise ratio of 4 in each temporal bin.

We used Xspec (v12.6) to fit the PC-mode spectrum between $3\times10^{-3}$ and 0.35\,d, assuming a 
photoelectrically absorbed power law model (\texttt{tbabs $\times$ ztbabs $\times$ pow}) and a 
Galactic neutral hydrogen column density of $N_{\rm H, MW}
= 1.1\times10^{20}$\,cm${}^{-2}$ \citep{kbh+05}, fixing the source redshift at $z=6.0$ (see 
Sections \ref{text:photoz} and \ref{text:results}).
Our best-fit model has a photon index of $\Gamma = 1.86^{+0.14}_{-0.11}$
($68\%$ confidence intervals, C-stat = 151 for 180 degrees of freedom). We
found no evidence for additional absorption with a $3\sigma$ upper limit of $N_{\rm H, int} \lesssim
6.6\times10^{22}$ \,cm${}^{-2}$, assuming solar metallicity.

To assess the impact of the uncertain intrinsic absorption, we fit a PC-mode spectrum with the
intrinsic $N_{\rm H}$ fixed to this $3\sigma$ upper limit and found $\Gamma = 2.03\pm0.26$.
Next, we fixed the intrinsic
absorption to zero and 
found $\Gamma = 1.77\pm0.21$.
The two light curves differ by less than $5\%$. In the following analysis, we assume $N_{\rm H, int}
= 0$ and use the corresponding computed 0.3 -- 10\,keV light curve, together with $\Gamma=1.77$ to
compute the 1\,keV flux density (Table \ref{tab:120521C:data:xrt}).

\subsection{Optical and Near-IR}
We obtained \textit{riz}-band imaging of the XRT error circle beginning about 40\,min after the 
BAT trigger using ACAM on the William Herschel Telescope (WHT) and MOSCA on the Nordic Optical
Telescope (NOT). We analyzed the data using standard procedures within
IRAF\footnote{IRAF is distributed by the National Optical Astronomy Observatory, which is operated 
by the Association of Universities for Research in Astronomy (AURA) under cooperative agreement 
with the National Science Foundation.} and astrometrically aligned and photometrically 
calibrated the images using SDSS stars in the field. We found a brightening point source in the WHT 
$z$-band images within the revised XRT error circle at the position RA(J2000) = 14$^{\rm h}$ 
17$^{\rm m}$ 08.82$^{\rm s}$, Dec(J2000) = +42$^{\circ}$ 08\arcmin 41.6\arcsec, with $z = 
23.5\pm0.3$\,mag\footnote{All magnitudes are in the AB system and not corrected for Galactic 
extinction, unless otherwise mentioned.} (at $\Delta t\approx0.04$\,d), $i \gtrsim 23.8$\,mag 
($3\sigma$), and $r \gtrsim 24.3$\,mag ($3\sigma$; Table \ref{tab:120521C:data:IR}).

Given the red color of the afterglow, $r-z \gtrsim 0.8$\,mag, we considered this to be a possible 
high redshift source, and thus triggered a sequence of optical and infrared imaging with 
the Gemini-North Multi-Object Spectrograph (GMOS) on Gemini-North ($iz$), the Low Resolution 
Imaging Spectrometer (LRIS) on the W.M.~Keck telescope ($gI$) and the Wide-Field Camera 
(WFCAM) on the United Kingdom Infrared Telescope (UKIRT; $JHK$). We reduced the data in the 
standard manner, using the instrument pipelines for GMOS and WFCAM. We performed aperture 
photometry using the Graphical Astronomy and Image Analysis tool (GAIA). We placed the aperture 
with reference to the GMOS $z$-band image with the highest signal-to-noise detection of the 
afterglow, and used an aperture size appropriate to the seeing FWHM. We determined the level and 
variance of the sky background from a large number of same-sized apertures placed on sky regions 
proximate to the burst location. We calibrated the optical photometry to SDSS and the $JHK$ 
photometry using 2MASS stars in the field.

\begin{figure}
\begin{tabular}{ccc}
\centering
 \includegraphics[width=0.30\columnwidth]{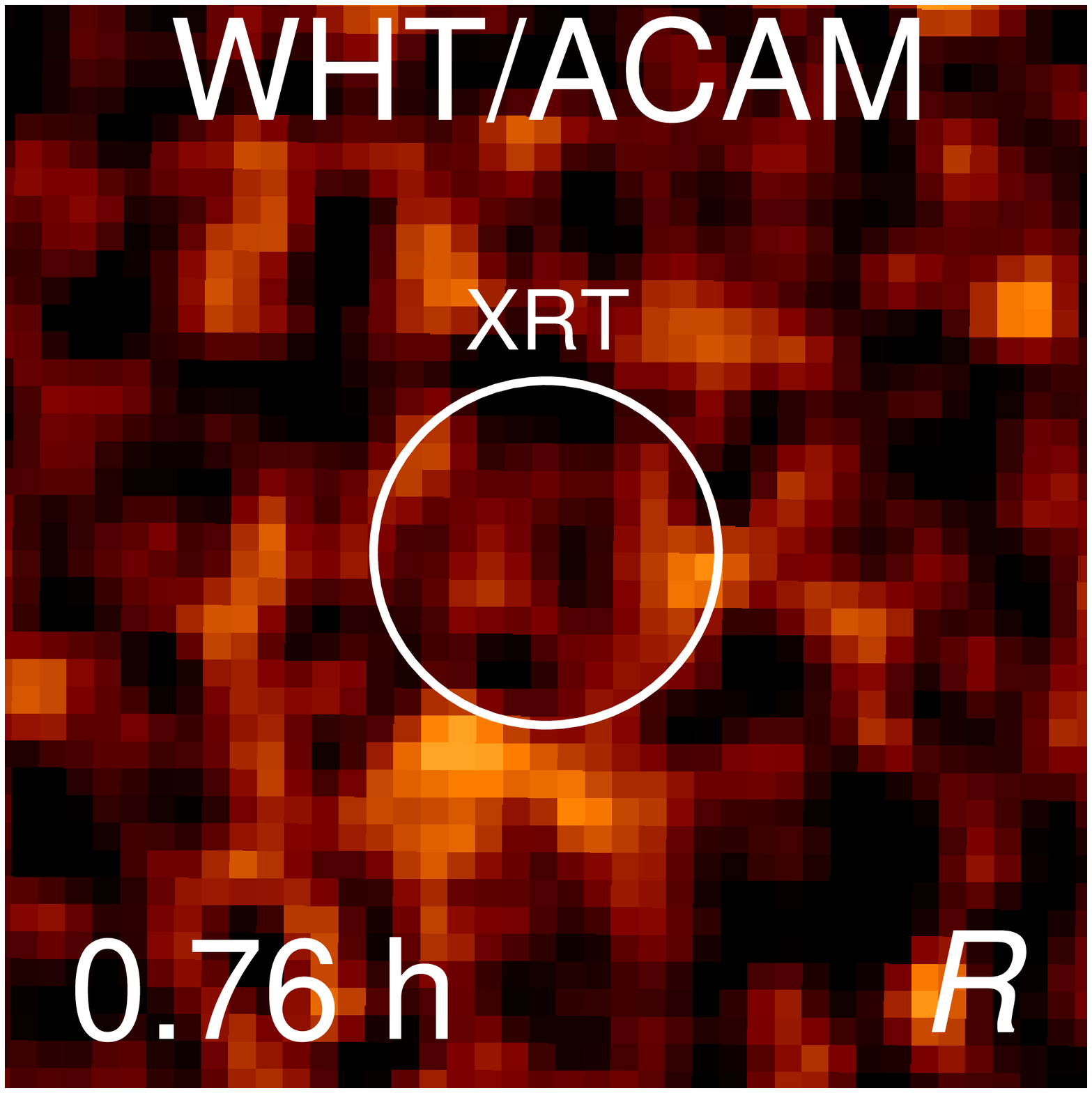} &
 \includegraphics[width=0.30\columnwidth]{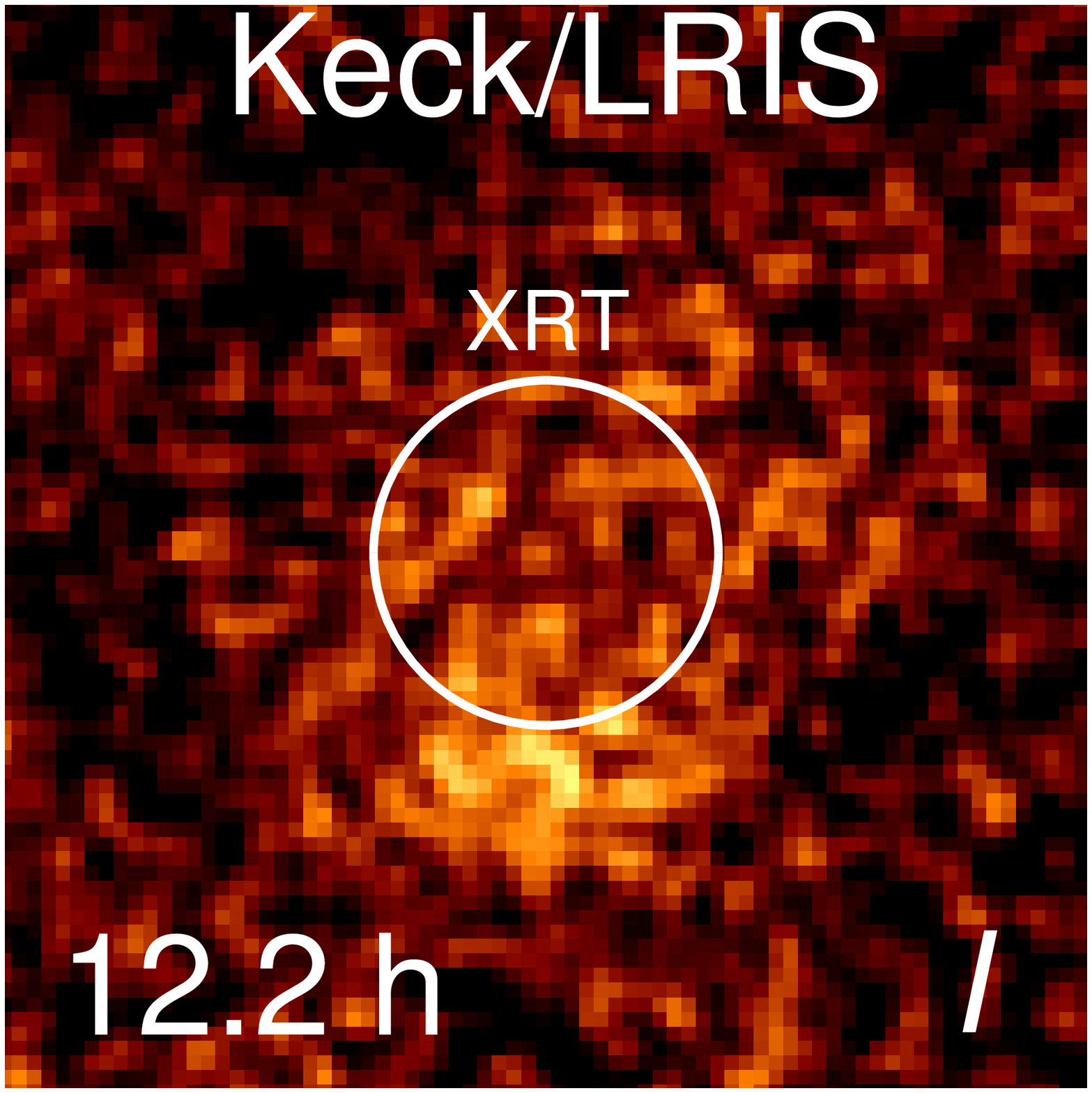} &
 \includegraphics[width=0.30\columnwidth]{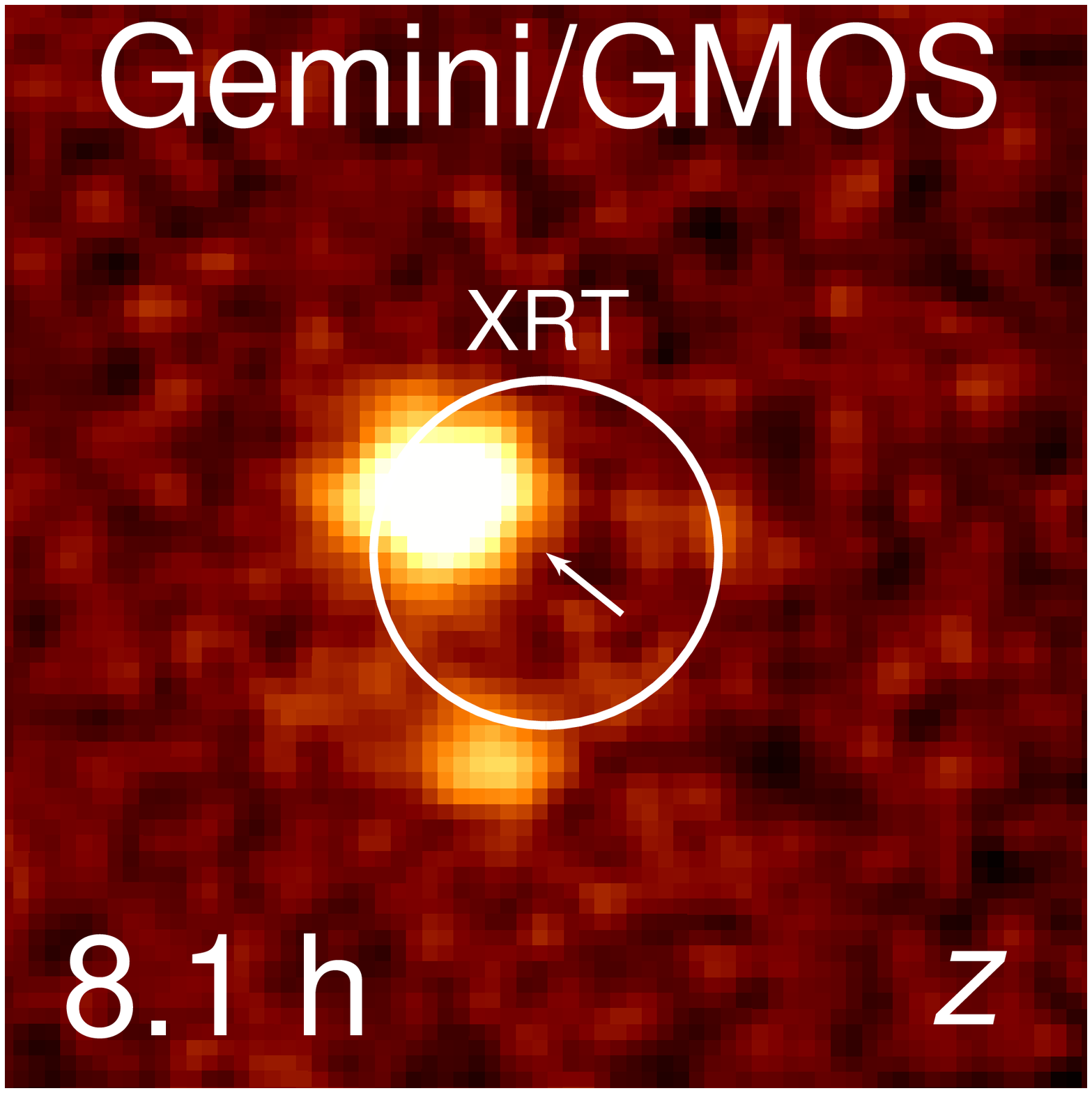} \\
 \includegraphics[width=0.30\columnwidth]{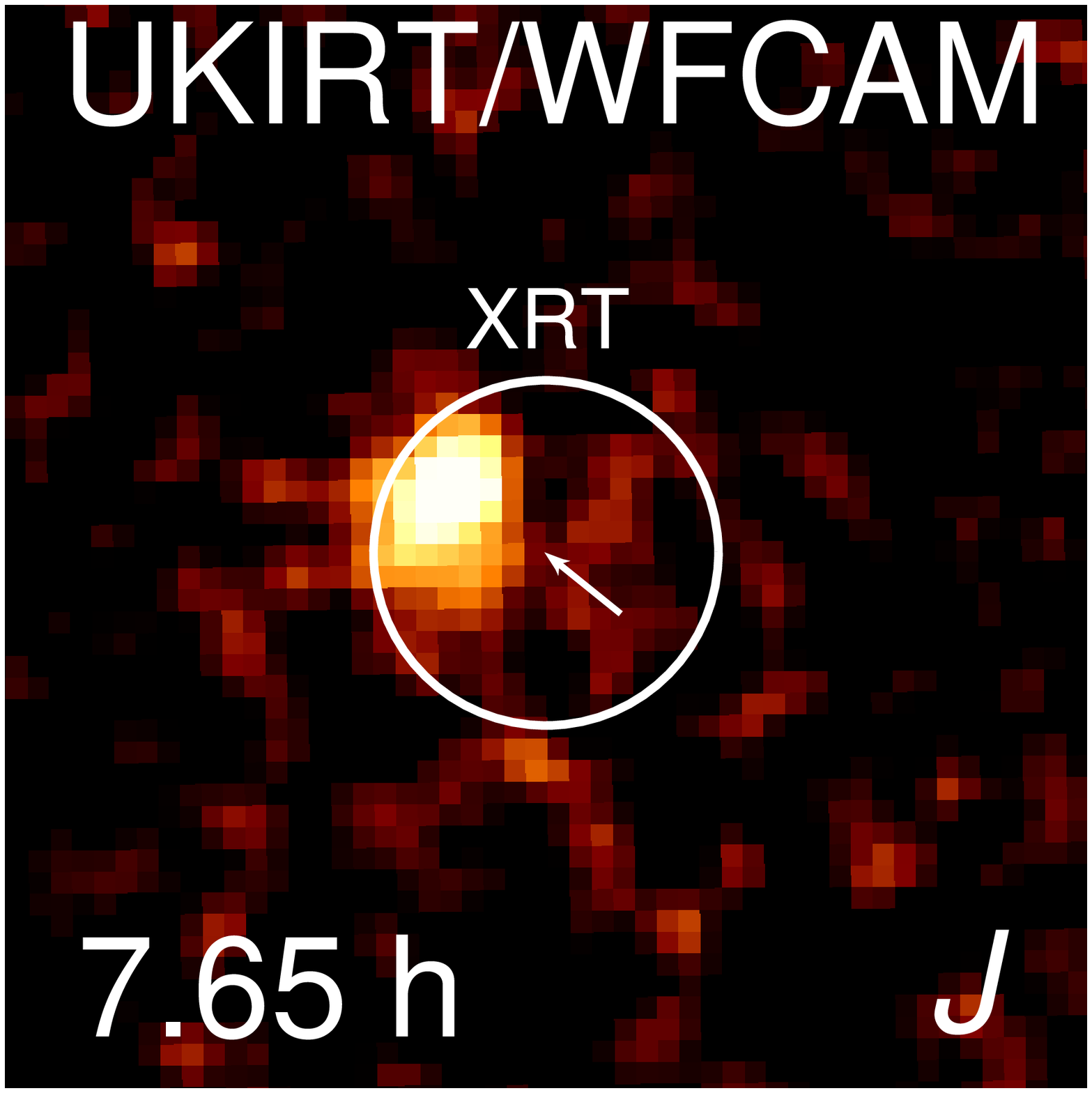} &
 \includegraphics[width=0.30\columnwidth]{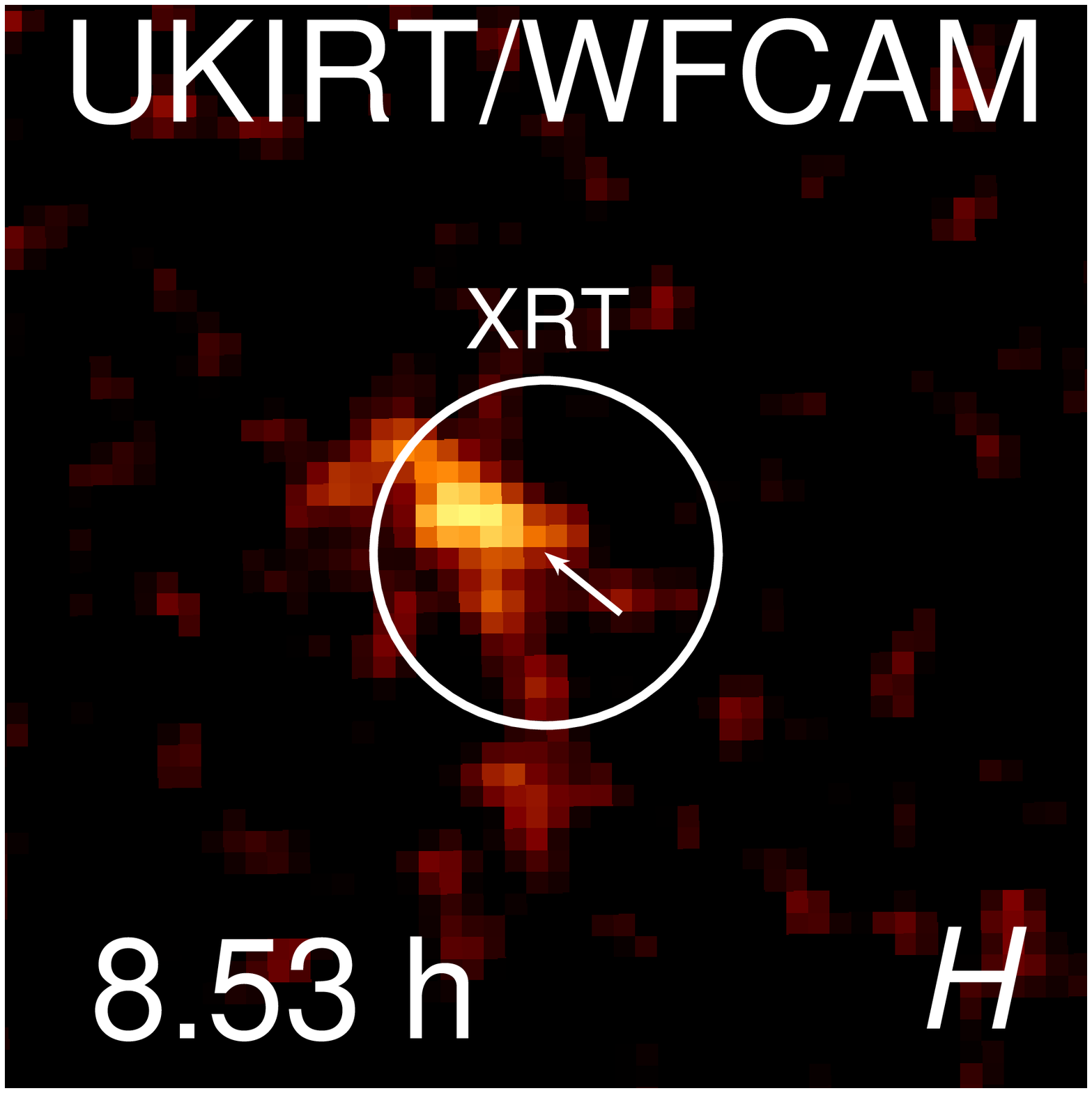} &
 \includegraphics[width=0.30\columnwidth]{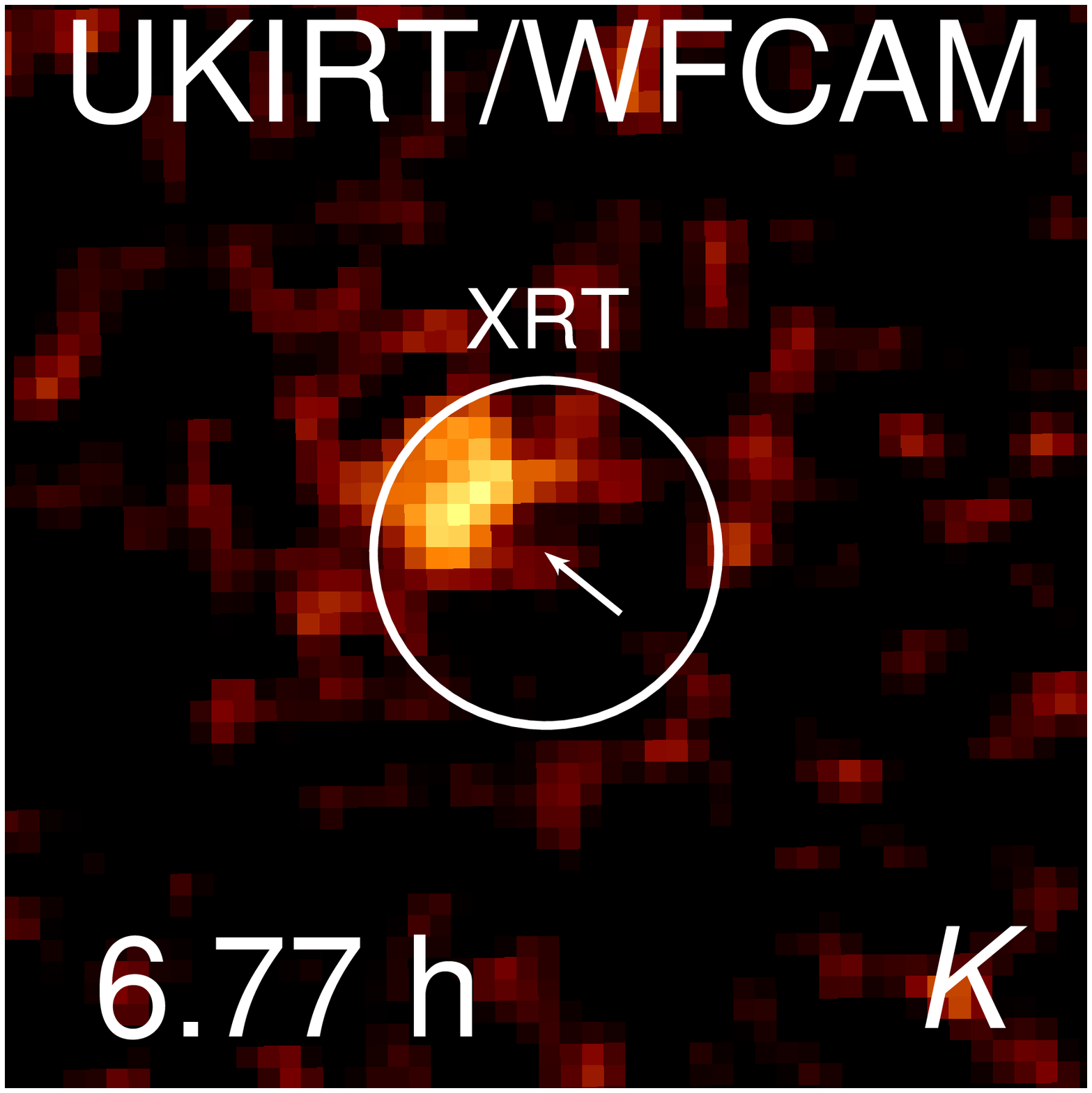} \\
\end{tabular}
\caption{Optical and near-infrared observations of GRB~120521C.
The refined XRT position is marked by the white circle ($1.6$\arcsec\ radius). The afterglow is 
detected in $z$-band with Gemini/GMOS and WHT/ACAM and in $JHK$ imaging with UKIRT/WFCAM (Table
\ref{tab:120521C:data:IR}), but is undetected at both $R$- and $I$-band.}
\label{fig:120521C_NIRim}
\end{figure}

We detected the afterglow in all filters redward of $z$-band, and obtained non-detections with deep
limits in the optical filters (\textit{gri}) at the level of $F_{\nu} \lesssim 0.45\,\mu$Jy 
(3$\sigma$; Figure \ref{fig:120521C_NIRim} and Table \ref{tab:120521C:data:IR}). On the other hand, 
the infrared colors were relatively blue: $J-H=0.13\pm0.21$\,mag and $J-K=0.12\pm0.21$\,mag. This 
suggested that reddening due to dust was negligible, and that the red $r-z$ color was due to the 
Ly$\alpha$ break falling within the $z$-band, implying a photometric redshift of $z\sim6$. We 
perform a full analysis to determine a photometric redshift in Section 
\ref{text:photoz}.

The \Swift\ UV/Optical Telescope (UVOT) began observing the field 77\,s after the burst. No optical
counterpart was detected at the location of the X-ray afterglow \citep{gcn13331}. We performed
photometry using the HEASOFT task \textsc{uvotsource} at the location of the NIR afterglow, and
report our derived upper limits in Table \ref{tab:120521C:data:uvot}.

\begin{figure}
\centering
 \includegraphics[angle=-90,width=\columnwidth]{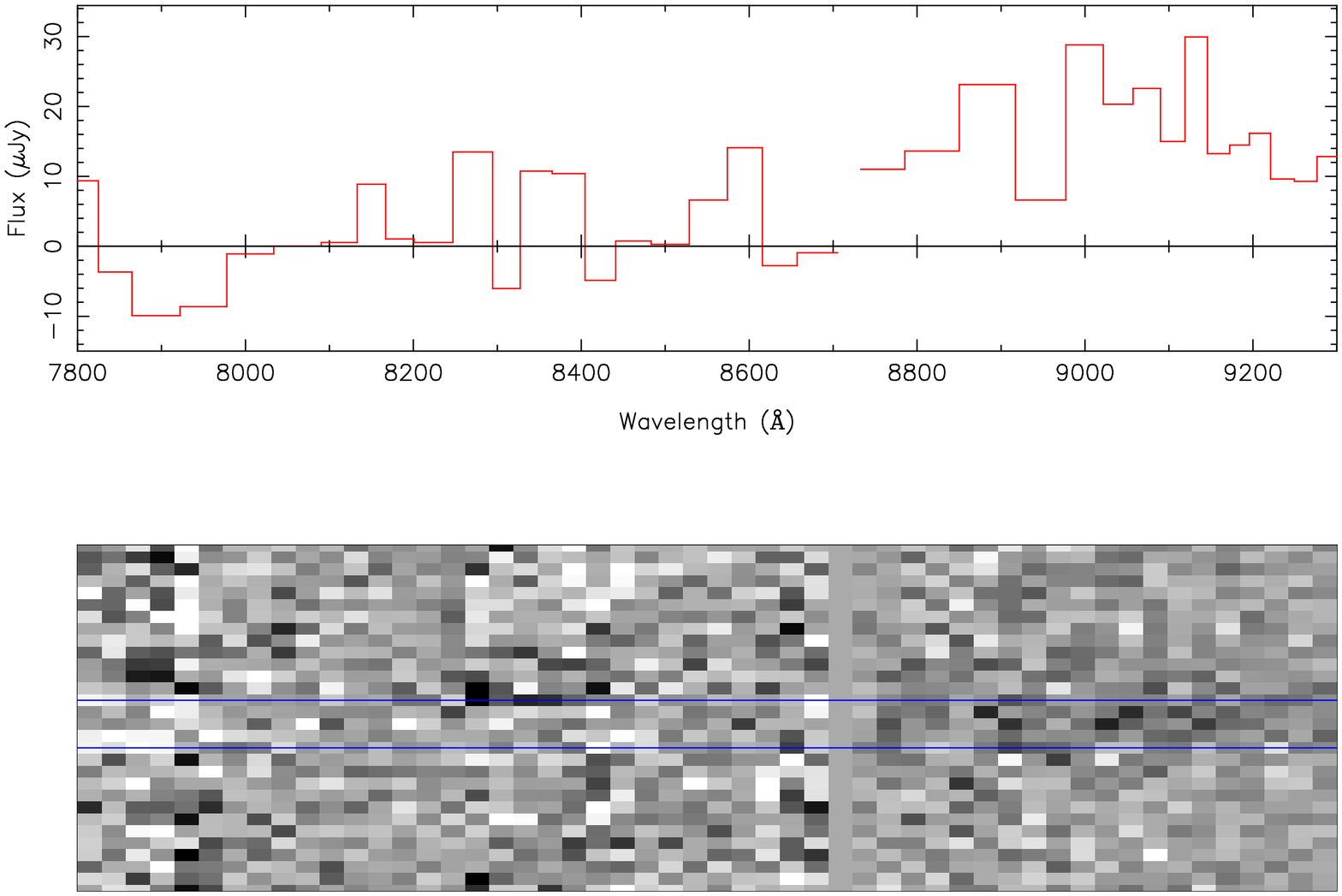}
\caption{1D (top) and 2D (bottom) Gemini-North/GMOS spectrum of GRB~120521C obtained 1.03\,d. 
after the burst. The blue box indicates the extraction region in the 2D spectrum, located using the 
trace of a reference star. The flux from the afterglow disappears blueward 8700\,\AA, coincident 
with a chip gap, and is weakly detected at redder wavelengths. Assuming this break is due to 
Ly$\alpha$, we find a redshift of $z\sim6.15$.}
\label{fig:120521C_spec}
\end{figure}

We obtained spectroscopic observations of the afterglow with Gemini-North/GMOS beginning 1.03\,d
post-burst for a total exposure of 3600\,s, by which time the source had faded to $z\approx23.2$
mag. We used the R400 grism and a slit width of 1\arcsec, providing a wavelength coverage of 
5850--10140\,\AA\ and a resolution of $R\approx1900$. The data were reduced using the GMOS pipeline. 
A faint trace of the afterglow was visible at the red end of the spectrum. The trace disappears 
around 8700\AA, which unfortunately coincides with the gap between the GMOS CCDs. Assuming this 
break is due to Ly$\alpha$, we deduce $z\approx6.15$, consistent with the red $r-z$ color. We plot 
the extracted spectrum in Figure  \ref{fig:120521C_spec}, adaptively re-binned to produce 
approximately the same noise in each bin.

\subsection{Radio}
We observed GRB\,120521C with the Karl G. Jansky Very Large Array (VLA) beginning on 2012 May 
22.12 UT at mean frequencies of 5.8~GHz (lower and upper sideband 
frequencies set at 4.9 and 6.7 GHz, respectively) and 21.8~GHz (lower and upper sideband 
frequencies of 19.1 and 24.4 GHz, respectively). We employed 3C286 as a flux and bandpass 
calibrator 
and interleaved observations of J1419+3821 repeatedly for calculating time-dependent antenna gains. 
All observations utilized the VLA WIDAR correlator \citep{pcb+11}. We excised radio frequency 
interference from the data, resulting in final effective bandwidths of $\approx$1.5 GHz at 5.8~GHz 
and $\approx$1.75 GHz at 21.8 GHz. We performed all data calibration and analysis with the 
Astronomical Image Processing System \citep[AIPS;][]{gre03} using standard procedures for VLA data 
reduction.

\begin{figure}
\begin{tabular}{cc}
\centering
 \includegraphics[width=0.23\columnwidth]{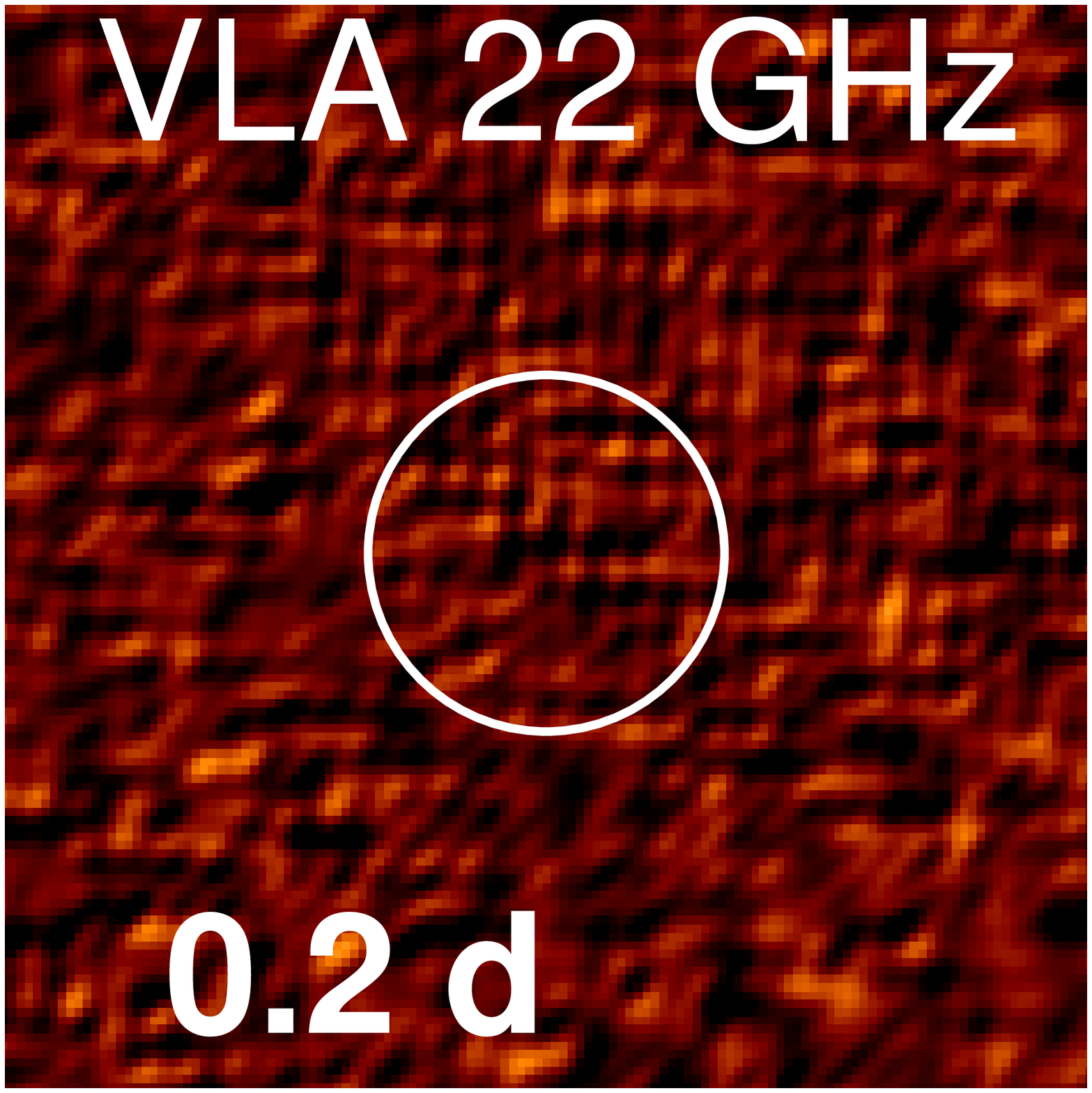} &
 \includegraphics[width=0.23\columnwidth]{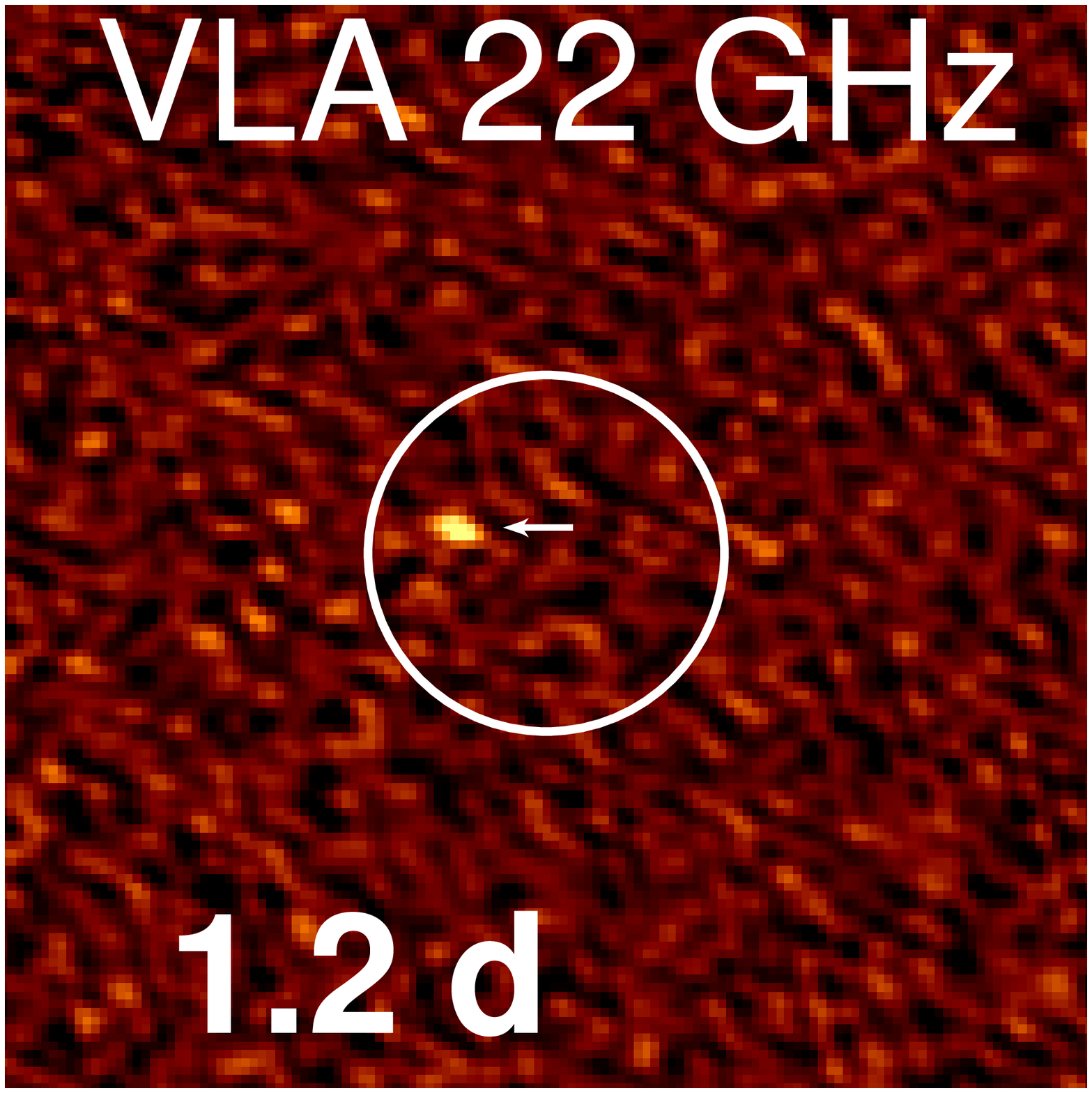} \\
 \includegraphics[width=0.23\columnwidth]{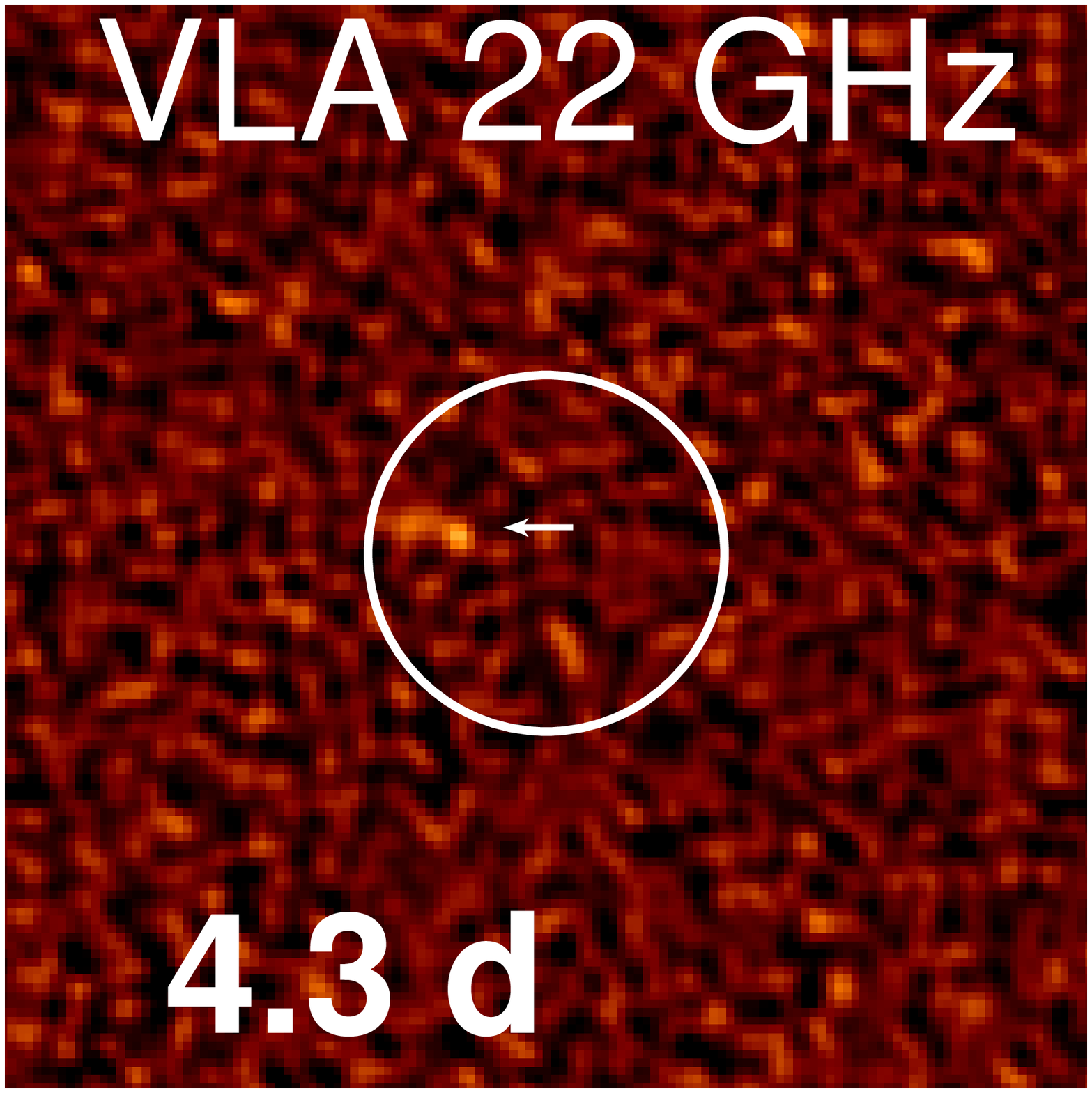} &
 \includegraphics[width=0.23\columnwidth]{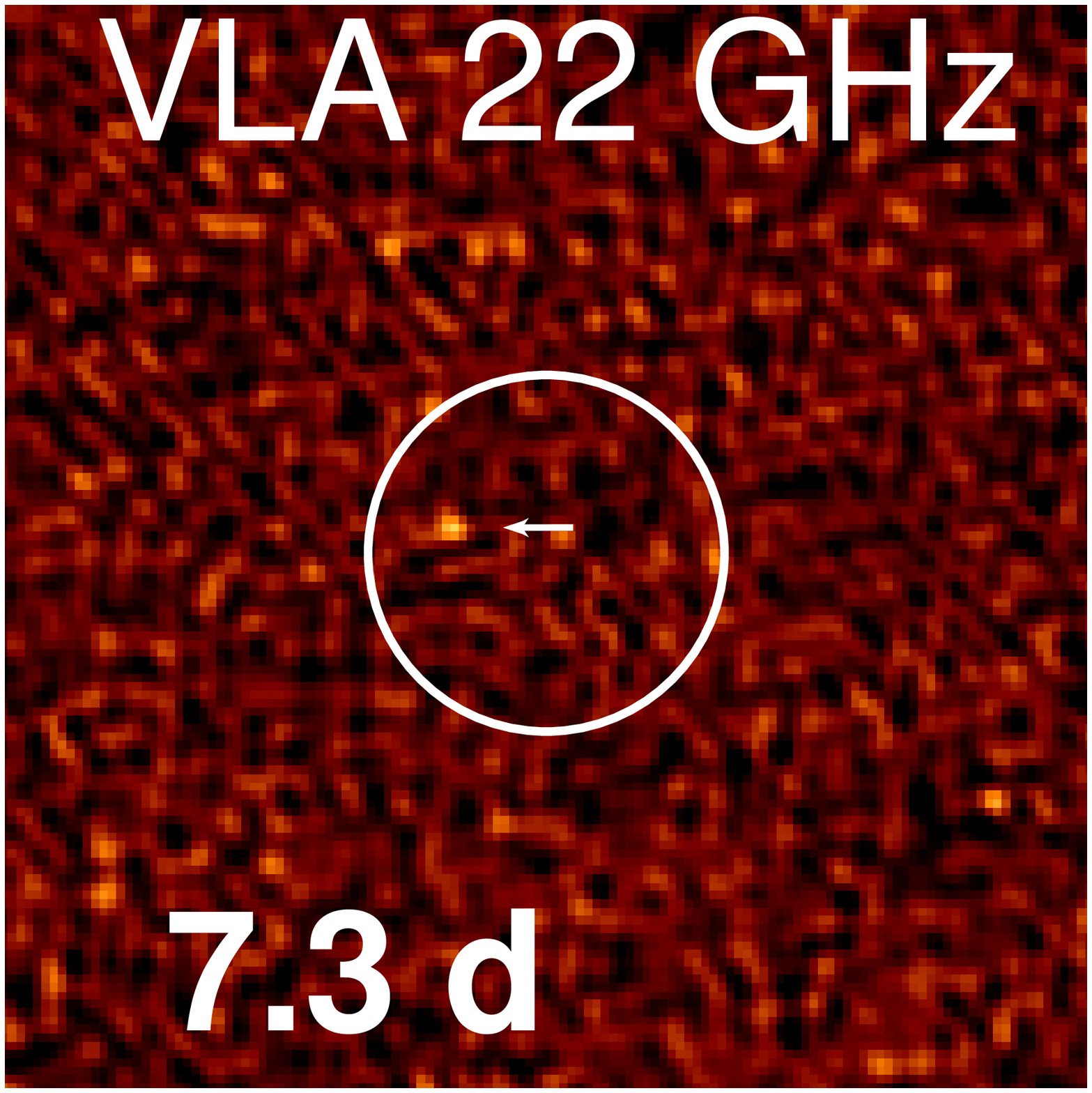} \\
 \includegraphics[width=0.23\columnwidth]{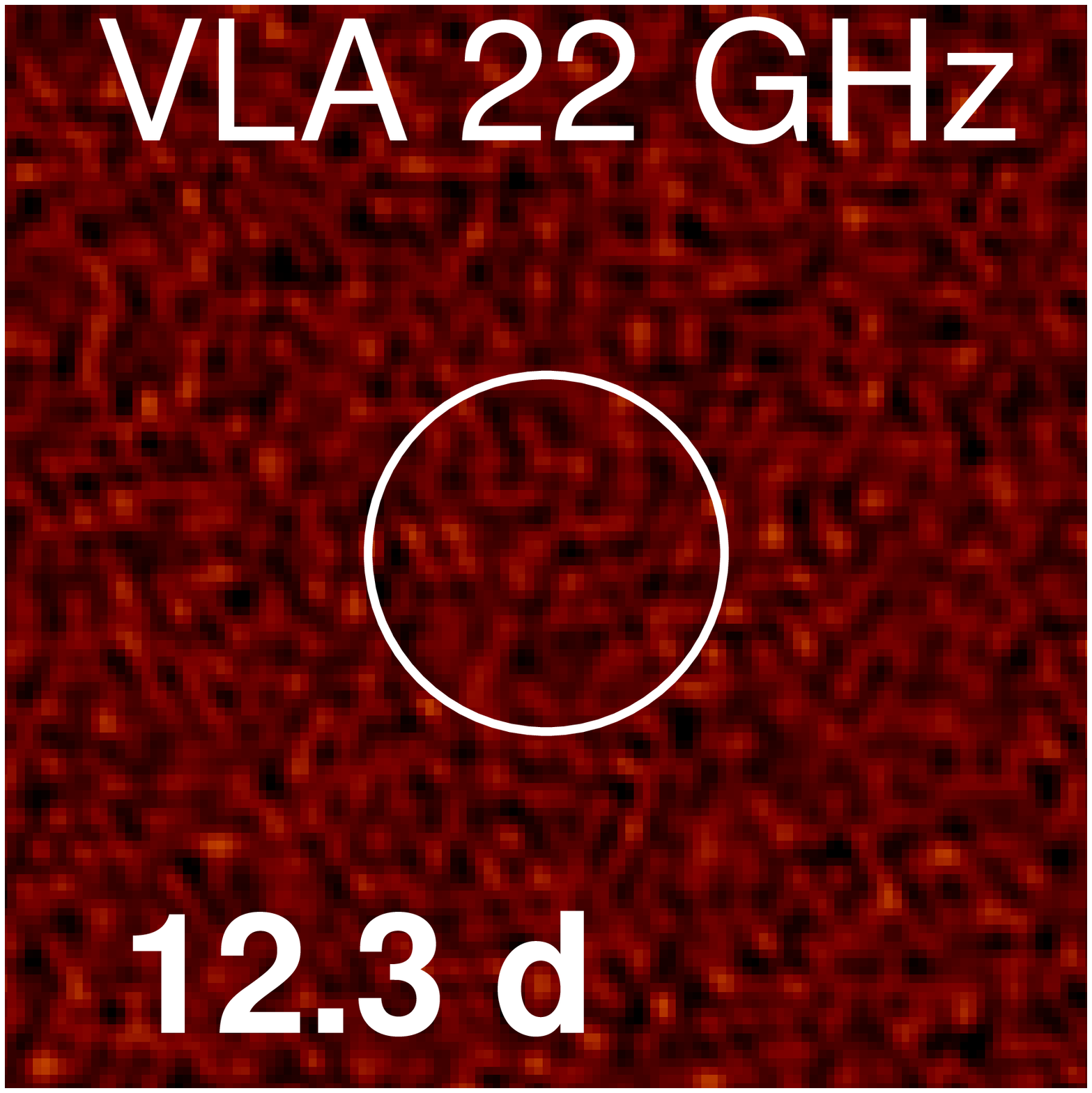} &
 \includegraphics[width=0.23\columnwidth]{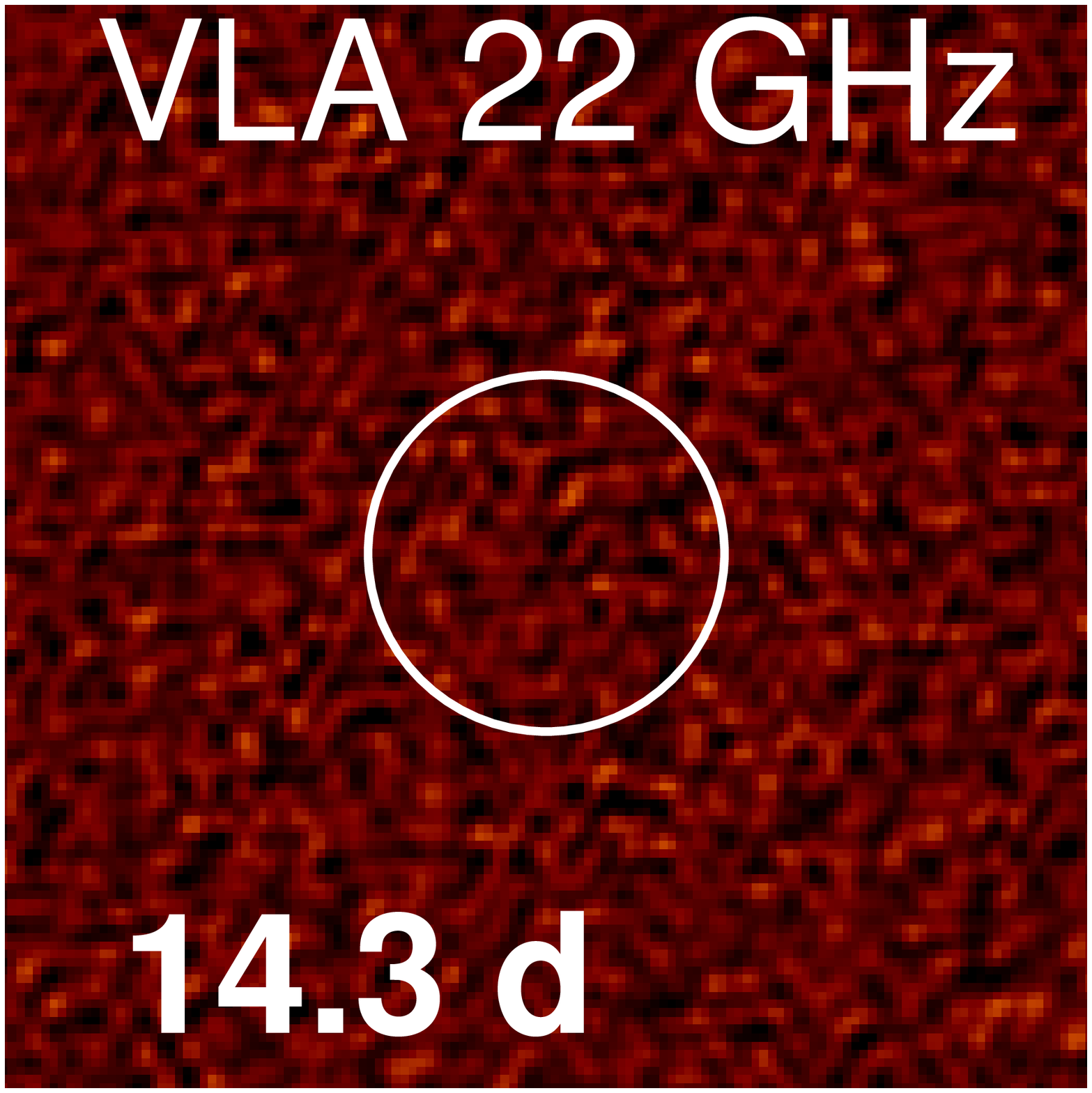} \\
 \includegraphics[width=0.23\columnwidth]{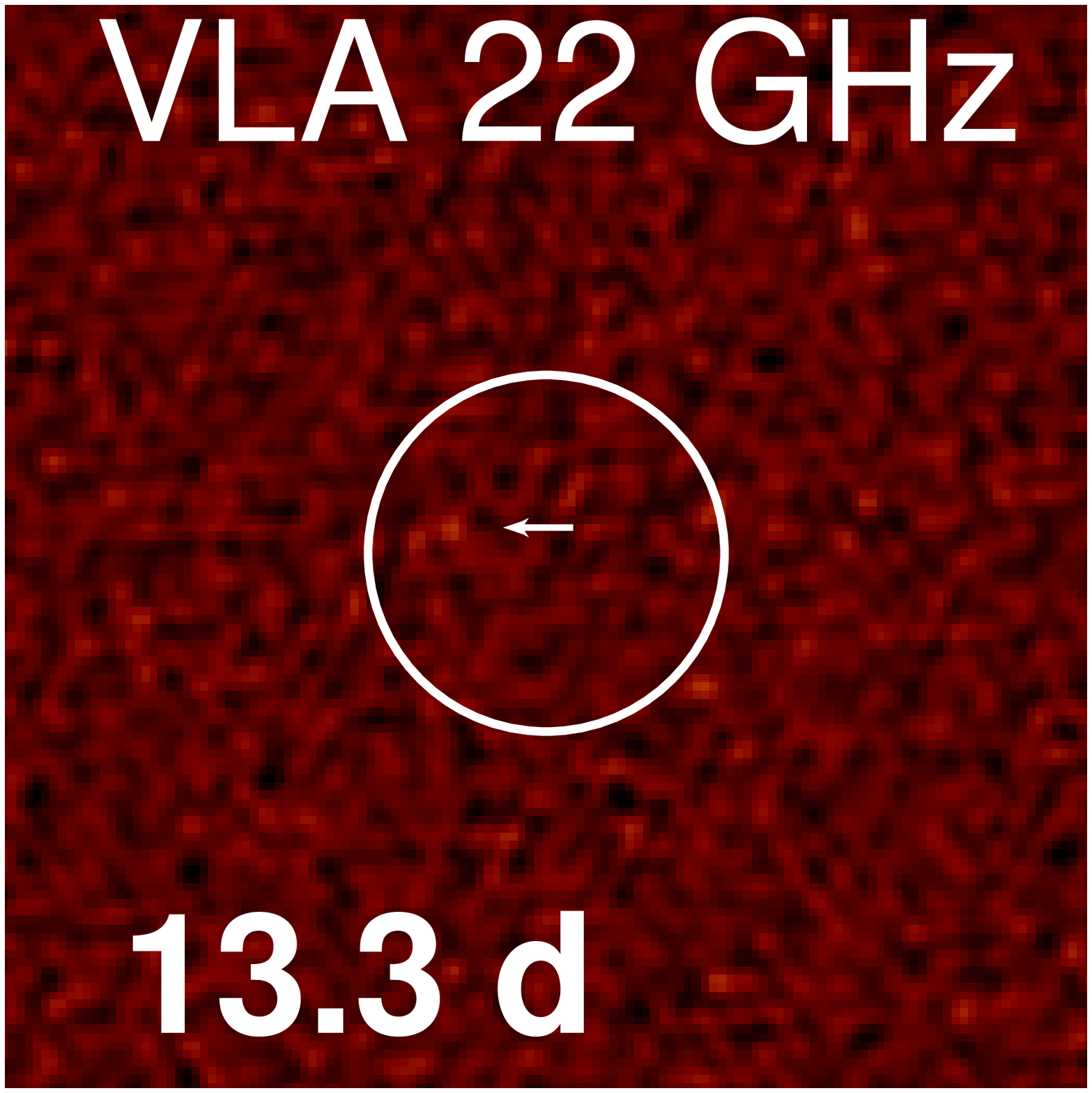} & \\
\end{tabular}
\caption{VLA observations of GRB~120521C at a mean frequency of 21.8\,GHz. The refined XRT position 
is indicated by the white circle (1.6\arcsec\ radius). The arrow marks the radio afterglow when 
detected. The last image is a stack of the data at 12.3 and 14.3\,d with a marginal detection at 
$\sim3\sigma$ (see Table \ref{tab:120521C:data:VLA} 
for details).}
\label{fig:120521C_radioim_K}
\end{figure}

\begin{figure}
\begin{tabular}{cc}
\centering
 \includegraphics[width=0.24\columnwidth]{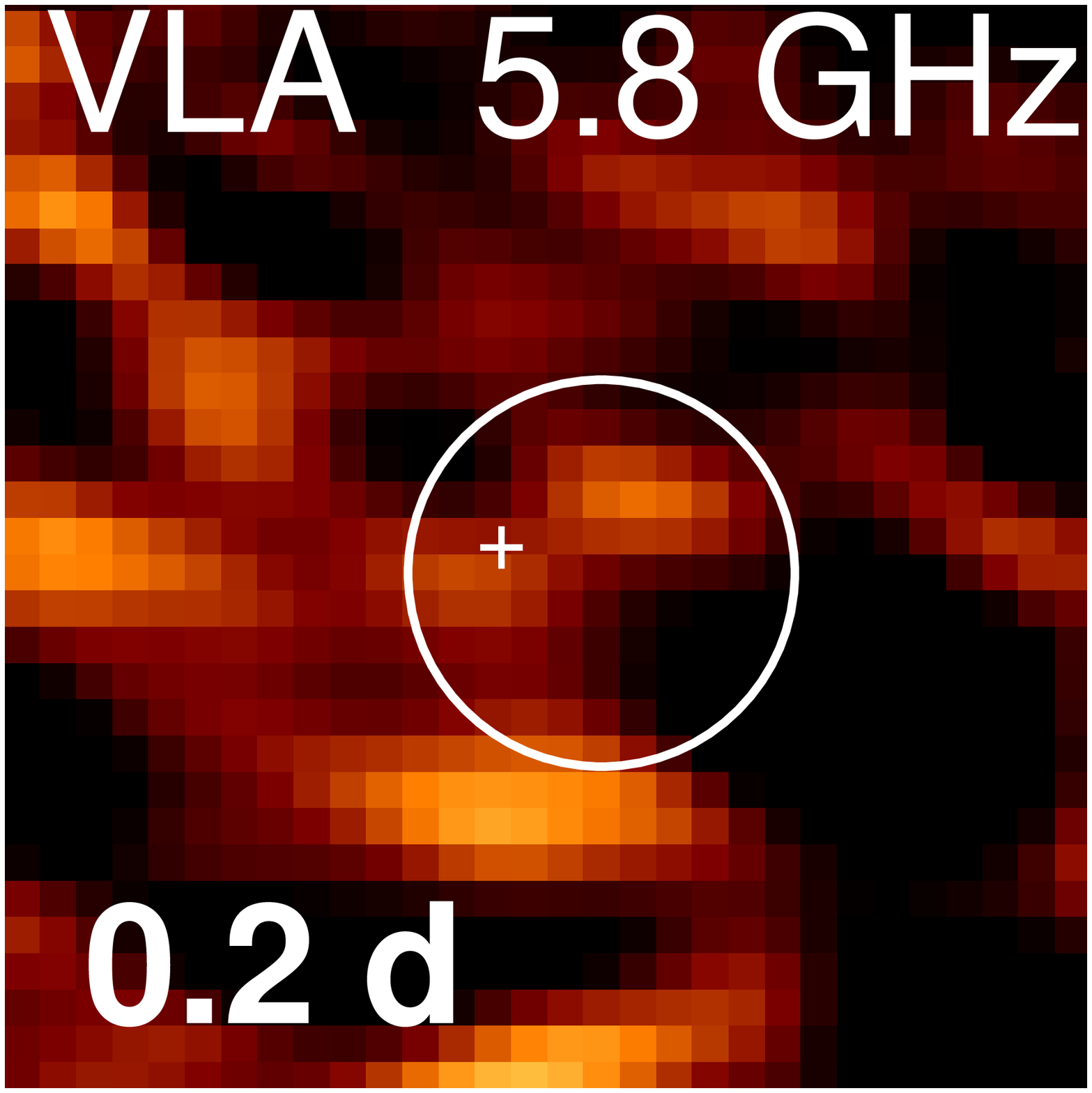} &
 \includegraphics[width=0.24\columnwidth]{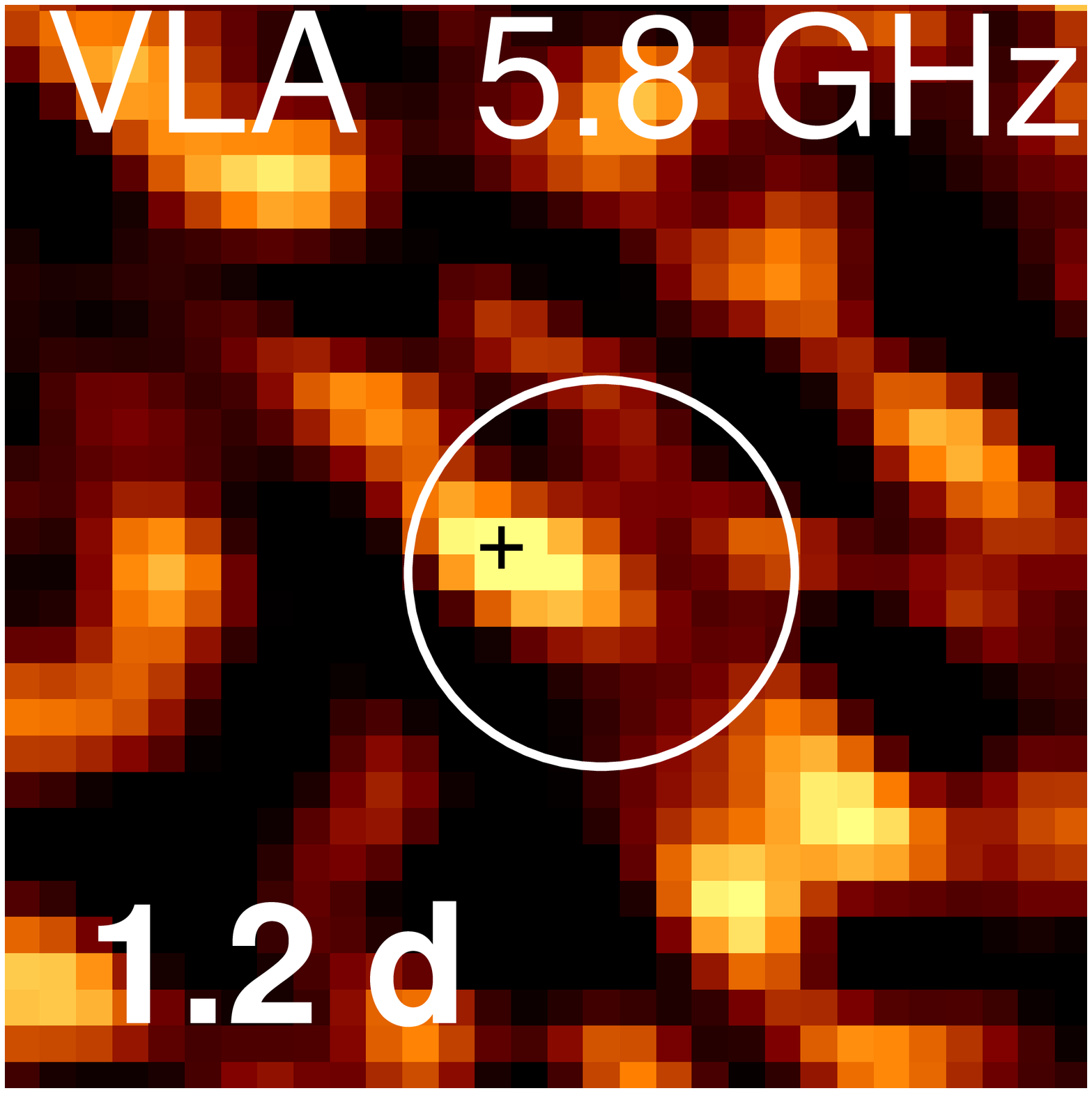} \\
 \includegraphics[width=0.24\columnwidth]{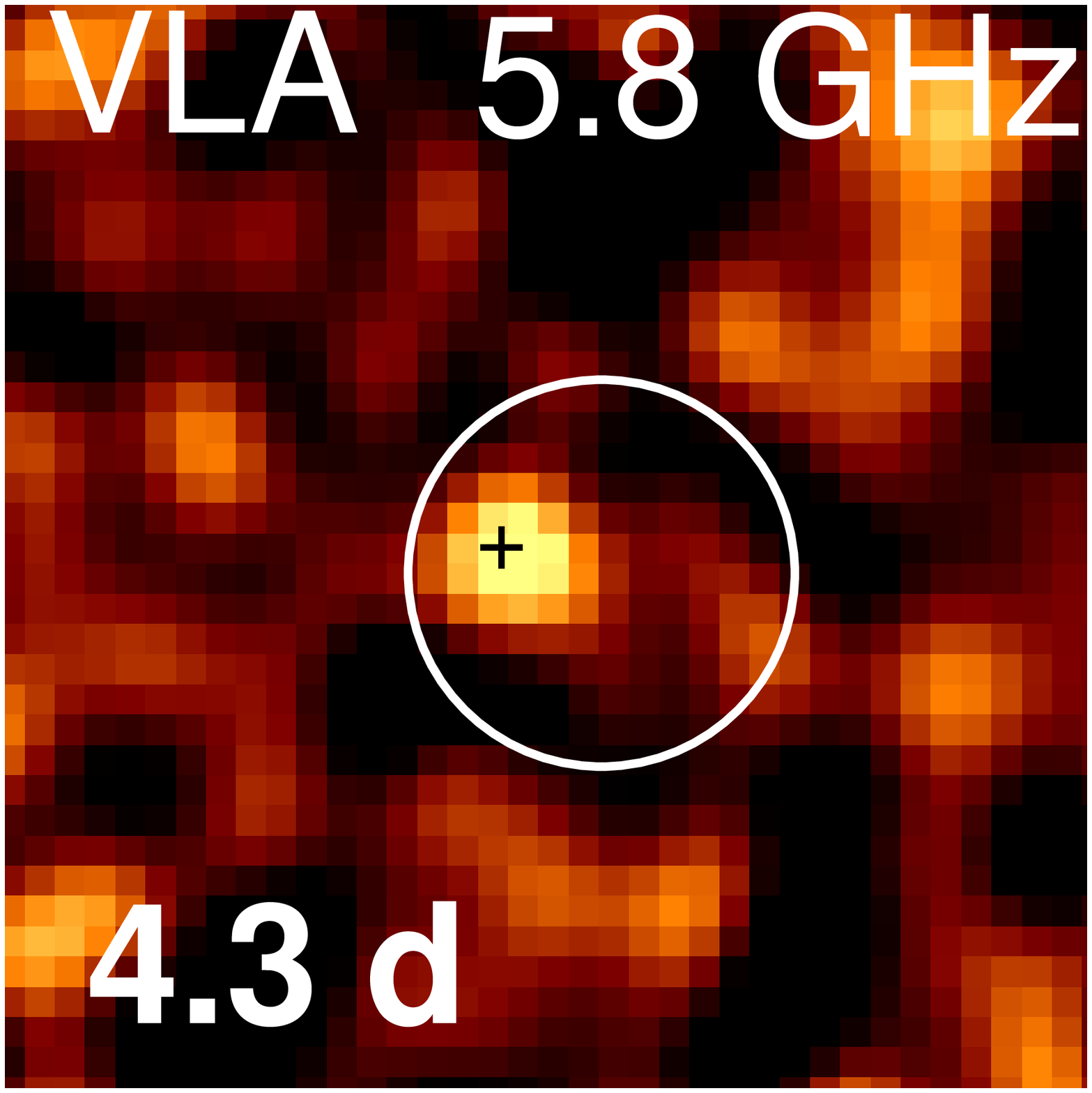} &
 \includegraphics[width=0.24\columnwidth]{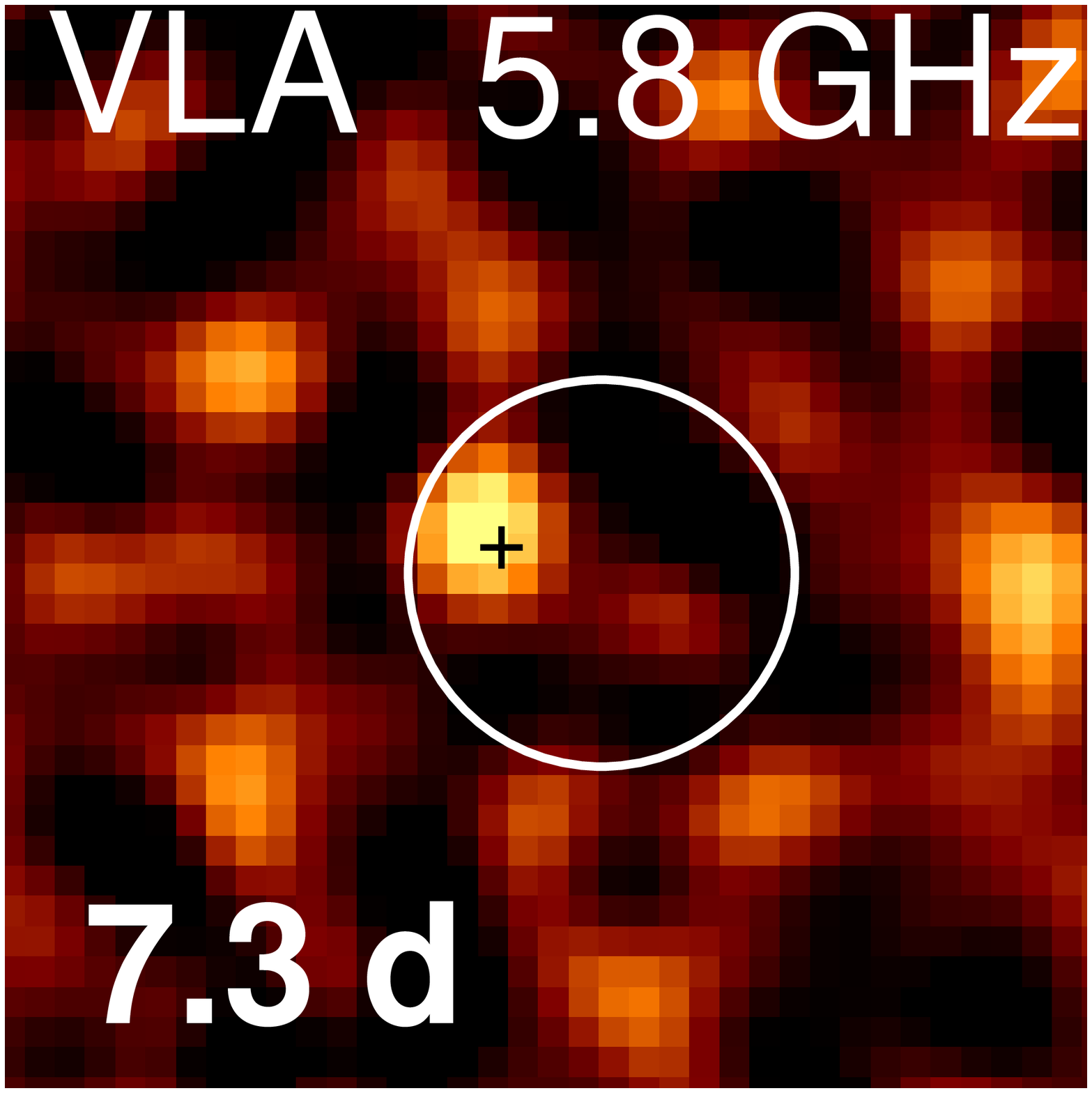} \\
 \includegraphics[width=0.24\columnwidth]{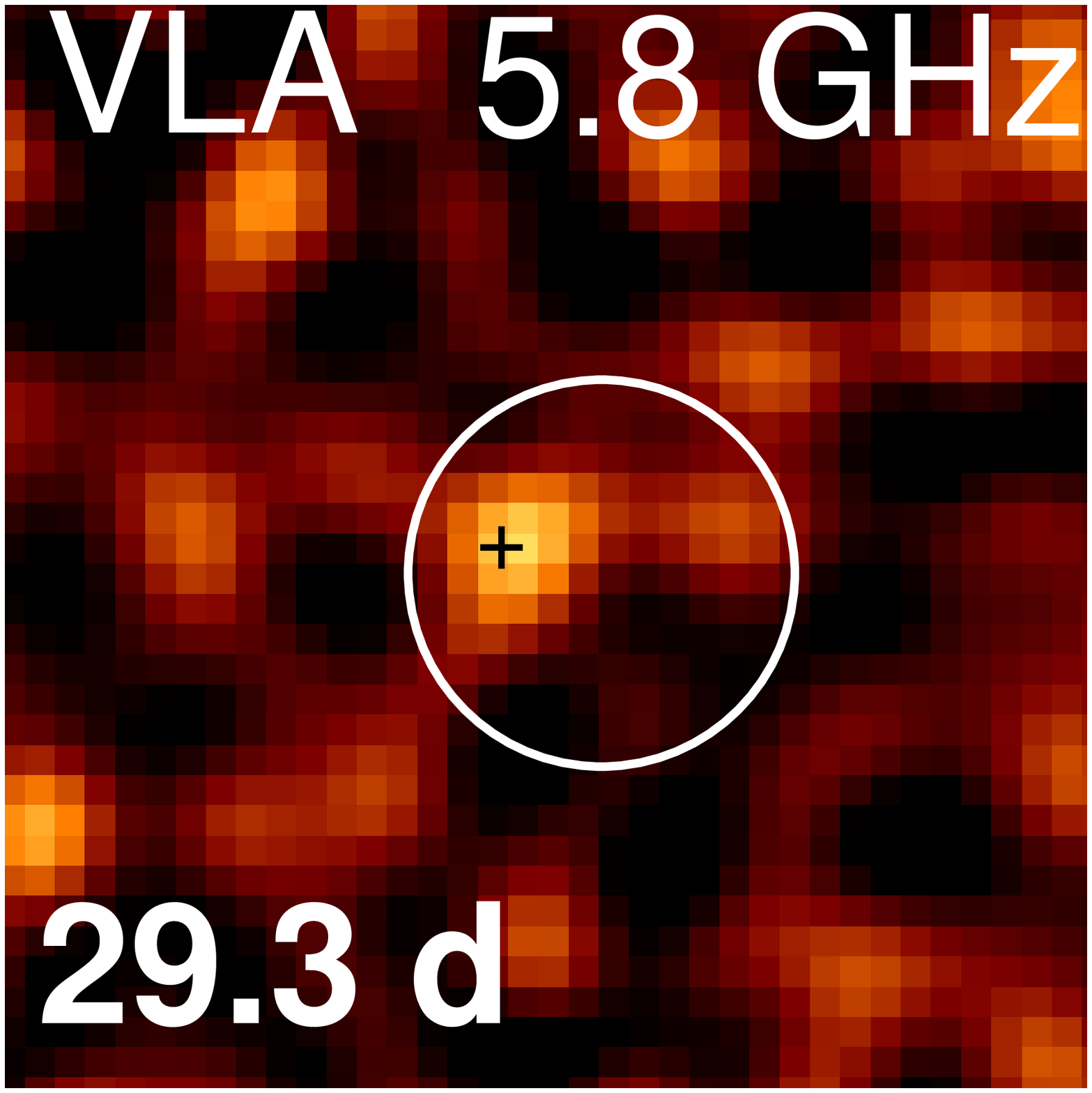} &
 \includegraphics[width=0.24\columnwidth]{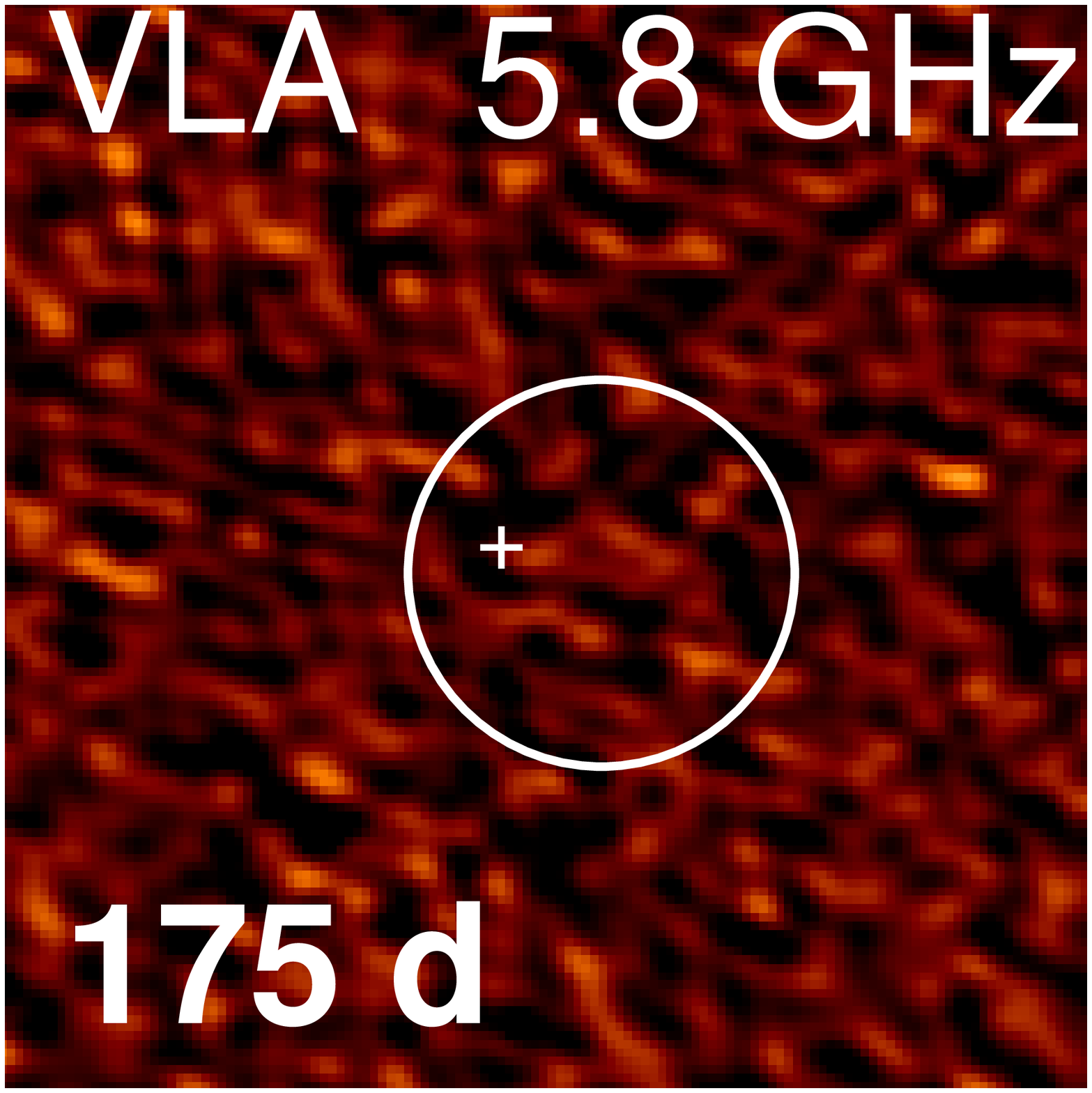} \\
\end{tabular}
\caption{VLA observations of GRB~120521C at a mean frequency of 5.8\,GHz. The refined XRT
position is marked by the white circle ($1.6$\arcsec\ radius). Crosses indicate the mean position 
of the GRB from our 21.8\,GHz observations (see Figure \ref{fig:120521C_radioim_K}).}
\label{fig:120521C_radioim_C}
\end{figure}

In our first epoch at 21.8~GHz (0.15\,d after the burst), we did not detect any significant radio 
emission within the refined \Swift\ XRT error circle to a $3\sigma$ limit of $50\,\mu$Jy (Table 
\ref{tab:120521C:data:VLA}). However, we detected a radio source in the second epoch at 1.15\,d 
after the burst (Figure \ref{fig:120521C_radioim_K}). This source subsequently faded, confirming it 
as the radio afterglow. We also detected the afterglow at 6.7~GHz in our observations taken between 
$4.25$ and $29.25$\,d after the burst; however, we did not find significant radio emission at 
4.9~GHz (Figure \ref{fig:120521C_radioim_C}). We treat these two side-bands separately in our 
analysis.

We used the AIPS task JMFIT to determine the positional centroid and integrated flux of the radio
afterglow in each epoch by fitting a Gaussian at the position of the source and fixing the
source size to the restoring beam shape. The weighted mean position of the source, determined by
combining all 21.8\,GHz detections is RA(J2000)  = 14$^{\rm h}$ 17$^{\rm m}$ 08.803$^{\rm s}$ $\pm$ 
0.002$^{\rm s}$, Dec(J2000) = +42$^{\circ}$ 08' 41.21" $\pm$ 0.03" ($1\sigma$).
We summarize the results of the radio observations in Table \ref{tab:120521C:data:VLA}.
GRB~120521C was also observed by the Arcminute Microkelvin Imager Large Array at 15.75\,GHz 
\citep[AMI-LA;][]{stf+13} and we include the reported upper limits in our analysis.

\section{Photometric Redshift}
\label{text:photoz}
\begin{figure}[ht]
\centering
\includegraphics[width=\columnwidth]{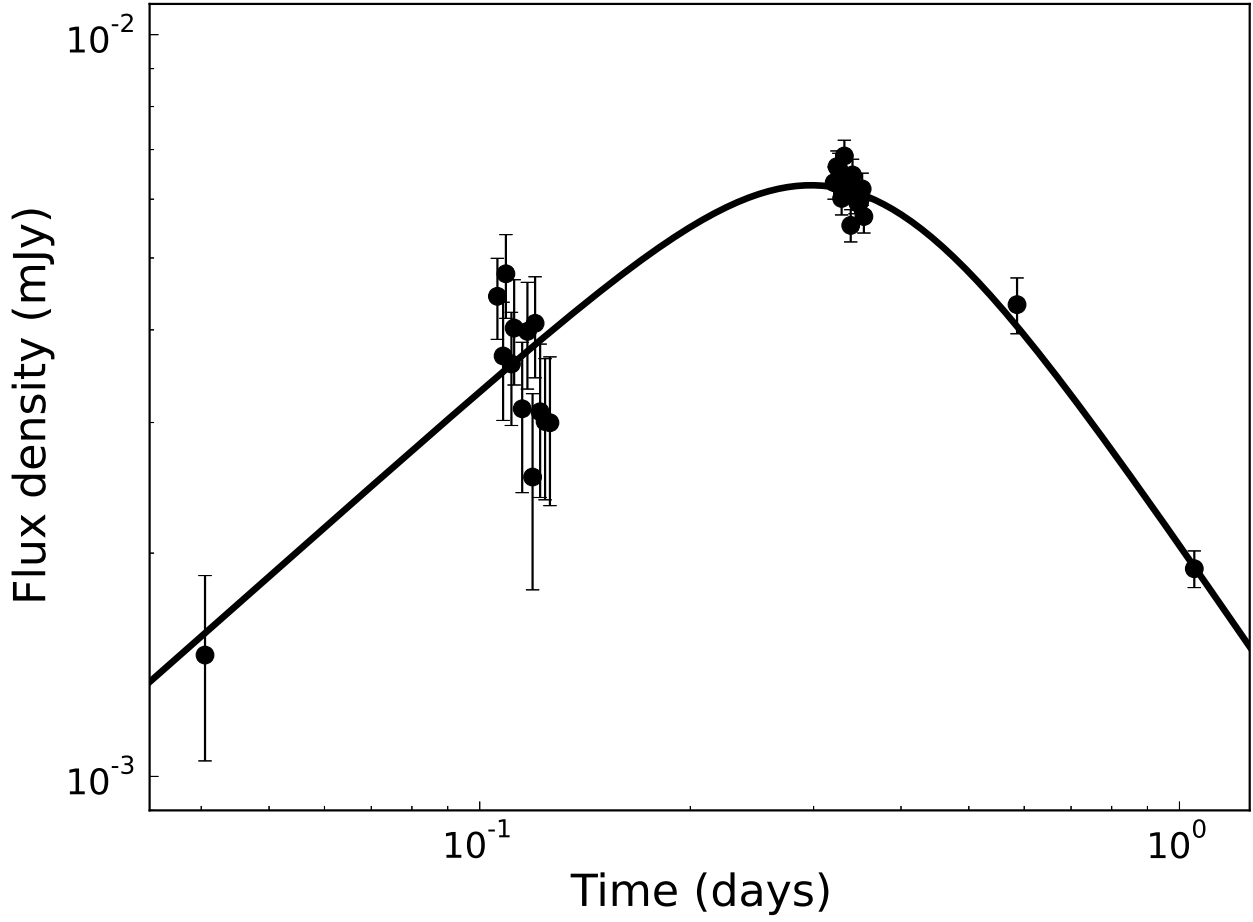}
\caption{$z$-band light curve of GRB~120521C. The solid line is the best-fit broken-power law
model described in Section \ref{text:zlc}.
\label{fig:120521C_zlcfit}}
\end{figure}

To determine a photometric redshift, we interpolate the optical and NIR observations to a
common time. To minimize this interpolation, we select a time of 8.1\,hr after the burst
when we obtained near-simultaneous $zJHK$ photometry. We perform a weighted sum of the GMOS $z$-band
observations at $7.7$\,hr $< \Delta t < $ $8.5$\,hr and find $F_{\nu}=6.22\pm0.05\,\mu$Jy at 
$\Delta t \approx 8.1\,$h. Since the NIR light curves are not well-sampled before $1$\,d, we use the
$z$-band light curve to extrapolate the NIR fluxes. We first fit the $z$-band light curve with
a broken power-law of the form
$F_{\nu} = F_{\rm b} \left(
\frac{(t/t_{\rm b})^{-s\alpha_1}+(t/t_{\rm b})^{-s\alpha_2}}{2}
\right)^{-1/s}$, where $t_{\rm b}$ is the break time, $F_{\rm b}$ is the flux at the break time, 
$\alpha_1$ and $\alpha_2$ are the temporal decay rates before and after the break, respectively, 
and $s$ is the sharpness of the break\footnote{We impose a floor of 5\% on the uncertainty of each 
data point, as explained in Section \ref{text:modelling}.}. We use the Python function 
\texttt{curve\_fit} to estimate these model parameters and the associated covariance matrix. Our 
best-fit parameters are: $t_{\rm b} = (0.34\pm0.07)\,\rm{d}$, $F_{\rm b}=6.89\,\mu$Jy, $\alpha_1 = 
0.83\pm0.31$, $\alpha_2 = -1.38\pm0.43$, and $s = 1.7\pm1.6$ (Figure \ref{fig:120521C_zlcfit}). 
Using this model to extrapolate the $JHK$ photometry, we obtain $F_{\nu}=11.1\pm1.1\,\mu$Jy, 
$12.8\pm1.4\,\mu$Jy, and $12.4\pm\,1.3\,\mu$Jy, at $J$, $H$, and $K$ band, respectively, at the 
common time of $8.1$\,hr. The uncertainties are statistical only and do not include the systematic 
uncertainties introduced by the interpolation, which are less than $2\%$.

After obtaining NIR fluxes at a common time, we build a composite model for the afterglow 
SED. We use a sight-line-averaged model for the optical depth of the intergalactic medium 
(IGM) as described by \citet{mad95}, accounting for Ly$\alpha$ absorption by neutral hydrogen along 
the line of sight and photoelectric absorption by intervening systems. We also include Ly$\alpha$ 
absorption by the host galaxy, for which we assume a column of 
$\log{(N_{\rm H}/{\rm cm}^{-2})}=21.1$, the mean value for GRBs at $z\sim1$ \citep{fjp+09}. 
The free parameters in our model are the redshift of the GRB, the extinction along the line of 
sight within the host galaxy (\AV), and the spectral index ($\beta$) of the afterglow SED, $F_{\nu} 
\propto \nu^{\beta}$. In order to not bias our results, we assume a flat prior for the redshift 
and the extinction. We further use the distribution of extinction-corrected spectral slopes, 
$\beta_{\rm ox}$ from \citet{gkk+11} as a prior on $\beta$. We use a Markov Chain Monte Carlo 
(MCMC) algorithm to explore the parameter space, integrating the model over the filter bandpasses 
and computing the likelihood of the model by comparing the resulting fluxes with the observed 
values. Details of our MCMC implementation are described in Section \ref{text:mcmc}.

We find $z=5.93^{+0.11}_{-0.14}$, $\beta=-0.16^{+0.34}_{-0.25}$, and $A_{\rm
V}=0.11^{+0.22}_{-0.10}$ mag, where the uncertainties correspond to $68\%$ credible intervals about
the median\footnote{Credible intervals are summary statistics for posterior density functions and 
are Bayesian analogues to the `confidence intervals' used in frequentist statistics. In this 
article, we use credible intervals based on percentiles of the posterior density, defined such that 
the probability of the parameter lying below and above the interval are equal. Such an interval 
includes the median of the posterior density by construction.}. The parameters of the 
highest-likelihood model are $z=6.03$, $\beta = -0.34$, and
$A_{\rm V} = 0$\,mag, consistent with the $68\%$ credible intervals derived from the posterior 
density functions (Table \ref{tab:120521C_sedfit}). We note that the median values differ from the 
highest-likelihood values. This is a standard feature of Monte Carlo analyses whenever the 
likelihood function is asymmetric about the highest-likelihood point. In this case, this occurs 
because the extinction is constrained to be positive, resulting in a truncation of parameter space. 
The best-fit model and a model with the median parameters are plotted in Figure 
\ref{fig:120521C_photoz}, while the full posterior density function for the redshift is shown in 
Figure \ref{fig:120521C_zhists}. We can rule out a redshift of $z\lesssim5.6$ at $99.7\%$ 
confidence. The corresponding 99.7\% confidence upper limit is $z \lesssim 6.2$.

We note that this constraint on the redshift relies on the assumed prior for $\beta$.
Using broad-band modeling we can locate the synchrotron break frequencies (explained in
the next section) and thereby constrain $\beta$ independent of the redshift. Therefore, in the 
subsequent multi-wavelength modeling we leave the redshift as a free parameter 
and fit for it along with the parameters of the explosion. For the optical and NIR frequencies, we
integrate the model over the filter bandpasses to take into account absorption by the intervening
IGM and the ISM of the host galaxy.

\section{Multi-wavelength Modeling}
\label{text:modelling}

\subsection{Synchrotron Model}
\label{text:synchrotron}
In the standard synchrotron model of GRB afterglows, the spectral energy distribution consists 
of multiple power-law segments delineated by `break-frequencies', namely the synchrotron cooling 
frequency (\nuc), the typical synchrotron frequency (\numax), and the self-absorption frequency 
(\nua). The location and evolution of these break frequencies, and the overall normalization of the 
spectrum depend upon the physical parameters of the explosion: the energy ($E_{\rm K,iso}$), the 
circumburst density (\dens, or the normalized mass-loss rate in a wind environment, $A_*$), the 
power-law index of the electron energy distribution ($p$), the fraction of the blastwave energy
transferred to relativistic electrons (\epse) and to the magnetic fields (\epsb), and 
the half-angle of the collimated outflow ($\theta_{\rm jet}$). For further details of the 
synchrotron model, see \citet{spn98}. 

We have developed Python software for broad-band modeling of GRB afterglows. Our software 
implements the full afterglow model with smoothly-connected power law segments presented in 
\citet[][henceforth GS02]{gs02}. The model includes synchrotron cooling and self-absorption for 
both ISM and wind-like environments. The full treatment of the synchrotron model including local 
electron cooling results in five different spectral regimes with 11 definitions of the break 
frequencies, corresponding to different orderings of the synchrotron frequencies. Depending on 
the circumburst density profile and the combination of physical parameters, the spectrum evolves 
from fast cooling ($\nu_{\rm c} < \nu_{\rm m}$) to slow cooling ($\nu_{\rm c} > \nu_{\rm m}$), 
transitioning through the various spectral regimes (Figure 2 in GS02). 

Given a set of explosion parameters, we compute the location of each of the 11 break frequencies 
using the expressions in GS02. Owing to slightly different normalizations of the break frequencies 
between the five spectral regimes, a sharp transition from one spectrum to another sometimes 
introduces discontinuities in the light curves. This is exacerbated by the fact that the transition 
times between spectra are not uniquely defined (see Table 3 in GS02). To overcome this and to 
establish a consistent framework, we add a linear combination of all spectra through which the 
spectrum evolves for a given set of physical parameters, with time-dependent weights. These weights 
are chosen such that each spectrum dominates in its own regime of validity, while allowing for the 
light curves to remain smooth when break frequencies cross each other at spectral transitions. 
A detailed description of our weighting scheme is provided in appendix \ref{text:weights}.

The hydrodynamics presented in GS02 assume spherical expansion. While this is a
good approximation in the early phase of the afterglow evolution when the Lorentz factor of the 
ejecta is $\Gamma \gg \theta_{\rm jet}$ and only a small fraction of the jet is visible to an 
observer on Earth, deceleration of the jet to $\Gamma \lesssim \theta_{\rm jet}$ results in a steep 
decline in the observed flux density at all frequencies at later times. We account for this `jet 
break' by changing the evolution of the break frequencies after the break time, \tjet, using the
prescription in \citet{sph99}, smoothing over the transition with a smoothing parameter
\footnote{We arbitrarily set $s = 5$ for the jet break, the precise value having negligible impact 
on derived physical parameters.} (for further discussion of the jet break based on numerical 
simulations, see \citealt{vem12} and \citealt{lvdhvew13}).

Our software also accounts for possible contributions in the optical and NIR from the host galaxy, 
as well as absorption and reddening of the afterglow light by dust in the host. For the former, 
we add the contribution of the host to the model afterglow light curve and fit for the flux density 
of the host in each waveband separately \footnote{Wherever light curves do not show any signature 
of flattening at late times, or when the last data point in a light curve is a deep non-detection, 
we assume the host flux is negligible and set it to zero to avoid biasing the model.}. For the 
latter, we use the Small Magellanic Cloud (SMC) extinction curve from \citet{pei92} and fit for the 
$B$-band extinction in the rest frame of the host galaxy. We use the optical $B$-band rather than 
$V$-band to normalize our model, since the extinction curves of \citet{pei92} are normalized in 
$B$-band. We find that using a Large Magellanic Cloud extinction model does not significantly 
affect the derived value of $A_{\rm B}$ and we therefore use the SMC model throughout for 
consistency. We convert $A_{\rm B}$ to $A_{\rm V}$ using $A_{\rm V} = 0.83 A_{\rm B}$
\citep{pei92}.

Radio observations can be strongly affected by scintillation, particularly at low frequencies
(below $\sim15$\,GHz). We account for scintillation in our modeling by calculating the modulation
index (the expectation value of the rms fractional change in flux density) in the direction of
the source and adding the expected flux variation in quadrature to the measured uncertainty. The
details of our method are described in Appendix \ref{app:scintillation}.

We note that several observations, particularly those in the optical/NIR, have high signal-to-noise 
ratios approaching $\sim50$, implying photometry precise to the $\sim 2\%$ level. However, the 
relative calibration of different instruments is generally not expected to be better than about
$5$\%. In addition, the synchrotron model is by its nature a simplification of a complex physical 
process and we therefore cannot expect the model to accurately represent the data at the 
$\lesssim5\%$ level. To account for this source of systematic uncertainty, we enforce a floor of 
$5\%$ on the reported uncertainties prior to fitting.

To determine the best-fit model, we compute the likelihood function using a Gaussian error model. 
The likelihood function for a data set comprised of both detections and non-detections is given by 
\citep[e.g.][]{law02,hel05}
\begin{equation}
L = \prod p(e_i)^{\delta_i} F(e_i)^{1-\delta_i}%\\
\end{equation}
where $e_i$ are the residuals (the difference between the measurement or
3$\sigma$ upper limit and the predicted flux from the model), $\delta_i$ is an
indicator variable (equal to 0 for an upper limit and 1 for a detection),
$p(e_i)$ is the probability density function of the residuals, and $F(e_i)$ is
the cumulative distribution function of the residuals, equal to
$\mathrm{Prob}(e_i\le t)$ for a limit $t$.
For a Gaussian error model,
\begin{equation}
p(e_i) = \frac{1}{\sqrt{2\pi}\sigma}e^{-\nicefrac{e_i^2}{2\sigma_i^2}},
\end{equation}
where $\sigma_i$ are the measurement uncertainties, while
\begin{equation}
F(e_i) = \frac{1}{2}
\left[1+erf\left(\frac{e_i}{\sqrt{2}\sigma_i}\right)\right], 
\end{equation}
where $erf(x)$ is the error function.
We determine the best-fit parameters by maximizing the likelihood function using sequential least 
squares programming tools available in the Python SciPy package \citep{scipy}.

\subsection{Markov Chain Monte Carlo}
\label{text:mcmc}
To fully characterize the likelihood function over a broad range of parameter space and to obtain a
Bayesian estimate for the posterior density function of the free parameters (leading to estimates
for uncertainties in and correlations between the derived parameters), we carry out a Markov Chain 
Monte Carlo (MCMC) analysis using the Python-based code \emcee\ \citep{fhlg12}. By implementing
an affine-invariant MCMC ensemble sampler, \emcee\ works well for both highly-anisotropic 
distributions, and distributions with localized regions of high likelihood \citep{gw10}. This is 
especially useful in high-dimensional problems such as the one presented here, where traditional 
MCMC methods spend large amounts of time exploring regions of parameter space with low likelihoods. 
MCMC analyses also allow us to uncover degeneracies in the model parameters, which are
present whenever some of the properties of the synchrotron spectrum (e.g.,~\nua) are not
well-constrained.

We note that the parameters \epse\ and \epsb\ are generally not expected to be larger than 
their equipartition values of $\nicefrac{1}{3}$. Accordingly, we truncate the priors for these 
parameters at an upper bound of $\nicefrac{1}{3}$. In addition, we sometimes find degeneracies in 
the models that result in large probability mass being placed at extremely high energies $E_{\rm 
K, iso, 52}\gtrsim10^3$ and low densities $n_{\rm 0} \lesssim 10^{-6}\,{\rm cm}^{-3}$. To keep the 
solutions bounded, we restrict the prior on the isotropic-equivalent kinetic energy to $E_{\rm K, 
iso, 52} < 500$.

For our MCMC analysis, we set up between 100 and 10,000 Markov chains (depending on the complexity 
of the problem) with parameters tightly clustered around the best-fit parameters determined using 
least squares minimization. We run the ensemble sampler until the average likelihood across the 
chains reaches a stable value and discard the initial period as `burn-in'.
We plot the marginalized posterior density for all parameters and check for
convergence by verifying that the distributions remain stable over the length of the
chain following burn-in \footnote{When plotting histograms of the logarithm of a quantity, we
transform the width of the bins appropriately such that the height of the bin is
equal to the value of the posterior density.}. Since the distributions frequently exhibit long
tails, we employ quantiles (instead of the mean or mode) to compute summary
statistics and quote $68\%$ credible regions around the median. We also provide the values of the 
parameters corresponding to the highest likelihood (``best-fit'') solution for completeness. 
However, the parameter values comprising the ``best-fit'' solution need not (and frequently do not) 
individually correspond to the modes of their respective marginal probability density functions. 

\section{Broad-band model for GRB~120521C}
\label{text:results}
We employ the model and fitting algorithm described in Section \ref{text:modelling} to determine 
the properties of GRB~120521C. The X-ray light curve displays a steep decline before $\sim0.01$\,d, 
followed by a plateau phase extending to 0.25\,d, neither of which can be described by the standard 
paradigm of the Blandford-McKee model \citep{bm76}. Such behavior is ubiquitous in the X-ray light 
curves of GRBs \citep[e.g.][]{nkg+06, mzb+13} and is usually attributed to the high-latitude 
component of the prompt emission (\citealt{kp00,wggo10}) and energy injection (\citealt{nkg+06, 
zfd+06,dsg+11}), respectively. The models we employ only account for the emission from the 
afterglow blastwave shock, and we therefore only utilize X-ray data after 0.25\,d in the broad-band 
fit.

In addition, the $z$-band light curve exhibits a peak at $\sim 8$\,hr. with a flux
density of $\approx7\,\mu$Jy. If we interpret this peak as the passage of \numax\ through the 
$z$-band, then \numax\ should pass through 21.8\,GHz at $\approx200$\,d (evolving as $t^{-3/2}$, 
before a jet break) or at the very earliest around $40$\,days (evolving as $t^{-2}$, if we assume 
that a jet break occurred at $8$ hours). In addition, the peak flux in the radio must be less than 
(in the wind model) or equal to (in the ISM model) the peak flux in optical/NIR. However, the 
$22$\,GHz radio light curve peaks before 10\,d and all the radio observations are at a higher flux 
level than all of the optical and NIR detections. Thus, the optical/NIR and radio light curves are 
not compatible under the assumption that \numax\ passes through $z$-band at 8\,hr. 
We therefore do not include the $z$-band data before 0.25\,d in our broad-band fit. We return to 
the point of the X-ray and $z$-band light curves before 0.25\,d in Section \ref{text:zlc}.

We find that an ISM model adequately explains all observations after $\sim 0.25$\,d (Figure 
\ref{fig:120521C_multimodel_ISM}). The spectrum remains in the slow cooling phase throughout, with 
the standard ordering of the synchrotron frequencies (\nua\ $<$ \numax\ $<$ \nuc) and with a peak 
flux density of $F_{\nu,\rm m}\approx132\,\mu$Jy. At $\Delta t = 1$\,d, the synchrotron break 
frequencies are located at $\nu_{\rm m}\approx5.5\times10^{11}$\,Hz and 
$\nu_{\rm c}\approx1.2\times10^{16}$\,Hz. The self-absorption frequency lies below the frequencies 
covered by our radio observations, $\nu_{\rm a} \lesssim 5$\,GHz and is therefore not fully 
constrained. Correspondingly, the physical parameters \epse, \epsb, \dens, and $E_{\rm K,iso}$ 
exhibit degeneracies, with the unknown location of \nua\ being the dominant source of uncertainty 
(Figure \ref{fig:120521C_ISM_mcmcgrid}). Using the values of \numax, \nuc\ and $F_{\nu,\rm{max}}$ 
from our best-fit model and the functional dependence of the microphysical parameters, \epse, \epsb, 
\dens, and \E\ on the measured quantities \nua\, \numax, \nuc, and $F_{\nu,\rm{max}}$, we derive 
the following constraints: 
$\epsilon_{\rm e}\approx 0.15{\nu_{\rm a, 9}}^{\nicefrac{5}{6}}$,
$\epsilon_{\rm B}\approx 4.0\times10^{-3}{\nu_{\rm a, 9}}^{\nicefrac{-5}{2}}$,
$n_{\rm 0}\approx 0.44{\nu_{\rm a, 9}}^{\nicefrac{25}{6}}\,\rm{cm}^{-3}$, and
$E_{\rm K, iso, 52}\approx 6.7{\nu_{\rm a, 9}}^{\nicefrac{-5}{6}}$, 
where ${\nu_{\rm a, 9}}$ is the self-absorption frequency in units of $10^{9}$\,Hz. Imposing
the restriction that $\epsilon_{\rm e}$ be less than its equipartition value of $\nicefrac{1}{3}$, 
we can further restrict the self-absorption frequency to $\nu_{\rm a} \lesssim 
2.7\times10^{9}\,$Hz. 
This allows us to place an \textit{upper bound} on the circumburst density, $n_{\rm 0} \lesssim 
27\,{\rm cm}^{-3}$, and \textit{lower bounds} on the isotropic equivalent energy, $E_{\rm K, iso, 
52} \gtrsim 2.9$ and $\epsilon_{\rm B} \gtrsim 3.5\times10^{-4}$. Similarly, imposing 
$\epsilon_{\rm 
B} < \nicefrac{1}{3}$, we can place \textit{lower bounds} on the self-absorption frequency,
$\nu_{\rm a} \gtrsim 1.7\times10^{8}\,$Hz, the circumburst density, $n_{\rm 0} \gtrsim 
2.8\times10^{-4}\, {\rm cm}^{-3}$, and $\epsilon_{\rm e} \gtrsim 3.4\times10^{-2}$, and an
\textit{upper bound} on the isotropic equivalent energy, $E_{\rm K, iso, 52} \lesssim 29$. The 
parameters corresponding to the highest likelihood models are presented in Table \ref{tab:bestfit} 
and the complete results of the Monte Carlo analysis are summarized in Table \ref{tab:mcmc}.

\begin{figure}[ht]
\centering
\includegraphics[width=\columnwidth]{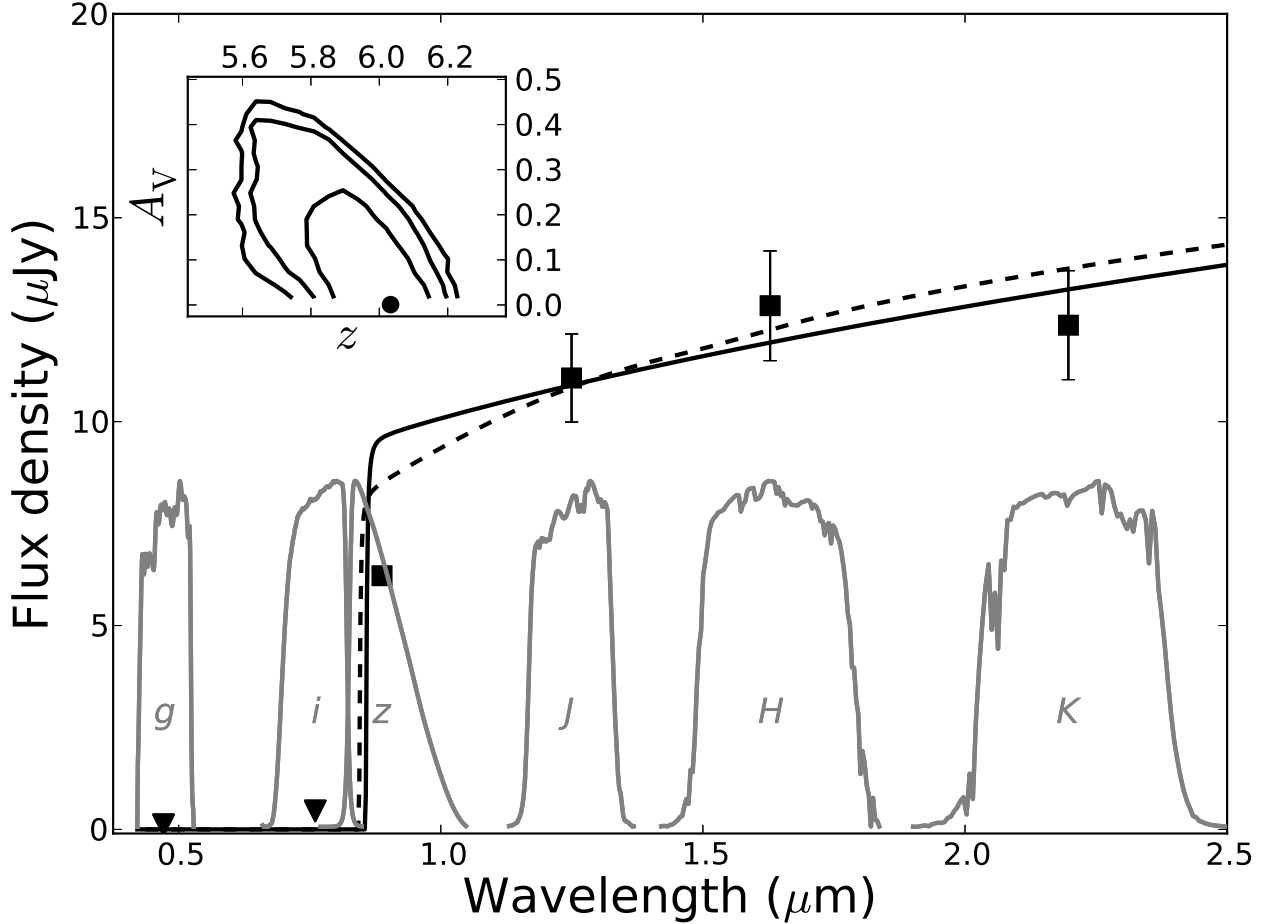}
\caption{The optical-to-NIR spectral energy distribution of GRB~120521C at $8.1\,$hr.
The $z$-band data point is a weighted average of all Gemini-North/GMOS frames taken at
7.7-8.5\,hr. (see Figure \ref{fig:120521C_zlcfit}). The $JHK$ photometry has been extrapolated from 
the nearest detections using the best-fit $z$-band light curve (Figure \ref{fig:120521C_zlcfit}), 
while the $g$ and $i$ upper limits are from Keck at $ \approx 12.2\,$h, used without 
extrapolation (Table \ref{tab:120521C:data:IR}). The data points have been placed at the centroid 
of the filter bandpass for clarity. The lines are models for the afterglow SED, including IGM and 
ISM absorption, using the best-fit (highest-likelihood) model (solid), and the median values of the 
parameter distributions (dashed, Table \ref{tab:120521C_sedfit}).
We show the  1$\sigma$, 2$\sigma$, and 3$\sigma$ contours for the correlation between extinction 
($A_{\rm V}$) and redshift ($z$) in the inset. The black dot indicates the best-fit model with no 
extinction and $z\approx 6.0$.
\label{fig:120521C_photoz}}
\end{figure}

\begin{figure}[ht]
\centering
\includegraphics[scale=0.5,bb=0 0 812 612]{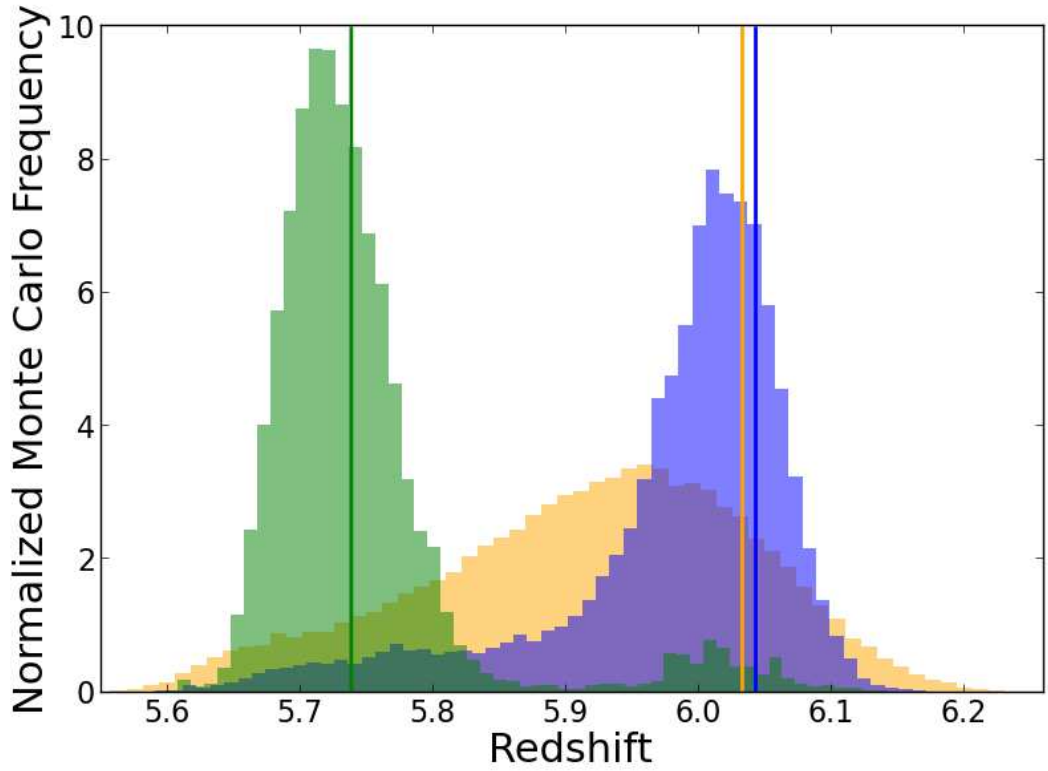}
\caption{Posterior density function for the redshift of GRB~120521C from fitting the SED at 
8.1\,hr. (orange, see Figure \ref{fig:120521C_photoz}), and from fitting all available afterglow
data with the redshift as a free parameter, using ISM (blue) and wind (green) models.
The vertical lines indicate the redshifts of the best-fit models.
\label{fig:120521C_zhists}}
\end{figure}

\begin{figure}[ht]
\centering
\includegraphics[width=\columnwidth]{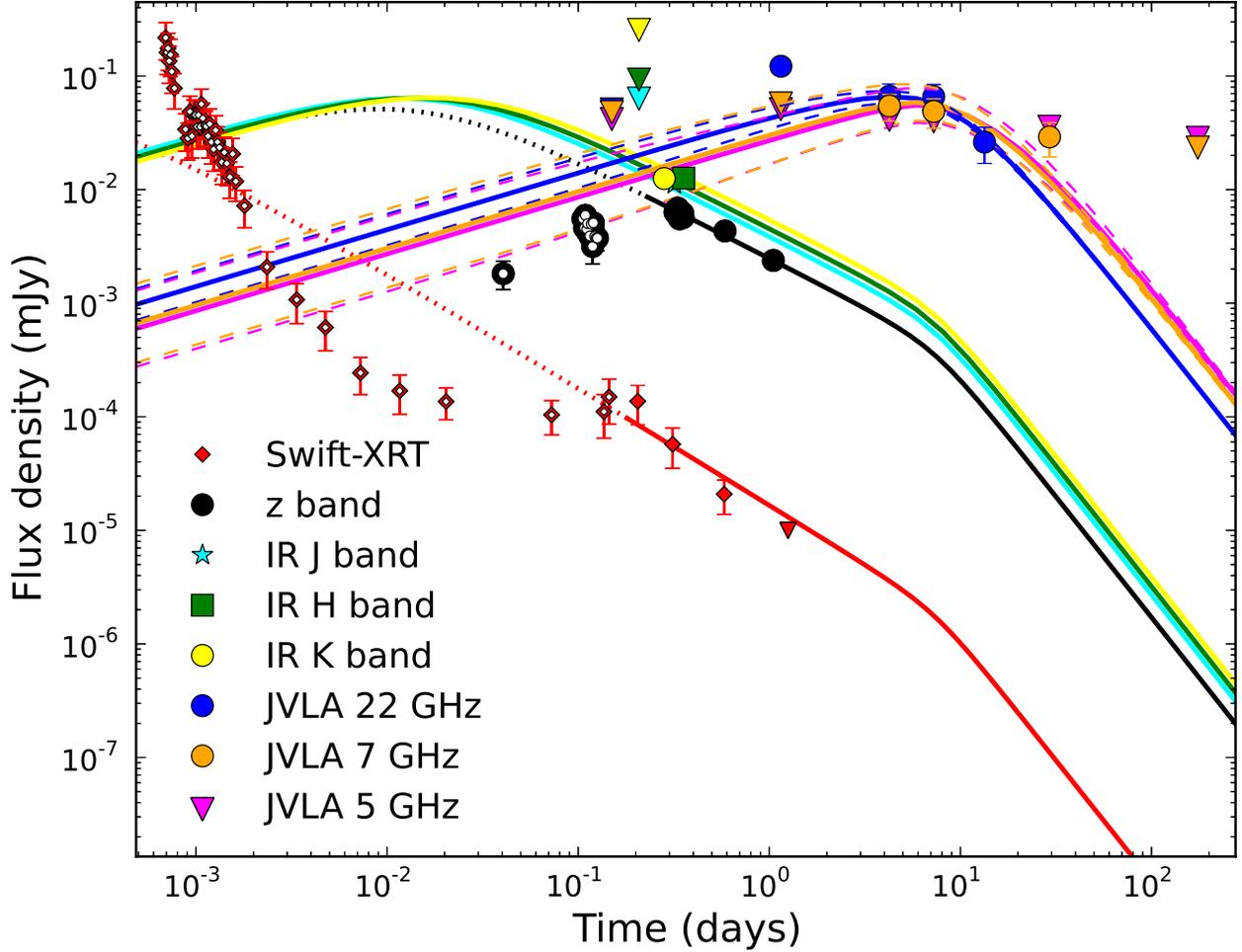}
\caption{Multi-wavelength modeling of GRB 120521C for a forward shock model with a homogeneous 
(ISM) 
environment \citep{gs02}. Triangles indicate $3\sigma$ upper limits and the dashed lines show 
the point-wise estimate of the 1$\sigma$ variation due to scintillation. Data 
excluded from the fit are shown as open symbols. We do not fit observations before 0.25\,d (see 
Section \ref{text:zlc}) and therefore the model before this time is shown as dotted lines. The 
$z$-band transmission functions of WHT/ACAM and Gemini-North/GMOS are substantially different and 
result in an expected suppression of the flux density of the WHT observations by a factor of 1.25 
compared to Gemini-North (see \S \ref{text:results} for details). For display purposes, the WHT 
$z$-band observations have been multiplied by 1.25 to bring them to the same scale as the GMOS 
observations. The black line is a light curve at the GMOS $z$-band frequency 
of $3.46\times10^{14}$\,Hz (887\,nm). The physical parameters of the burst derived from the 
best-fit solution are listed in Table 
\ref{tab:bestfit}.
\label{fig:120521C_multimodel_ISM}}
\end{figure}

\begin{figure}
\begin{tabular}{ccc}
\centering
 \includegraphics[width=0.31\columnwidth]{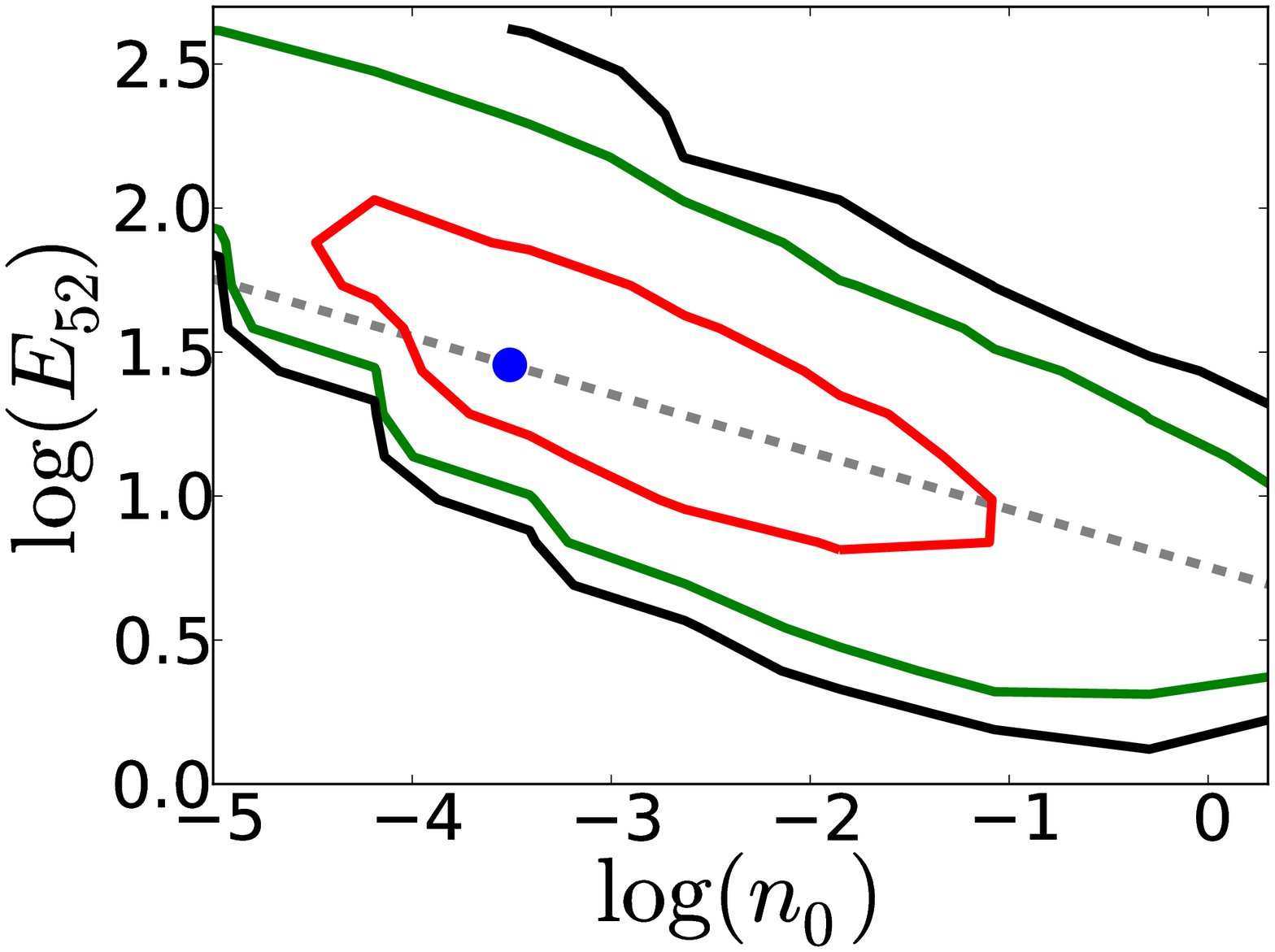} &
 \includegraphics[width=0.31\columnwidth]{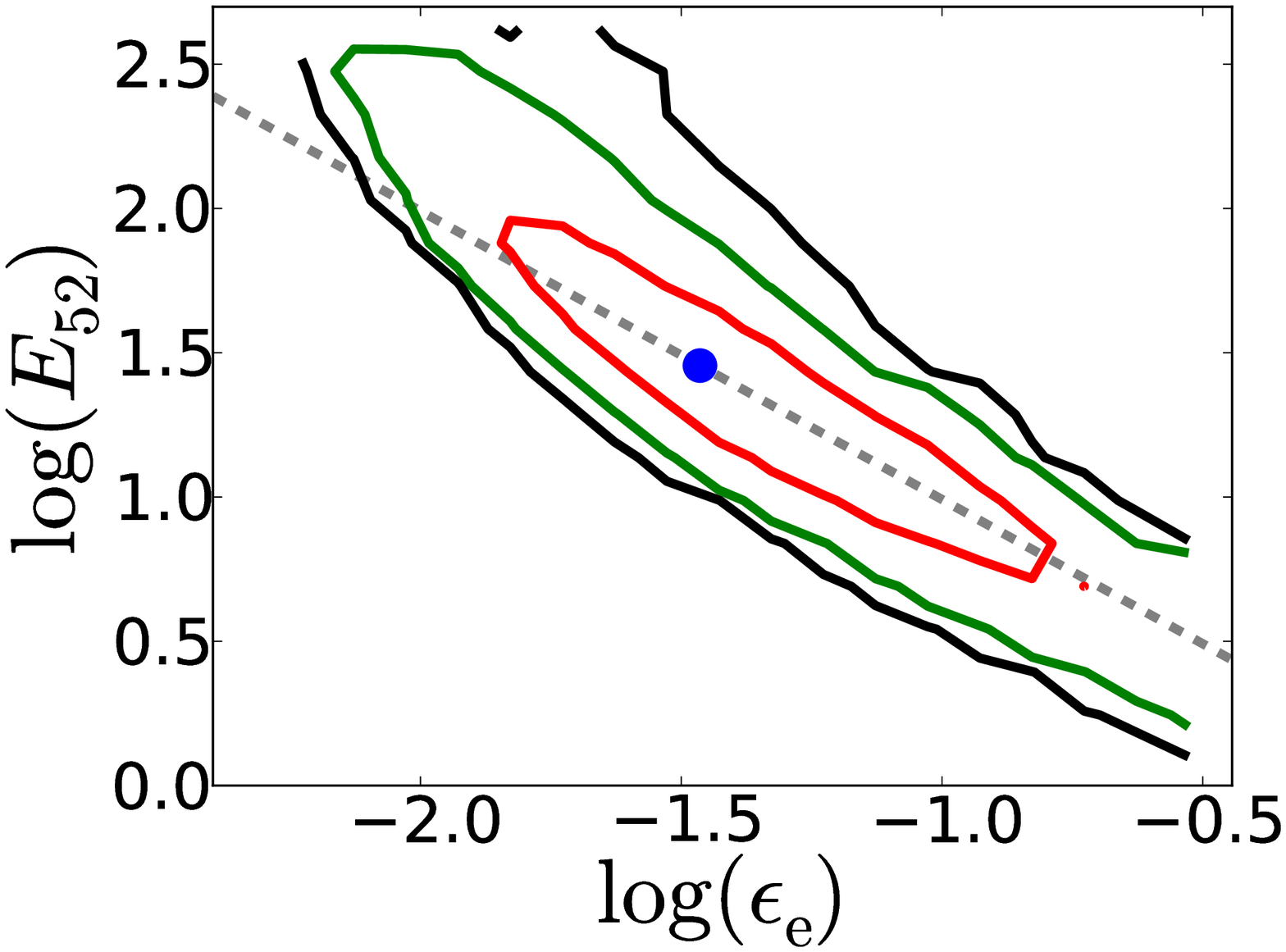} &
 \includegraphics[width=0.31\columnwidth]{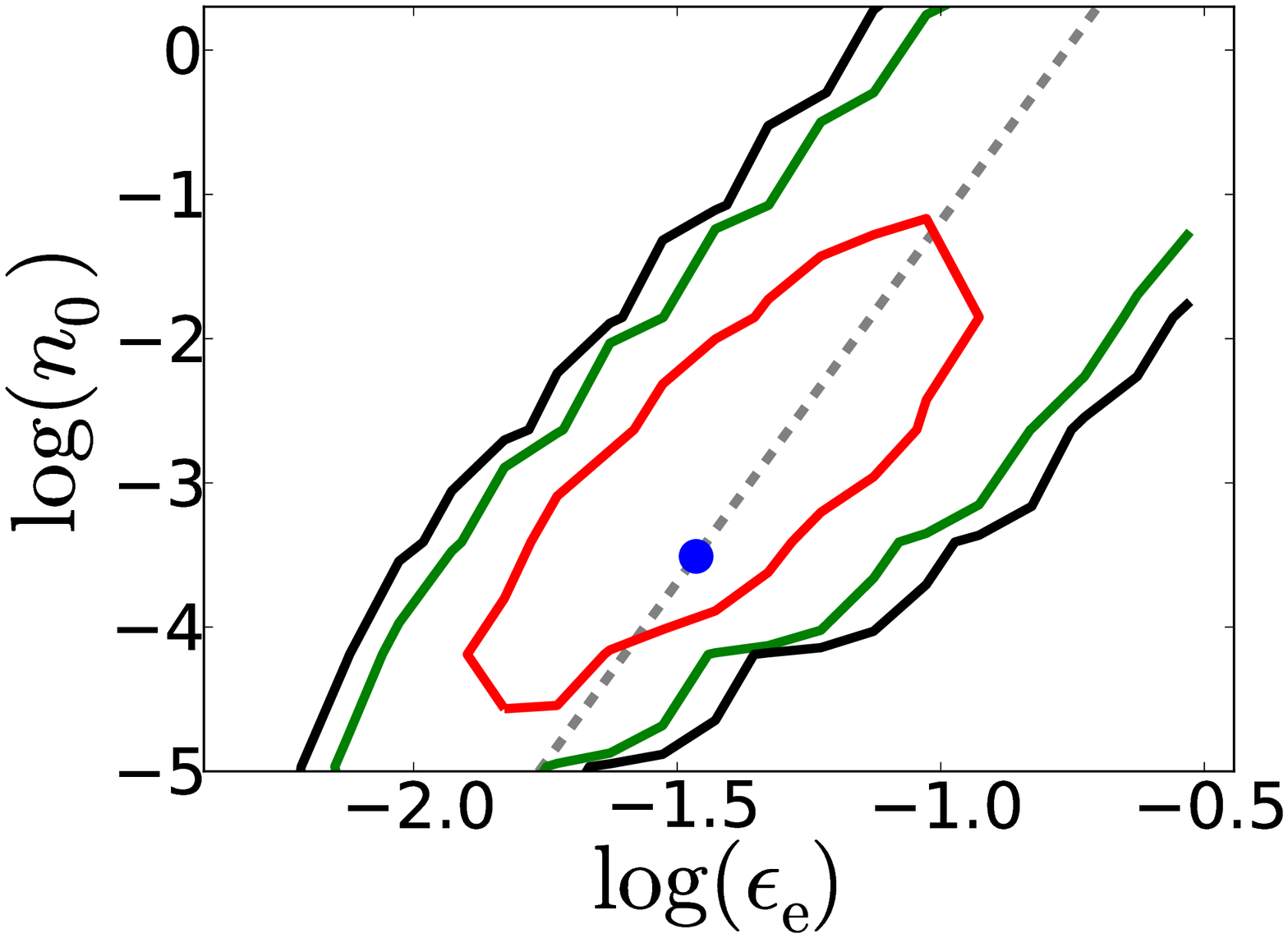} \\
 \includegraphics[width=0.31\columnwidth]{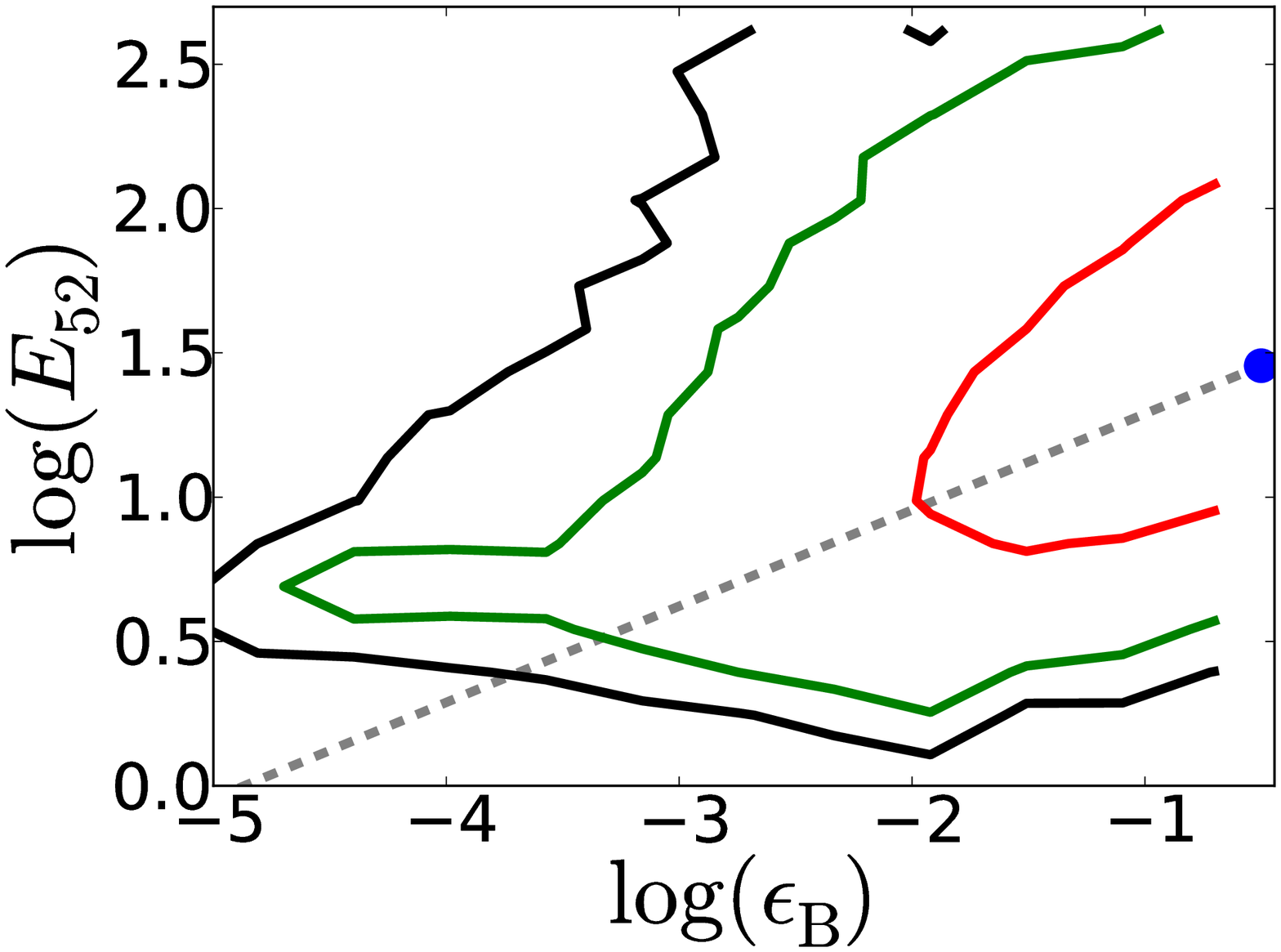} &
 \includegraphics[width=0.31\columnwidth]{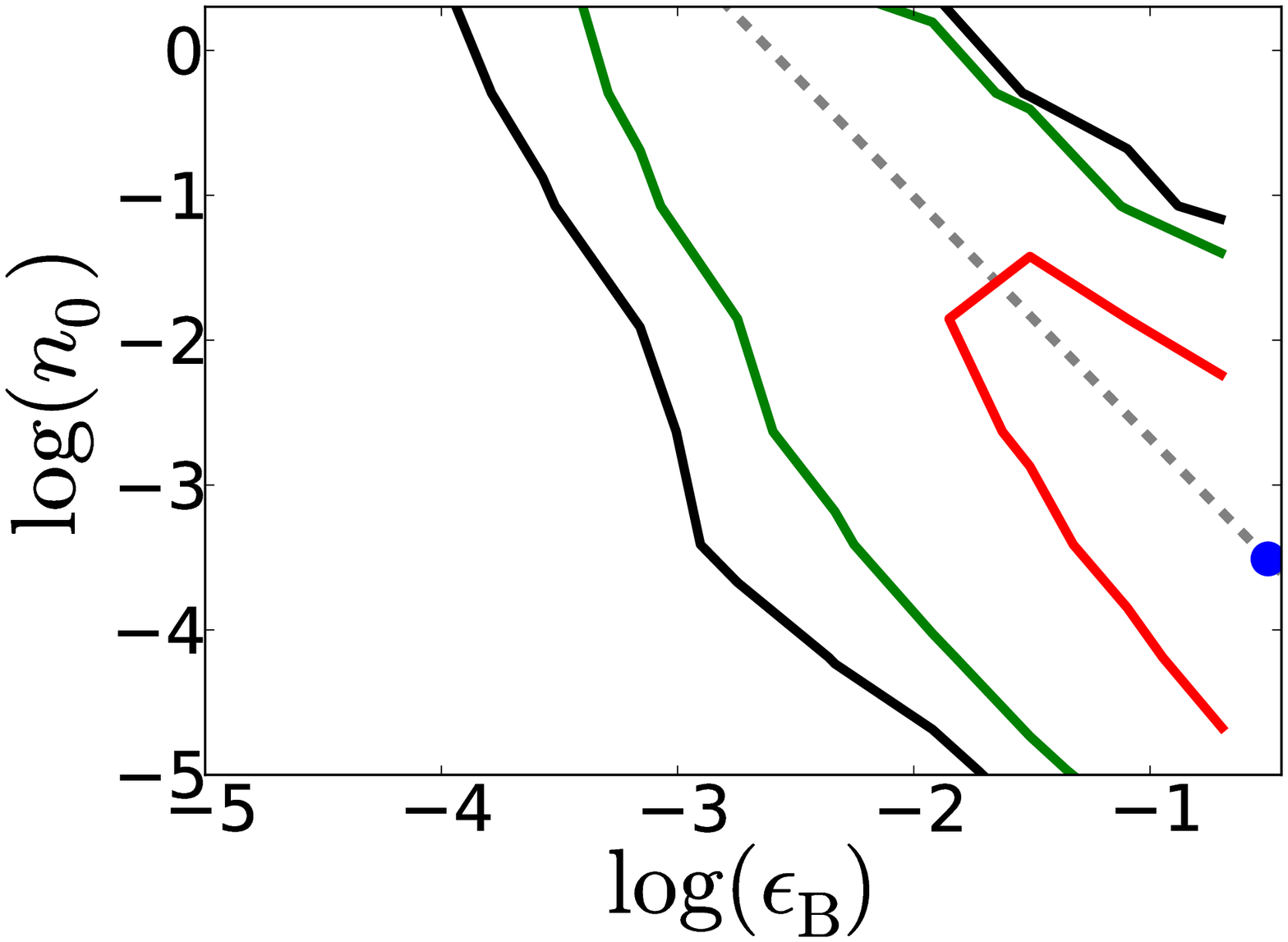} &
 \includegraphics[width=0.31\columnwidth]{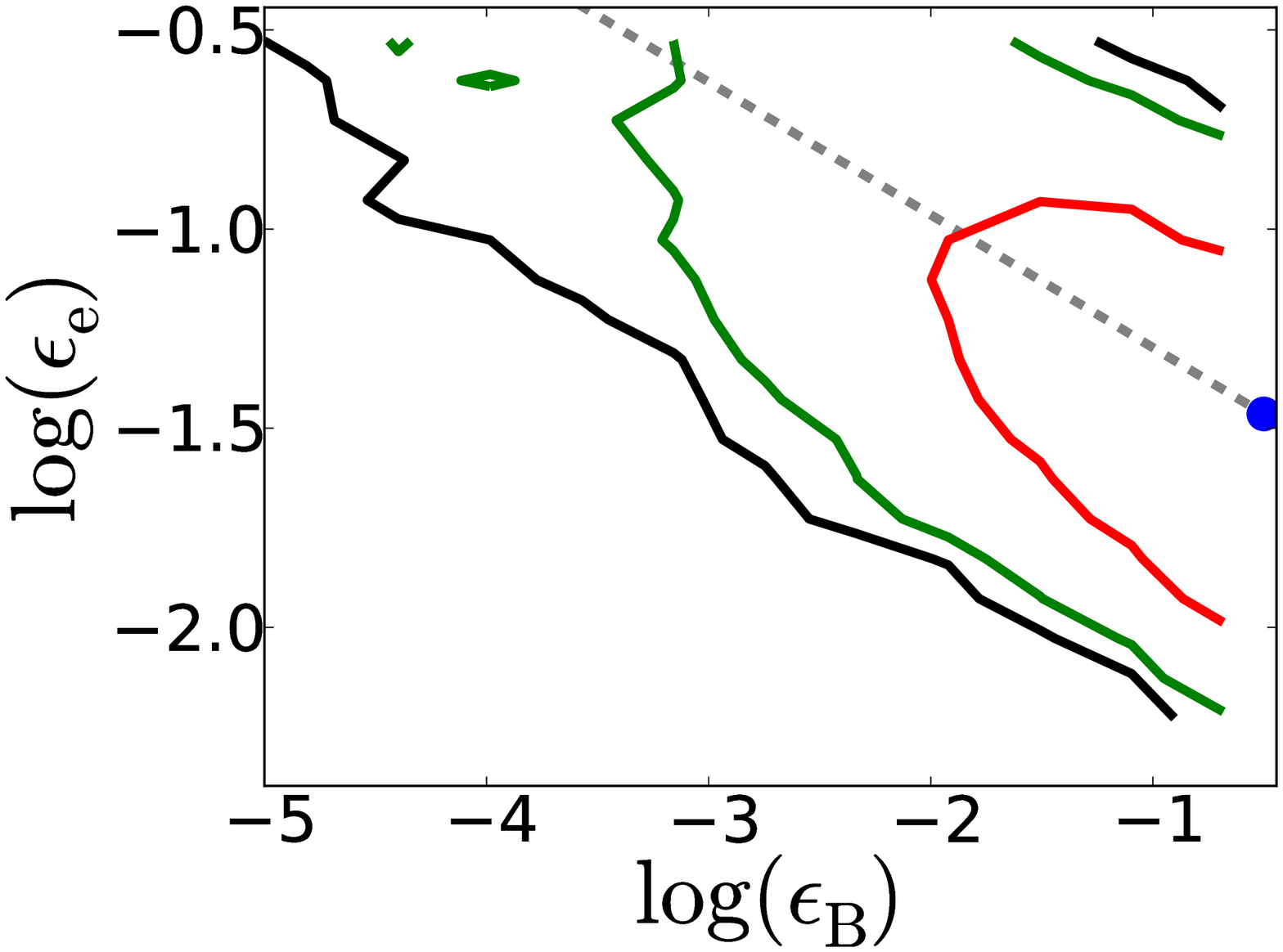} \\
\end{tabular}
\caption{1$\sigma$ (red), 2$\sigma$ (green), and 3$\sigma$ (black) contours for correlations
between the physical parameters, \E, \dens, \epse, and \epsb\ in the ISM model for GRB~120521C
from Monte Carlo simulations. We have restricted $E_{\rm K, iso, 52} < 500$, $\epsilon_{\rm e} <
\nicefrac{1}{3}$, and $\epsilon_{\rm B} < \nicefrac{1}{3}$. The dashed grey
lines indicate the expected relations between these parameters when \nua\ is not fully constrained:
$E_{\rm K, iso, 52}\propto n_{0}^{-1/5}$, $E_{\rm K, iso, 52}\propto \epsilon_{\rm e}^{-1}$,
$n_0\propto \epsilon_{\rm e}^{5}$,
$E_{\rm K, iso, 52}\propto \epsilon_{\rm B}^{1/3}$,
$n_{0}\propto \epsilon_{\rm B}^{-5/3}$,
$\epsilon_{\rm e}\propto\epsilon_{\rm B}^{-1/3}$,
normalized to pass through the highest-likelihood point (blue dot).
The contours lie parallel to these lines, indicating that the primary source of uncertainty in
the physical parameters comes from the poor observational constraint on \nua.
See the on-line version of this Figure for additional plots of correlations between these
parameters and $p$, $z$, $t_{\rm jet}$, $\theta_{\rm jet}$, and $A_{\rm V}$.
\label{fig:120521C_ISM_mcmcgrid}}
\end{figure}

Our MCMC analysis allows us to constrain the redshift to $6.01^{+0.05}_{-0.09}$ (the full posterior
density function is shown in Figure \ref{fig:120521C_photoz} as the blue histogram). This is 
consistent with the photometric redshift of $z=5.93^{+0.11}_{-0.14}$, which was based solely on the 
optical/NIR data and a prior on the spectral index (Section \ref{text:photoz}).
At this redshift, the \Swift/BAT $\gamma$-ray fluence, $F_{\gamma} = (1.1\pm
0.1)\times10^{-6}$\,erg\,cm$^{-2}$, corresponds to an isotropic energy release of 
$E_{\gamma, \rm iso} = (6.6\pm0.6)\times10^{52}$\,erg (104--1040 keV observer frame).
Since this burst was not observed by any wide-band $\gamma$-ray satellite, we do not have 
information about its $\gamma$-ray spectrum outside the \Swift\ 15--150\,keV band. We therefore use 
an average K-correction based on the observed \Swift/BAT fluence and computed $1$--$10^4$\,keV 
rest-frame isotropic-equivalent $\gamma$-ray energies of the other $z\gtrsim6$ GRBs: 050904, 080913, 
and 090423 \citep{gcn3938,gcn8222,gcn8256,gcn9204,gcn9251,agf+08}. We find that this K-correction 
ranges from a factor of about 1.8 (for GRBs 080913 and 090423) to 3.6 (for GRB 050904).
We infer an approximate value of $E_{\gamma,{\rm iso}}=(1.9\pm0.8)\times10^{53}$\,erg
for GRB~120521C, where the range accounts for the uncertainty in the K-correction. Our best 
estimate of the kinetic energy from the Monte Carlo analysis is $E_{\rm 
K,iso}=(2.2^{+3.7}_{-1.4})\times10^{53}$\,erg, indicating that the radiative efficiency, 
$\eta_{\rm rad} = E_{\gamma, \rm iso}/(E_{\gamma,\rm iso}+E_{\rm K, iso}) \approx 0.5$.

\begin{figure}
\begin{tabular}{ccc}
\centering
 \includegraphics[width=0.31\columnwidth]{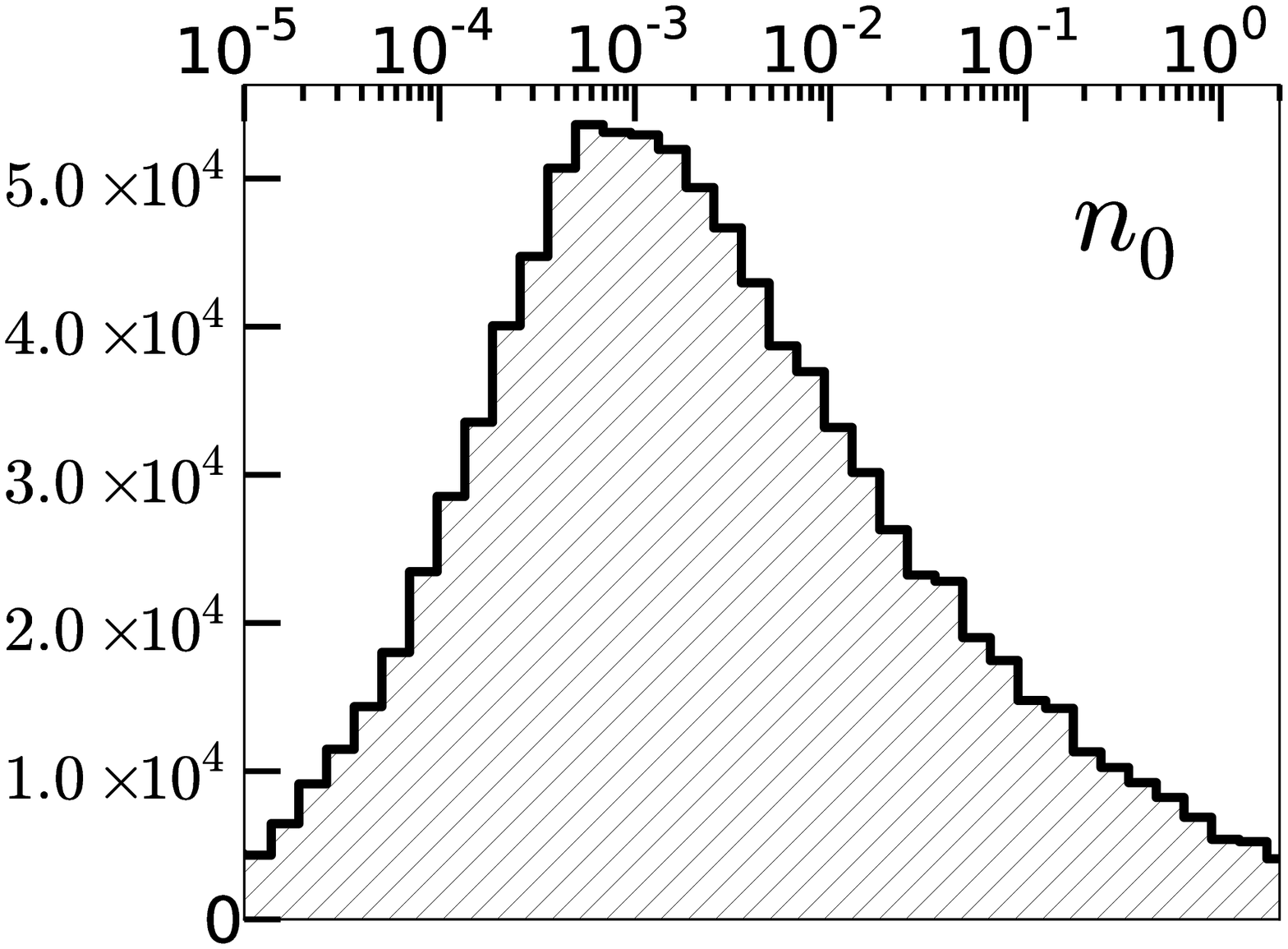} &
 \includegraphics[width=0.31\columnwidth]{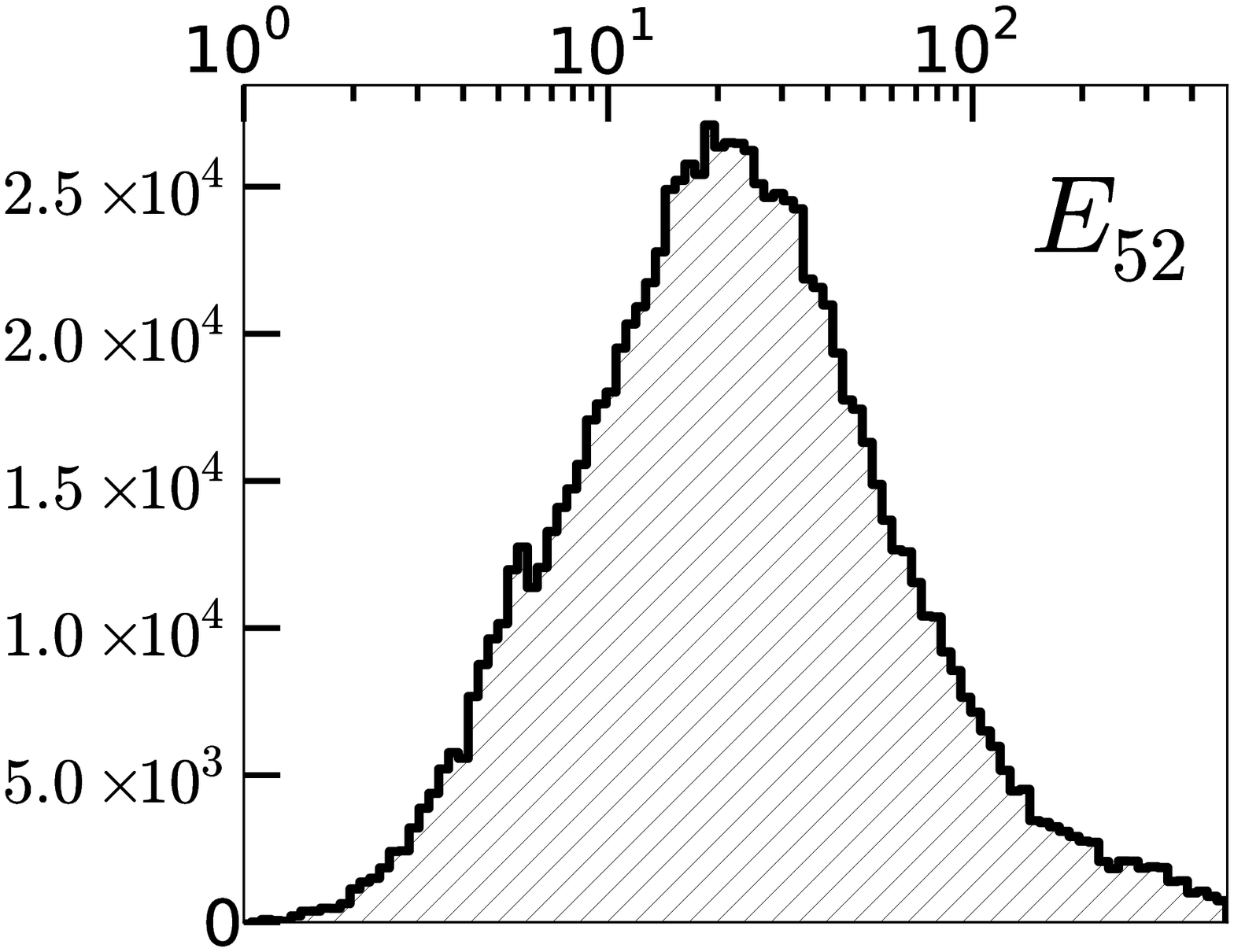} & 
 \includegraphics[width=0.31\columnwidth]{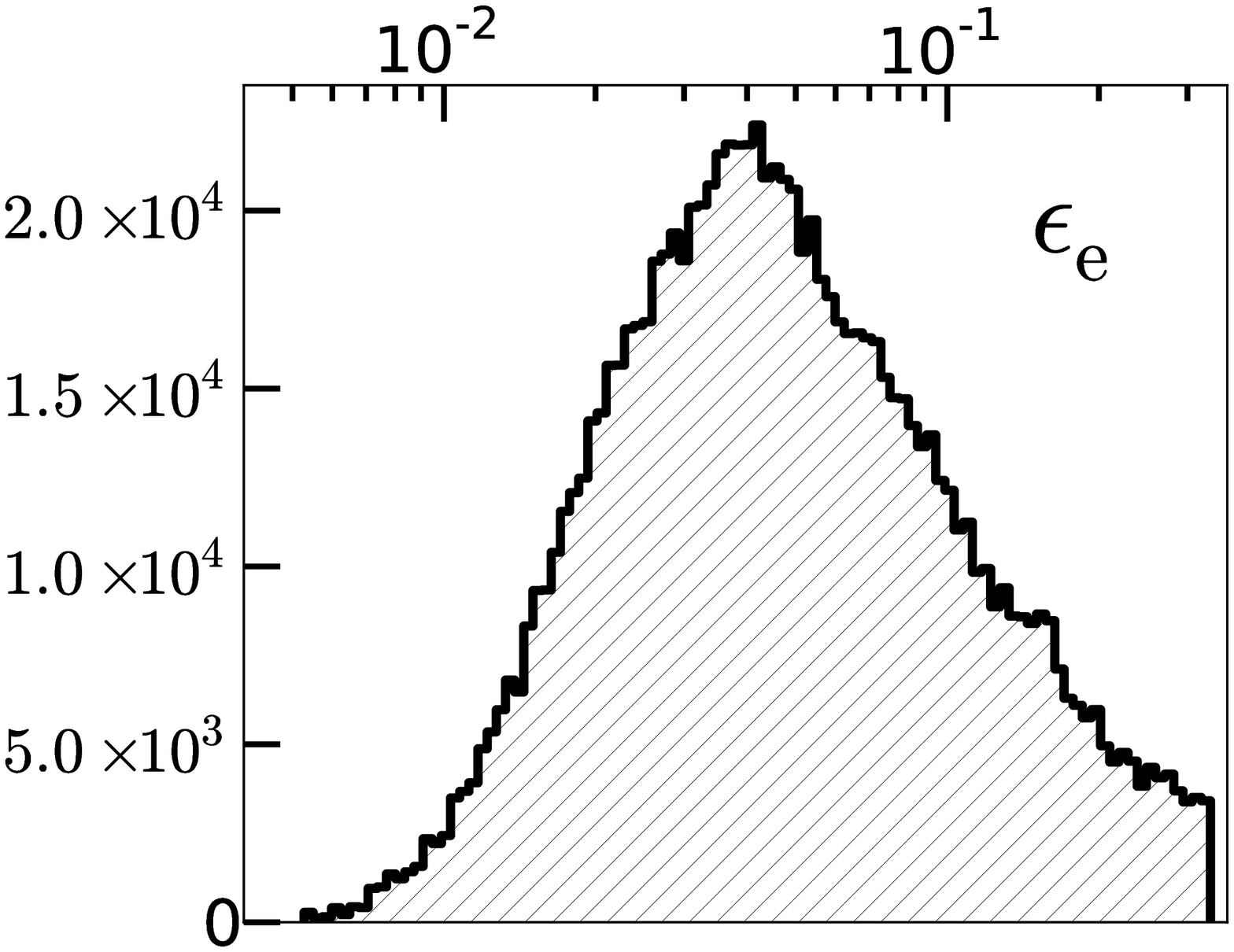} \\
 \includegraphics[width=0.31\columnwidth]{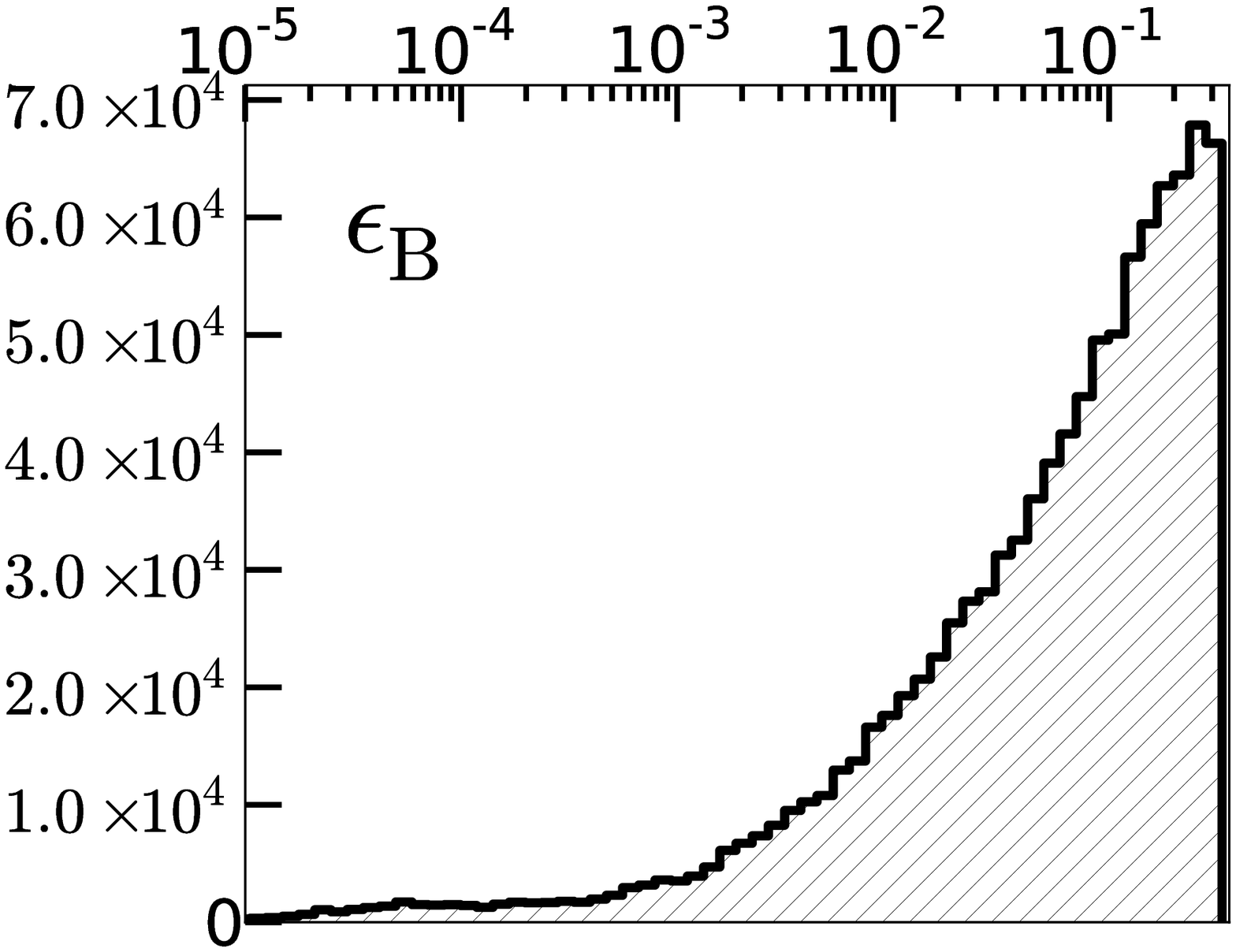} &
 \includegraphics[width=0.31\columnwidth]{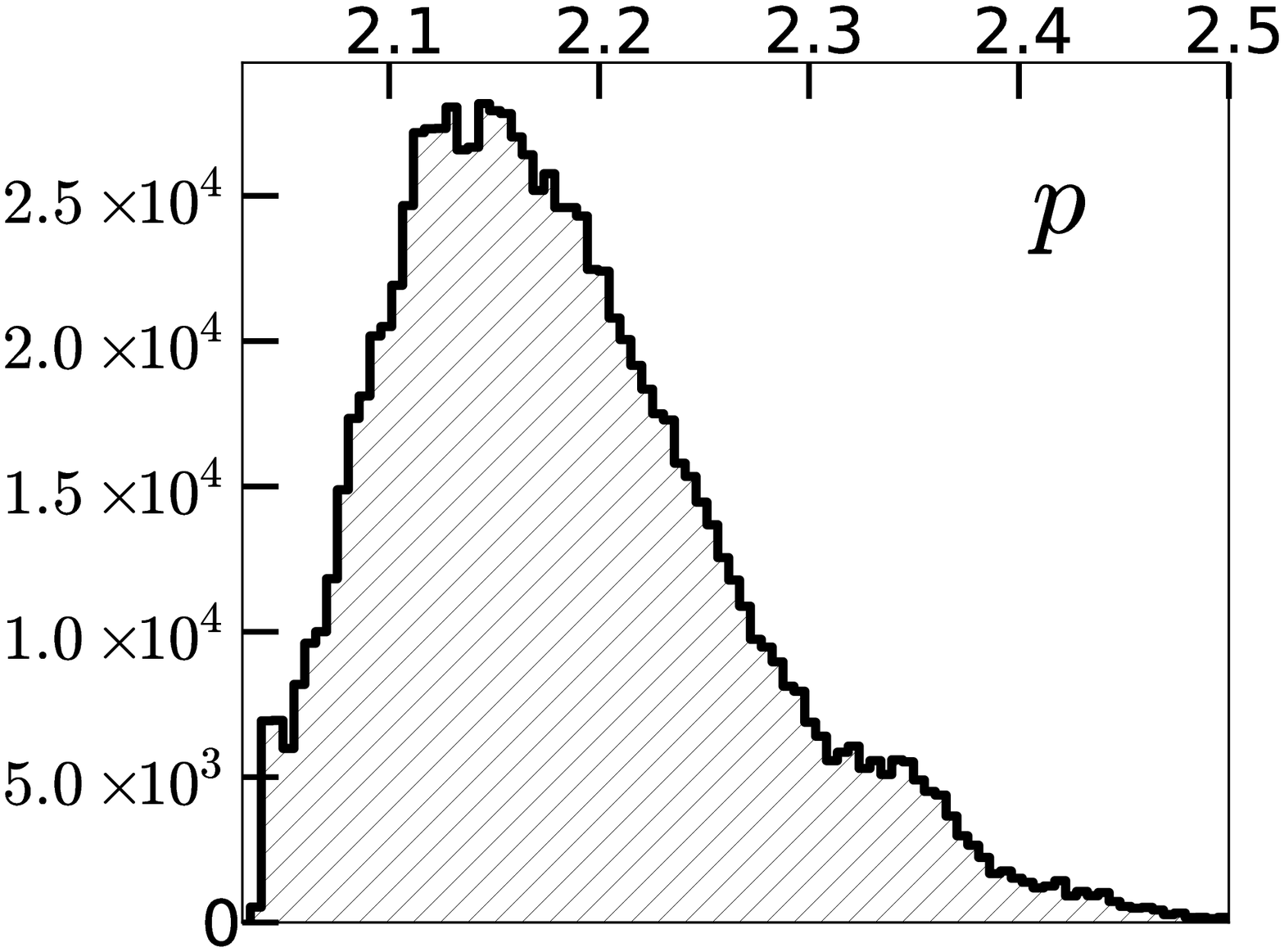} &
 \includegraphics[width=0.31\columnwidth]{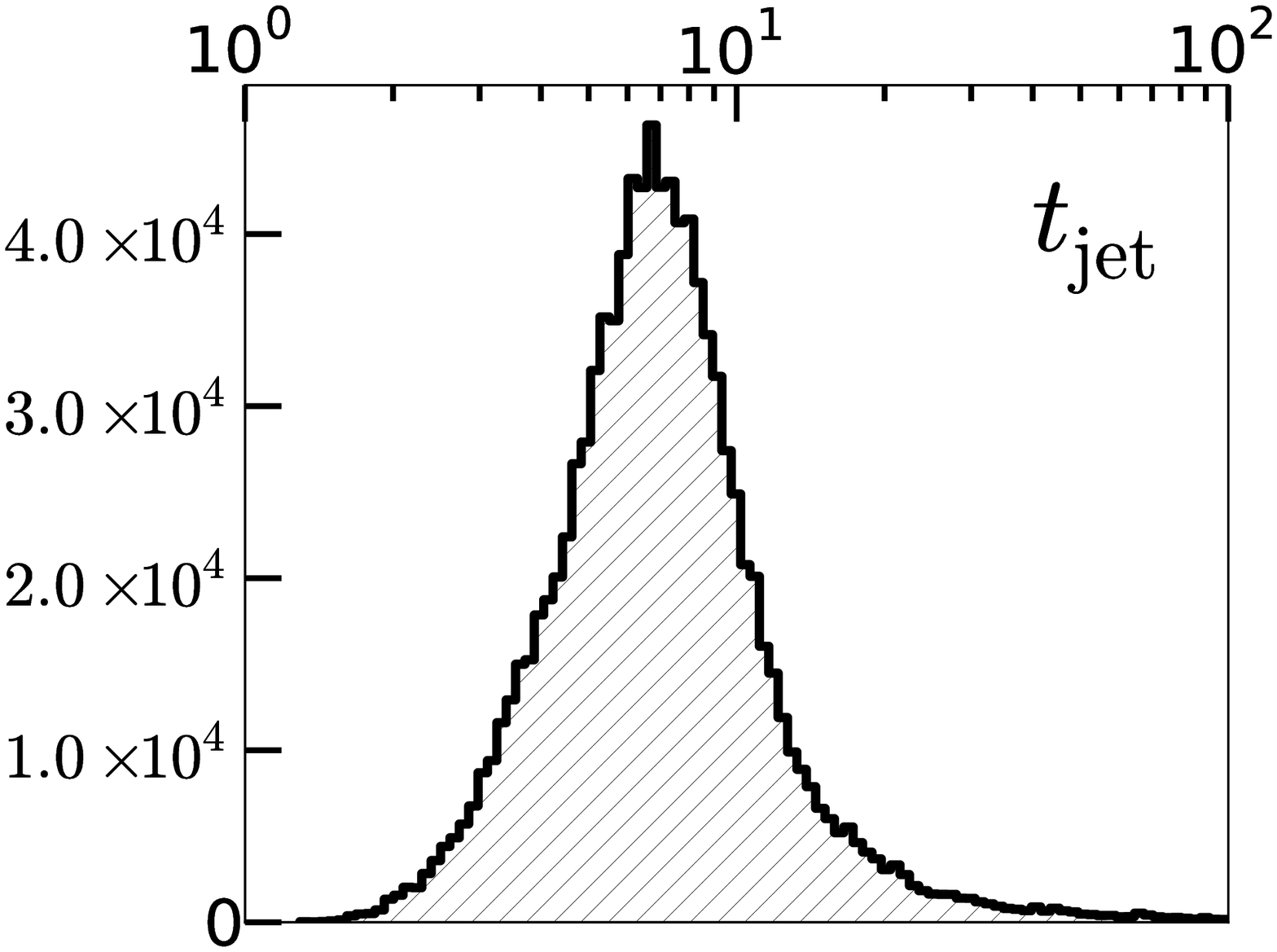} \\
 \includegraphics[width=0.31\columnwidth]{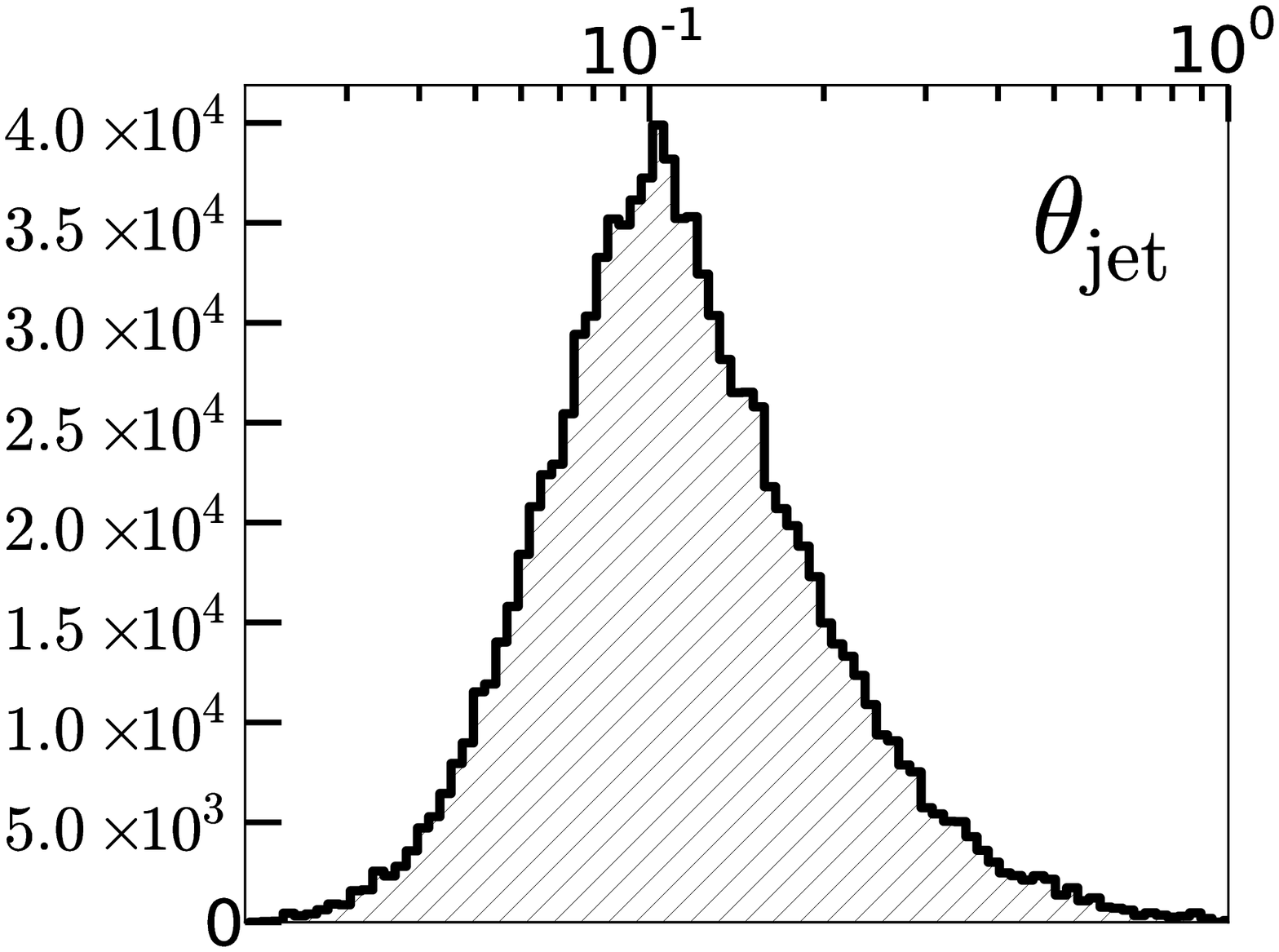} &
 \includegraphics[width=0.31\columnwidth]{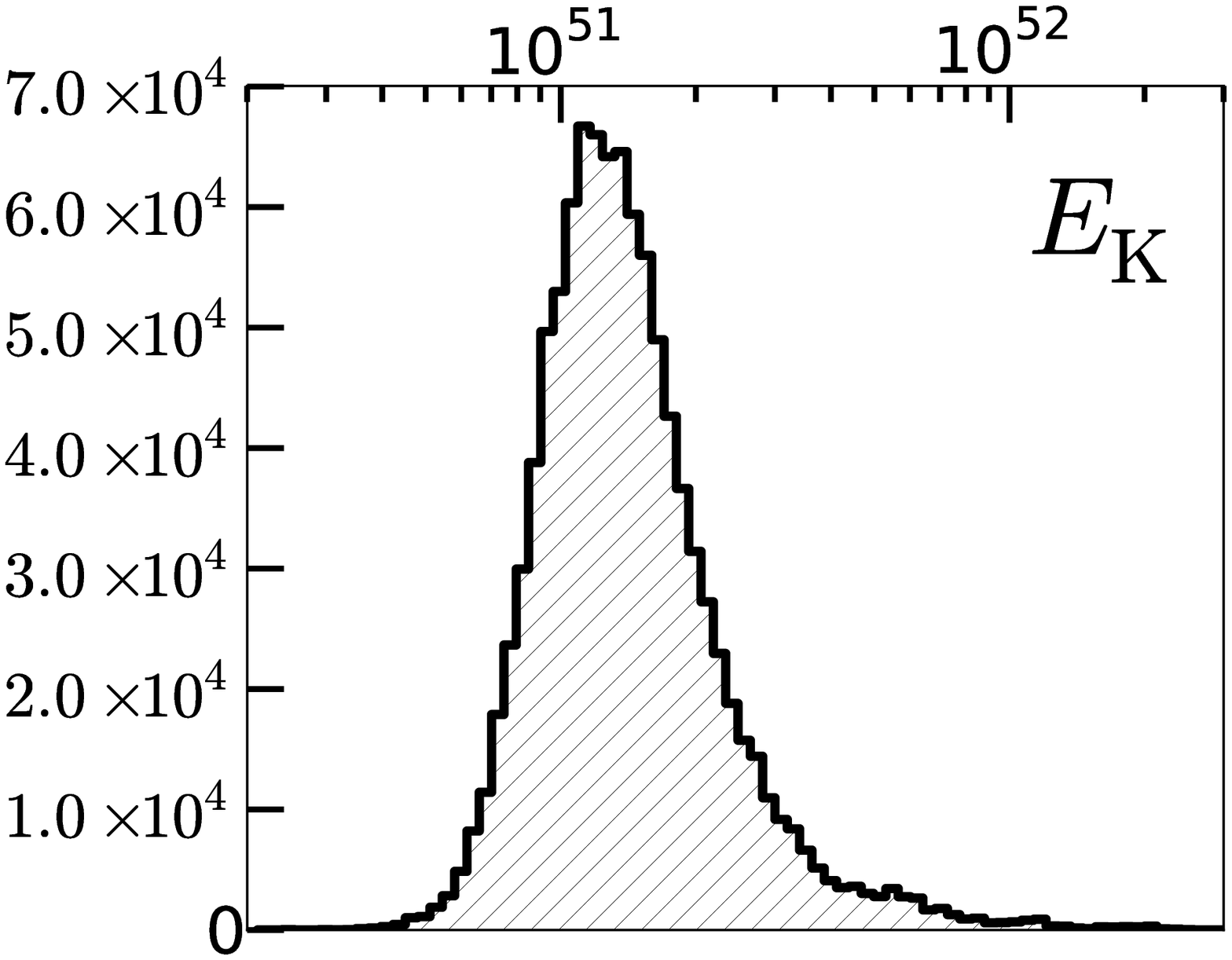} &\\
\end{tabular}
\caption{Posterior probability density functions of the physical parameters for GRB~120521C from
MCMC simulations. We have restricted $E_{\rm K, iso, 52} < 500$, $\epsilon_{\rm e} <
\nicefrac{1}{3}$, and $\epsilon_{\rm B} < \nicefrac{1}{3}$.
\label{fig:120521C_hists_ISM}}
\end{figure}

The $21.8$\,GHz radio light curve displays a plateau around 6\,d at a flux level of $f_{\nu,\rm 
m}\approx 70\,\mu$Jy (Figure \ref{fig:120521C_multimodel_ISM}). If we interpret this plateau as the 
passage of \numax\ through the 21.8\,GHz band, then we would expect \numax\ to pass through 
$6.7$\,GHz at around $12$\,d with a comparable flux density and for the 21.8\,GHz flux density to 
decline only modestly to about $50\,\mu$Jy (evolving as $t^{\nicefrac{(1-p)}{2}}\sim t^{-0.5}$). In 
addition, this would predict a flux density of $45\,\mu$Jy at 6.7\,GHz at the next epoch at $\Delta 
t = 29.3$\,d. However, the $6.7$\,GHz light curve does not rise as expected, while the 21.8\,GHz 
flux density plummets to about $26\,\mu$Jy at $\Delta t = 13.3$\,d. In addition, the $6.7$\,GHz 
observation at $\Delta t = 29.3$\, yields a detection at barely $3\,\sigma$ of $30\,\mu$Jy. 
This behavior indicates a departure from isotropic evolution and we find that a jet break 
at $\Delta t \approx 7$\,d adequately accounts for the radio observations after 10 days. The 
presence of a jet break means that the peak flux density of the broad-band spectrum declines with 
time, while the break frequencies evolve faster; this explains why 
the $6.7$\,GHz flux density does not rise to the level observed at 21.8\,GHz, and why the 
$21.8$\,GHz flux density rapidly declines following the plateau. Using the relation  $\theta_{\rm 
jet} = 0.1\left(\frac{E_{\rm K, iso, 52}}{n_0}\right) ^{\nicefrac{1}{8}} \left(\frac{t_{\rm 
jet}/(1+z)}{6.2\,\rm hr}\right)^{\nicefrac{3}{8}}$ for the jet opening angle \citep{sph99}, and the 
distributions of \Efte, \dens, $z$, and $t_{\rm jet}$ from our MCMC 
simulations (Figure \ref{fig:120521C_hists_ISM}), we find $\theta_{\rm jet} = 3.0^{+2.3}_{-1.1}$ 
degrees. Applying the beaming correction, 
$E_{\gamma} = E_{\gamma,\rm iso}(1-\cos{\theta_{\rm jet}})$, 
we find $E_{\gamma} = (2.6^{+4.4}_{-2.0})\times10^{50}$\,erg. Similarly, the beaming-corrected 
kinetic energy is $E_{\rm K} = (3.1^{+1.9}_{-0.9})\times10^{50}$\,erg.

The first radio detection in the $21.8$\,GHz band at $\Delta t=1.2$\,d ($1.22\pm0.02$\,mJy)
is a factor of $2.7$ times brighter than predicted by the model ($0.45\pm0.1$\,mJy, $1\sigma$
deviation from scintillation). Early-time excess radio emission in GRB afterglows has frequently
been attributed to the presence of a reverse shock component 
\citep[e.g.][]{kfs+99,sp99a,bsfk03,sr03,cff+10,lbz+13}. We investigate the potential contribution 
of a reverse shock and derive an estimate for the Lorentz factor of the ejecta in Appendix 
\ref{text:120521C_RS}.

\begin{figure}[ht]
\centering
\includegraphics[width=\columnwidth]{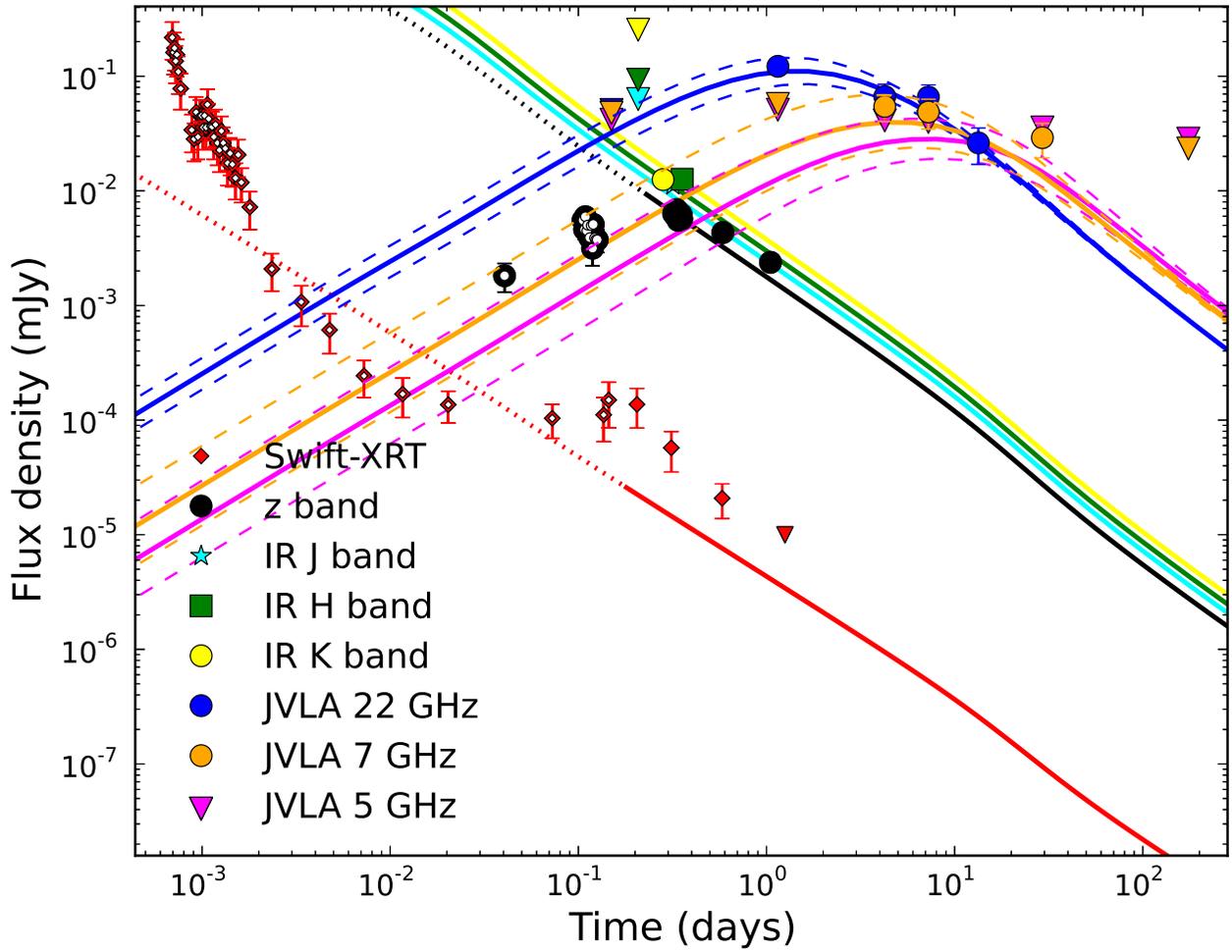}
\caption{Same as Figure \ref{fig:120521C_multimodel_ISM}, but for a wind environment. The model
matches the first $21.8$\,GHz radio observation, but under-predicts the X-ray data and is 
therefore disfavored. The physical parameters of the burst derived from the 
best-fit solution are listed in Table \ref{tab:bestfit}.
\label{fig:120521C_multimodel_wind}}
\end{figure}

We also perform the Monte Carlo analysis detailed in \S \ref{text:mcmc} for a wind-like environment.
The redshift distribution from the wind model is shown in Figure \ref{fig:120521C_photoz} as the
green histogram. Our best-fit wind model is plotted in Figure \ref{fig:120521C_multimodel_wind}. We
find that the model matches the radio observations (including the first radio detection, which is
missed by the ISM model), but under-predicts all X-ray data included as part of the fit. In this
model, \nua\ is constrained to lie between $7$ and $22$\,GHz at $\Delta t =1.15$\,d, breaking the
degeneracy encountered in the ISM model. We list the derived parameters in Tables \ref{tab:bestfit} 
and \ref{tab:mcmc}. However, since the X-ray data are not fit well, we do not consider the wind 
model 
as an adequate representation of the dataset.

\subsection{Potential Explanations for the $z$-band peak at $\approx8$\,hr}
\label{text:zlc}
We now return to the peak in the $z$-band light curve at $\Delta t\approx8$\,hr, which cannot be 
explained by the passage of the synchrotron peak frequency (see Section \ref{text:results}). One 
possible explanation for this peak is that the blastwave encounters a density jump, causing a 
long-lasting optical flare. \citet{ng07} showed that the greatest change expected in an optical 
light curve due to a density jump is bounded at $\Delta \alpha \lesssim 1$ (see also 
\citealt{gvem13}), whereas the temporal behavior of the $z$-band flux density indicates a change of 
$\Delta \alpha \sim 2.2$. Hence, the $z$-band light curve is unlikely to be the result of an 
inhomogeneous external medium.

Another way to suppress the $z$-band flux before 8\,hr is through absorption by neutral hydrogen 
in the vicinity of the progenitor. This is an attractive explanation in this case because 
the $z$-band straddles the Lyman break and the flux density in this band is therefore highly 
sensitive to small variations in the neutral hydrogen column along the line of sight. In particular, 
if the neutral hydrogen column were to decline with time due to destruction by the blastwave or by 
photo-ionization, it would lead to the observed behavior of the rising $z$-band flux density. Our 
first $z$-band detection is at $\approx8$\,min in the rest-frame of the burst, corresponding to a 
distance of $\sim 1$\,AU from the progenitor, while the $z$-band peak occurs at $\approx 1.2$\,hr 
in the rest frame, corresponding to a distance of $\sim$8 AU. We find that an additional neutral 
hydrogen column of $N_{\rm H}\sim2\times10^{22}\,{\rm cm}^{-2}$ at $z=6$ would be sufficient to 
suppress the first $z$-band point to the observed flux level and the ionization of this column would 
therefore lead to the observed increase in flux. For a path length of $\sim7$\,AU, this column 
corresponds to a density of $\approx 2\times 10^8\,{\rm cm}^{-3}$ or a mass of about 
$10^{-7}$\,$M_{\odot}$ (assuming a spherical cloud). Although the requisite mass is not very large, 
the inferred density is four orders of magnitude higher than a typical molecular cloud in the Milky 
Way \citep{sch01,mo07}. Thus ionization of a large neutral hydrogen column along the line of sight 
is a feasible explanation for the rising $z$-band light curve only if the densities of molecular 
clouds at $z\sim6$ can be much greater than observed locally.

Another possible explanation for the initial rise in $z$-band is the injection of energy into the 
blastwave shock by slower-moving relativistic ejecta catching up with the decelerating blastwave. 
If the injection is rapid enough it could create a rising light curve at $z$-band, which would then 
be expected to break into a fading power-law if \numax\ is located below $z$-band at the end of the
injection phase. Energy injection has been frequently invoked to explain the plateau phase of GRB 
X-ray afterglows (e.g. \citealt{nkg+06, zfd+06,dsg+11}. The X-ray light curve of GRB~120521C indeed 
shows such a plateau at 0.01--0.25\,d. 

To test whether the X-ray and NIR light curves can result from energy injection, we use our 
ISM model as an anchor at $\Delta t = t_{\rm end}\approx 8$\,hr, after which 
it is the best-fit model to the multi-wavelength data set (including the $z$-band and XRT 
observations). We then assume a period of energy injection between the start of the X-ray plateau 
at $t_{\rm start} \approx \text{few}\times10^{-2}$\,d and $t_{\rm end}$ and use a simple power-law 
prescription for the energy as a function of time,
\[
 E_{\rm K, iso}(t) =
  \begin{cases}
   E_{\rm K, iso,0}\left(\frac{t_{\rm end}}{t_{\rm start}}\right)^{\zeta}=\textit{const.}, 
                                                            & t < t_{\rm start}\\
   E_{\rm K, iso,0}\left(\frac{t}{t_{\rm start}}\right)^{\zeta}\propto t^{\zeta},
                                                            & t_{\rm start} < t < t_{\rm end} \\
   E_{\rm K, iso,0} = \textit{const.},  & t > t_{\rm end},
  \end{cases}
\]
where $E_{\rm K,iso,0}$ is the total isotropic-equivalent blastwave kinetic energy after 
energy injection is complete. We note that the XRT light curve displays a steep decline before the 
plateau with $\alpha_{\rm X} = -3.5\pm0.2$ at 90--345\,s (Figure \ref{fig:120521C_multimodel_ISM}), 
which cannot be explained by the afterglow forward shock and is likely related to the prompt 
emission (see also Section \ref{text:results}). We therefore add an additional power-law component 
with a fixed slope of $\alpha_{\rm X} = -3.5$ to the model X-ray light curve.

\begin{figure}[ht]
\centering
\includegraphics[width=\columnwidth]{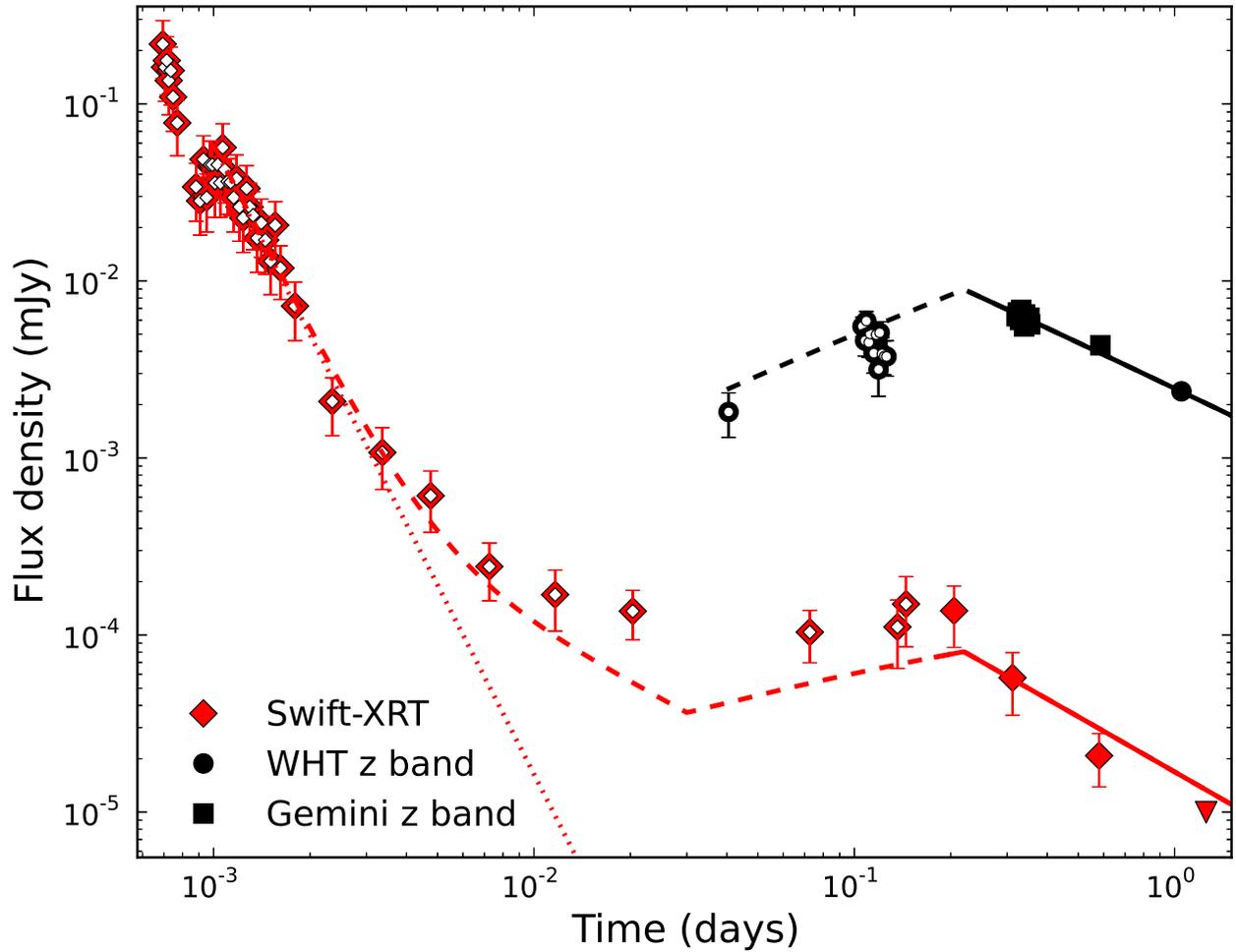}
\caption{Energy injection model for GRB~120521C (dashed lines), using the forward shock model (solid
lines) as fit to the observations after 0.25\,d (filled symbols). The dotted line is a power-law
fit ($\alpha = -3.5\pm0.2$) to the XRT data between 90\,s and 345\,s. The 
\textit{WHT} $z$-band observations have been scaled by a
factor of 1.25 as in Figures \ref{fig:120521C_multimodel_ISM} and \ref{fig:120521C_multimodel_wind}.
\label{fig:120521C_enj}}
\end{figure}

We set $E_{\rm K,iso,0} = 2.85\times10^{53}$\,erg using our highest-likelihood model (values in 
parentheses in Table \ref{tab:bestfit}) and vary $\zeta$, $t_{\rm start}$, and $t_{\rm end}$ to 
obtain a good match to the X-ray and $z$-band light curves. We find that in general we are able to 
model either the X-ray plateau or the $z$-band rise, but not both. Our best simultaneous match to 
both light curves is shown in Figure \ref{fig:120521C_enj} with the parameters, $t_{\rm start} \sim 
2.6\times10^3$\,s, $t_{\rm end} \sim 1.9\times10^4$\,s, and $\zeta\sim1.25$, corresponding to an 
increase in blastwave kinetic energy by a factor of $\left(\frac{t_{\rm end}}{t_{\rm 
start}}\right)^{\zeta} \sim 12$ over this period. Although the resulting light curves do not match 
perfectly, energy injection provides the most plausible explanation for the $z$-band peak. 
Finally, we note that there is some evidence for `flickering' in the form of 
statistically-significant scatter about the overall $z$-band rise (Figure \ref{fig:120521C_zlcfit}), 
but the observations do not sample these rapid time-scale flux variations well enough to allow us to 
comment on the nature or source of the variability.

\section{Other GRBs at $z\gtrsim6$ with Radio to X-ray Detections}
\label{text:othergrbs}
To place the physical properties of GRB~120521C derived above in the context of other high-redshift 
events, and to compare them in a uniform manner, we apply the above analysis to the other 
two GRBs at $z\gtrsim6$ with radio to X-ray detections reported in the literature: GRB~050904 at 
$z=6.29$ and GRB~090423 at $z=8.23$.

\subsection{GRB~050904}
\label{text:050904}
GRB~050904 was discovered with \Swift/BAT on 2005 September 4 at 1:51:44\,UT
\citep{gcn3910}. The burst duration was $T_{90} = 22.5\pm10$\,s \citep{gcn3938},
with a fluence of $F_{\gamma} = (5.4 \pm 0.2) \times10^{-6}$\,erg\,cm$^{-2}$
(15--150\,keV). A photometric redshift was reported by \citet{tac+05} and \citet{hnr+06}, and
spectroscopically confirmed by \citet{kka+06}, making GRB~050904 the highest
redshift GRB observed at the time.

We analyzed the XRT data for this burst 
in the same manner as described in section \ref{text:data_analysis:XRT}. In our spectral modeling, 
we assume 
$N_{\rm H, MW} = 4.53\times10^{20}$\,cm${}^{-2}$ \citep{kbh+05}. The best-fit neutral hydrogen 
column density intrinsic to the host is $N_{\rm H, int} = 5.61^{+2.98}_{-2.44}\times10^{22}$ 
\,cm${}^{-2}$ (68\% confidence intervals). In our temporally-resolved spectral analysis, we find 
that the X-ray photon index is consistent with $\Gamma = 2.03\pm 0.10$ ($68\%$ confidence interval) 
for all XRT data following 490\,s after the GRB trigger. We use this value of the photon index to 
convert the observed $0.3$--$10$\,keV light curve to a flux density at 1\,keV. The X-ray data 
before $1.7\times 10^3$\,s and at $3\times10^3$ -- $5\times10^4$\,s are dominated by multiple 
flares. We ignore XRT data in this time range in our analysis.

We compiled NIR observations of GRB~050904 in the $Y$, $J$, $H$, and $K$ bands
from the literature \citep{hnr+06,gfm07}, and corrected for Galactic
extinction along the line of sight assuming $E(B-V)=0.061\,$mag
\citep{sf11}.
Since $z$-band is located blueward of Lyman-$\alpha$ in the rest-frame 
of the GRB, flux within and blueward of this band is heavily suppressed by absorption by neutral 
hydrogen in the IGM and we do not include these bands in our multi-wavelength fit. 
This burst was observed over multiple epochs in the $8.46$\,GHz radio band with the VLA 
\citep{fck+06} and we use the individual observations and limits in our analysis. We list all 
photometry we use in our model in Table \ref{tab:050904:data:all}.

\begin{figure}[ht]
\centering
\includegraphics[width=\columnwidth]{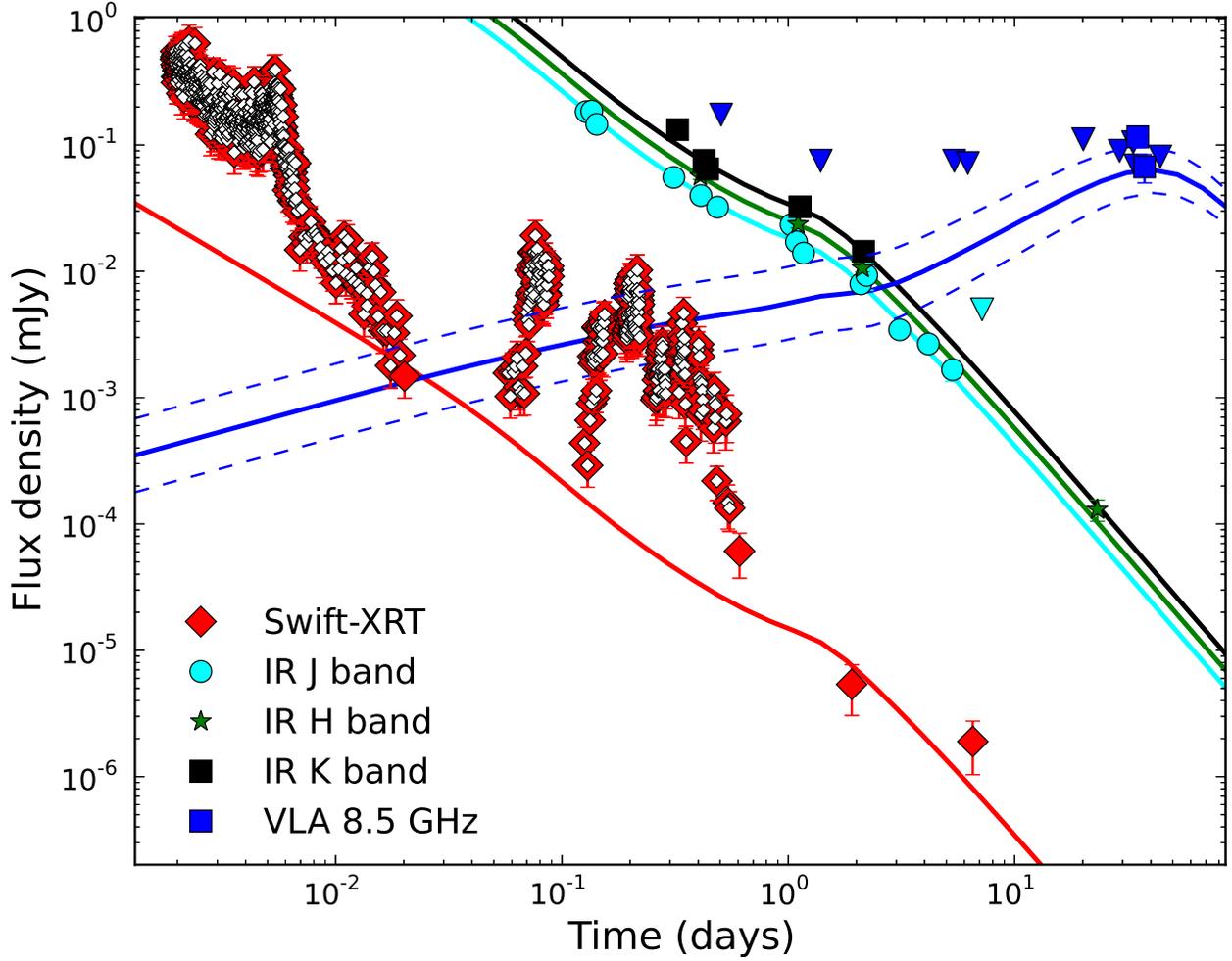}
\caption{Multi-wavelength modeling of GRB 050904 for a forward shock model with a 
homogeneous (ISM) environment \citep{gs02}. Triangles indicate $3\sigma$ upper limits and the dashed
lines show the point-wise estimate of the 1$\sigma$ variation due
to scintillation. $Y$-band data are included in the fit but are not shown  in the plot for clarity.
The X-ray data between 0.03 and 0.6\,d are dominated by large flares, while the
steeply-declining XRT light curve before 0.02\,d is likely associated with the prompt emission. We
ignore these segments in the afterglow model fit (open symbols). The physical parameters of the
burst derived from the best-fit solution are listed in Table
\ref{tab:bestfit}.
\label{fig:050904_mm}}
\end{figure}

As in previous studies of this burst \citep{fck+06,gfm07}, we find that an ISM model provides an 
adequate fit to the data. Our best-fit model is shown in Figure \ref{fig:050904_mm} and the 
corresponding physical parameters are listed in Table \ref{tab:bestfit}. The $8.5$\,GHz flux is 
severely suppressed by self absorption, with the self absorption frequency located around 
$280$\,GHz, above the characteristic synchrotron frequency, i.e., $\nu_{\rm m} < \nu_{\rm a}$. This 
requires a high-density circumburst environment, with \dens\ $\sim 10^3$\,$\rm{cm}^{-2}$, while
a jet break at $\sim2$\,d is required to explain the sharp drop in the NIR light curves.

\begin{figure}
\begin{tabular}{ccc}
\centering
 \includegraphics[width=0.31\columnwidth]{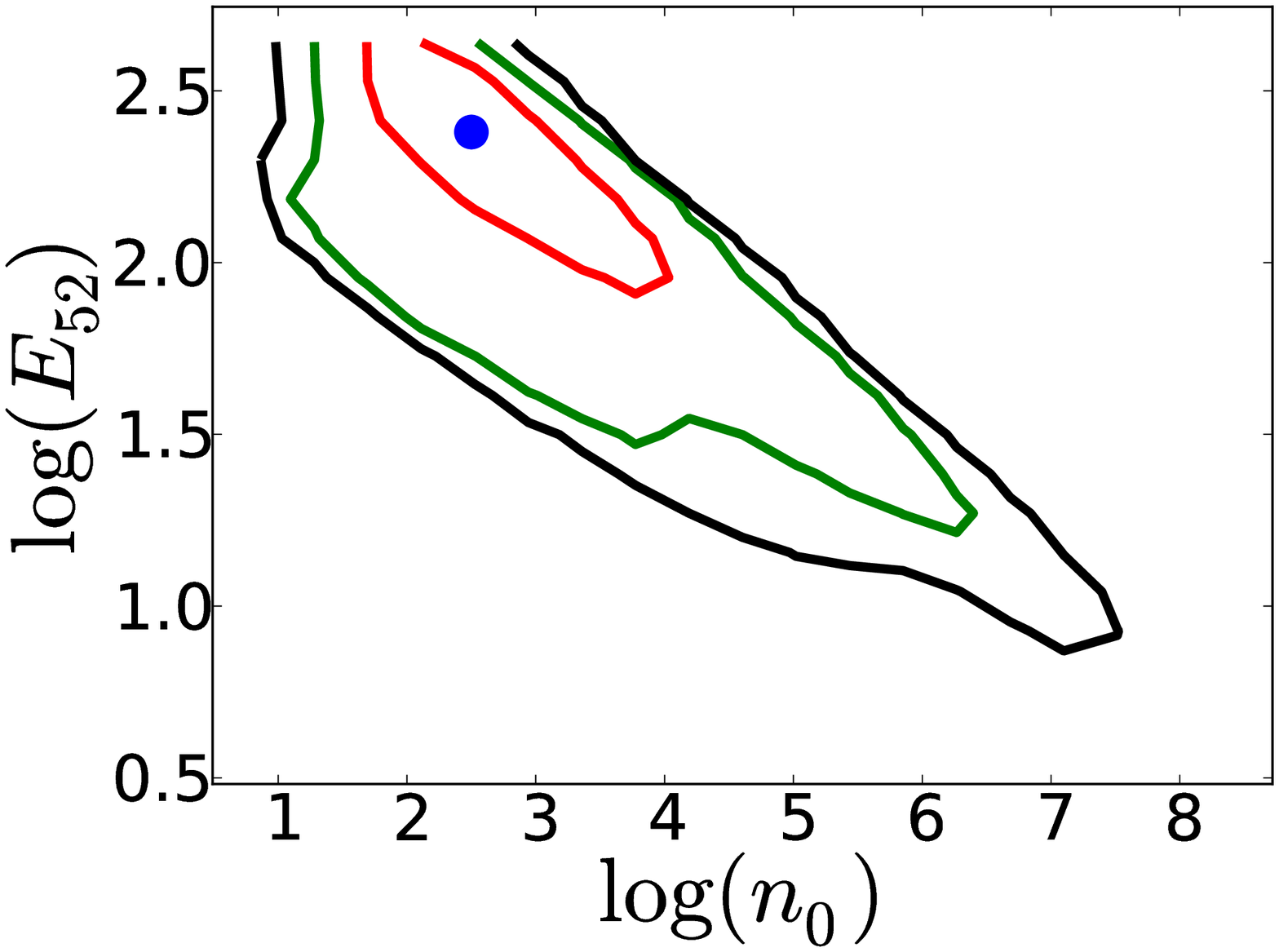} &
 \includegraphics[width=0.31\columnwidth]{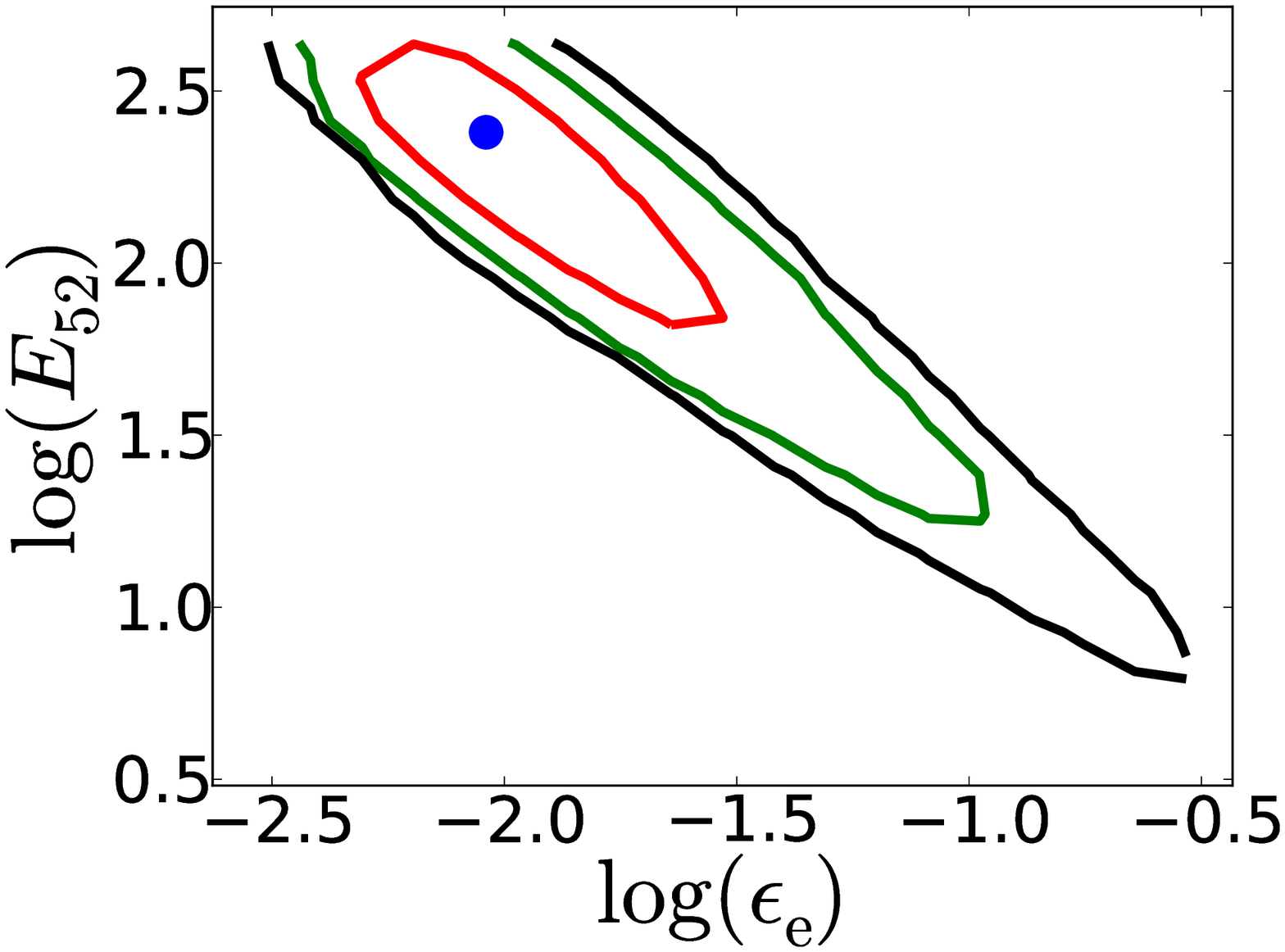} &
 \includegraphics[width=0.31\columnwidth]{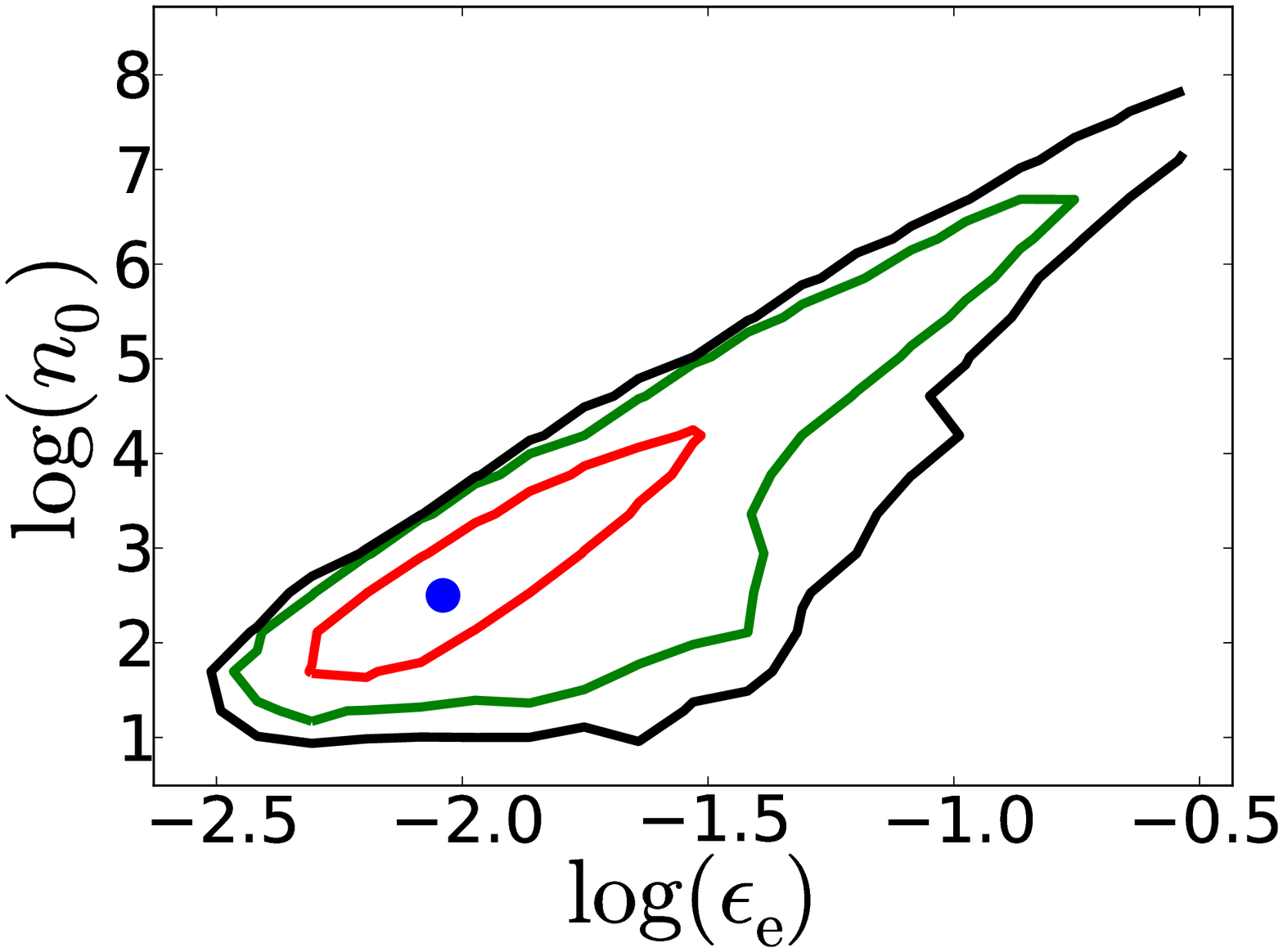} \\
 \includegraphics[width=0.31\columnwidth]{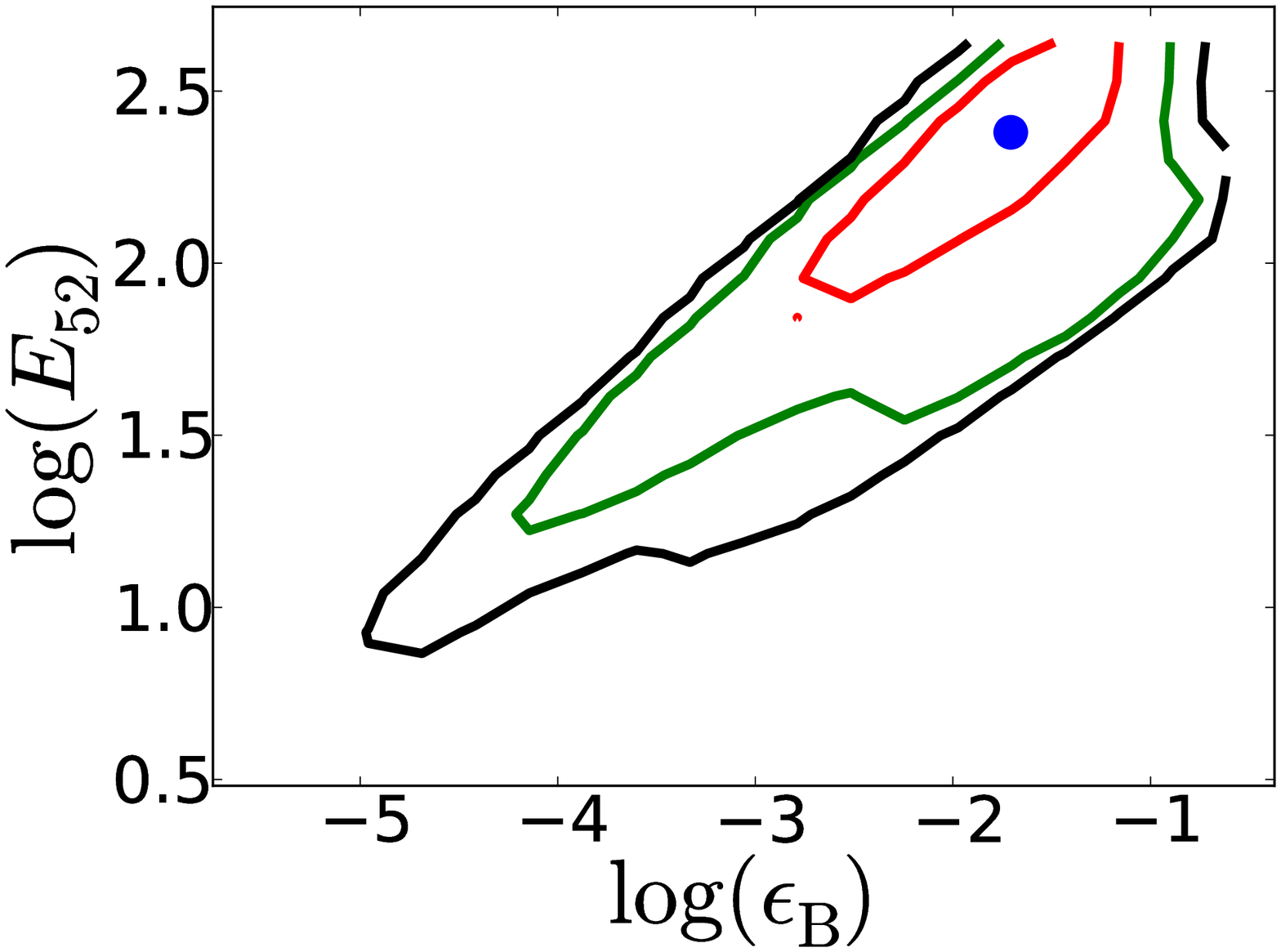} &
 \includegraphics[width=0.31\columnwidth]{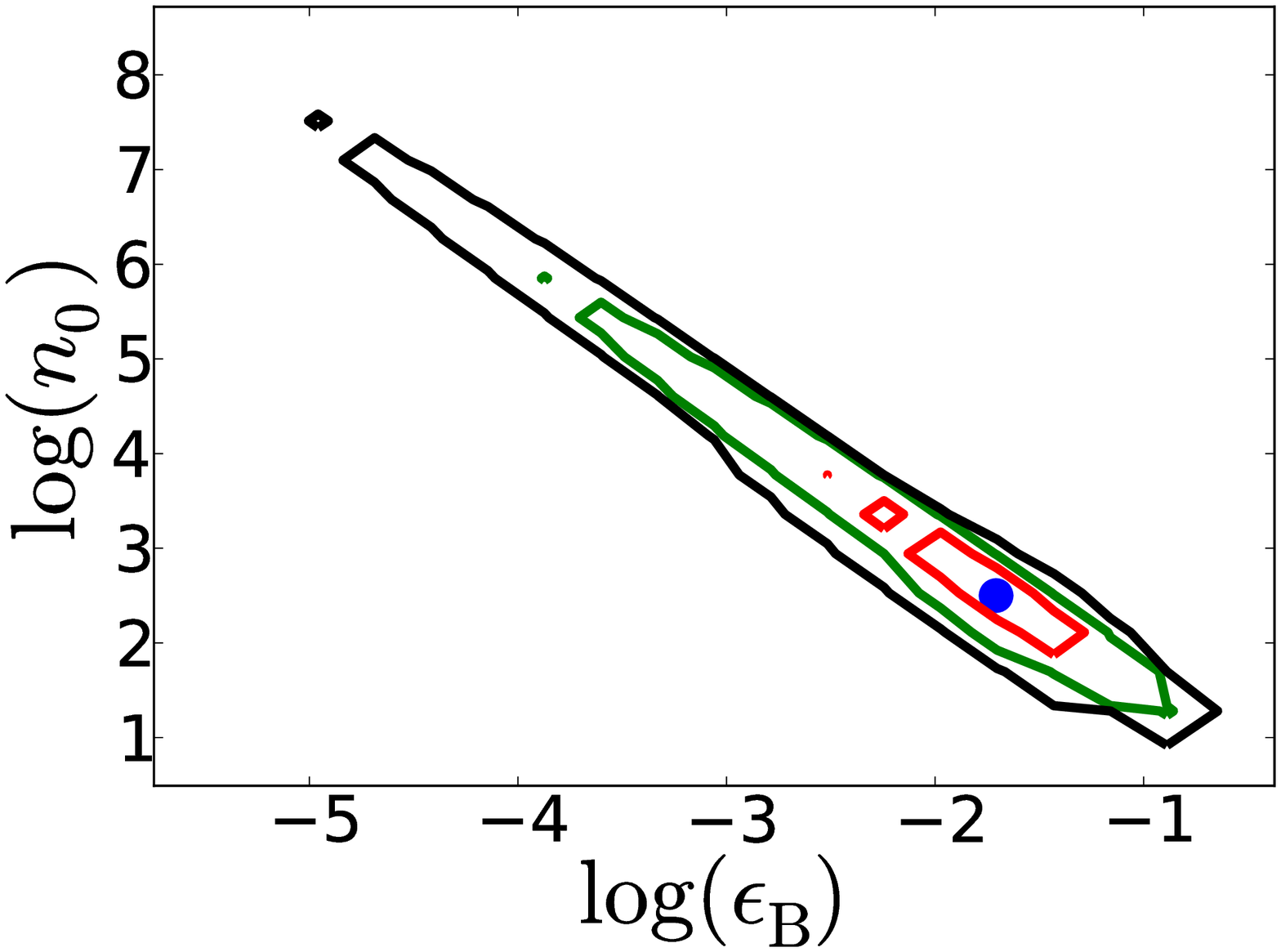} &
 \includegraphics[width=0.31\columnwidth]{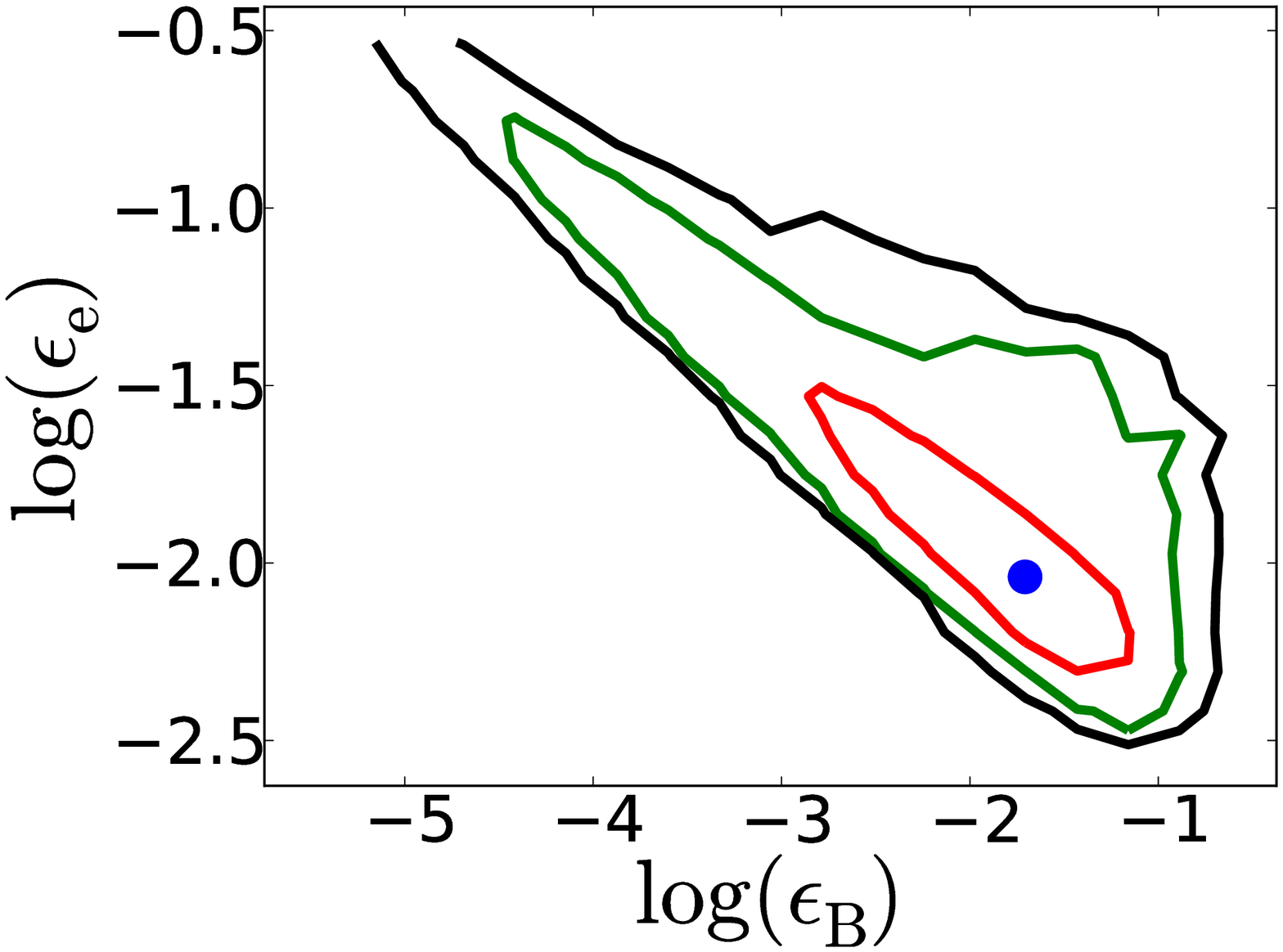} \\
\end{tabular}
\caption{1$\sigma$ (red), 2$\sigma$ (green), and 3$\sigma$ (black) contours for correlations
between the physical parameters, \E, \dens, \epse, and \epsb\ for GRB~050904 from
Monte Carlo simulations. We have restricted $E_{\rm K, iso, 52} < 500$, $\epsilon_{\rm e} <
\nicefrac{1}{3}$, and $\epsilon_{\rm B} < \nicefrac{1}{3}$.
See the on line version of this Figure for additional plots of correlations between these
parameters and $p$ and $\theta_{\rm jet}$. \label{fig:050904_mcmcgrid}}
\end{figure}

Using MCMC analysis, we confirm the high density of the circumburst environment, $\log{(n_{\rm 0})} 
= 2.8^{+1.1}_{-0.7}$, with \E$=(1.7^{+1.2}_{-1.0})\times10^{54}$\,erg,
\epse$=(1.2^{+1.5}_{-0.5})\times10^{-2}$, \epsb$=(1.3^{+2.2}_{-1.1})\times10^{-2}$, and 
$p=2.07\pm0.02$. The values of all the parameters are consistent with those reported by 
\citet{gfm07} within $\sim2\sigma$. We find a jet break time of $t_{\rm jet}=1.5^{+0.2}_{-0.1}$\,d 
which is earlier than $t_{\rm jet}\sim3$\,d reported previously \citep{tac+05,gfm07,kmk07}; however, 
our derived value of the jet opening angle, $\theta_{\rm jet}=6.2^{+3.3}_{-1.4}$\,deg is consistent 
with the value reported by \citet{gfm07}, who also performed a full multi-wavelength analysis. We 
compare our derived posterior density functions for $p$, \epse, \epsb, \dens, $E_{\rm K,iso}$, and 
$A_{\rm V}$ directly with those reported by \citet{gfm07} in Figure \ref{fig:050904_mcmcgrid}. Our 
distributions are similar, except that we find slightly smaller values for $p$. We note that we use 
different prescriptions for the synchrotron self-absorption frequency and evolution in the fast 
cooling regime. In addition, \citet{gfm07} include the effects of inverse Compton losses, which we 
ignore in our model. 

\begin{figure}
\begin{tabular}{ccc}
\centering
 \includegraphics[width=0.31\columnwidth]{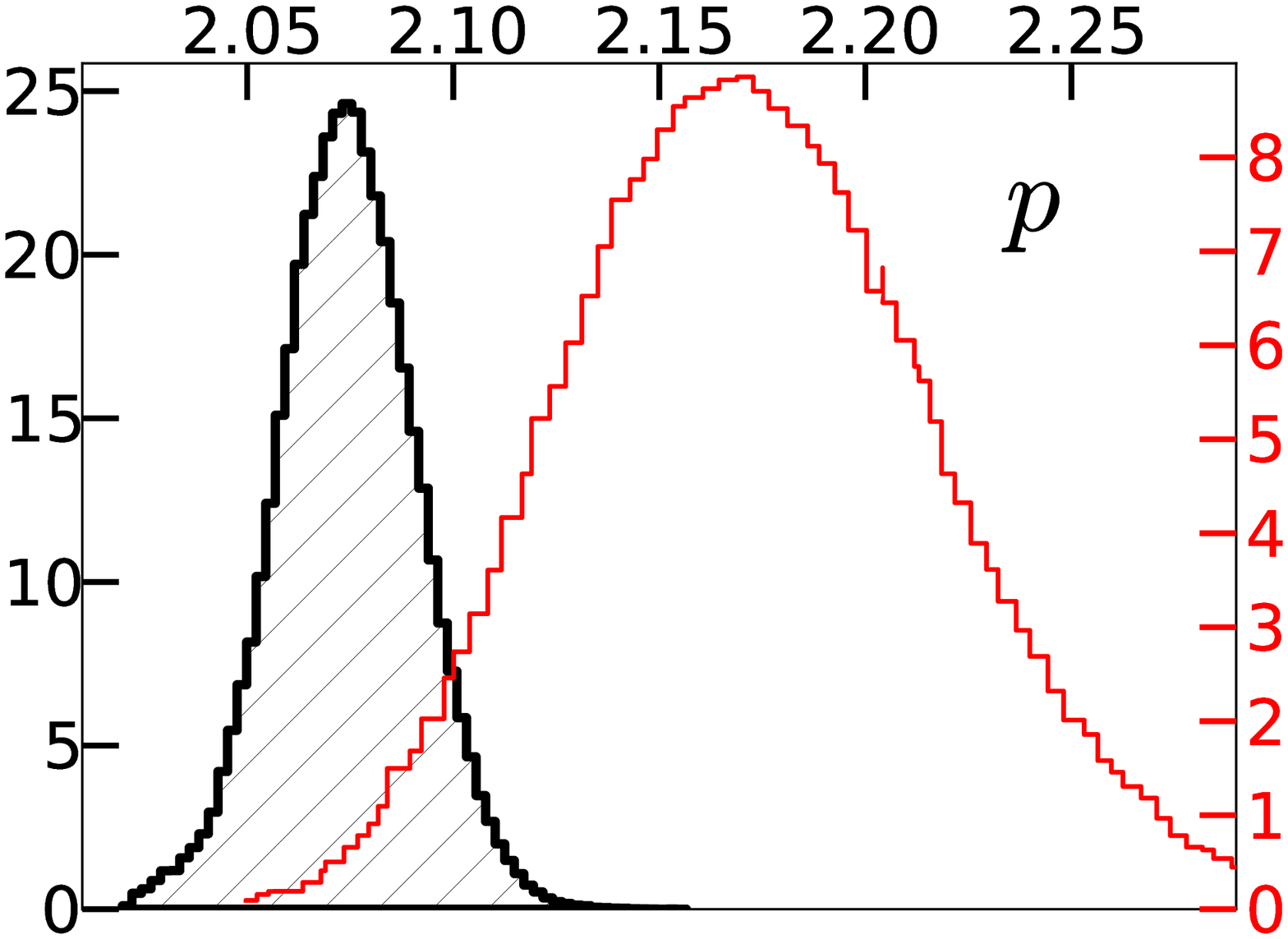} &
 \includegraphics[width=0.31\columnwidth]{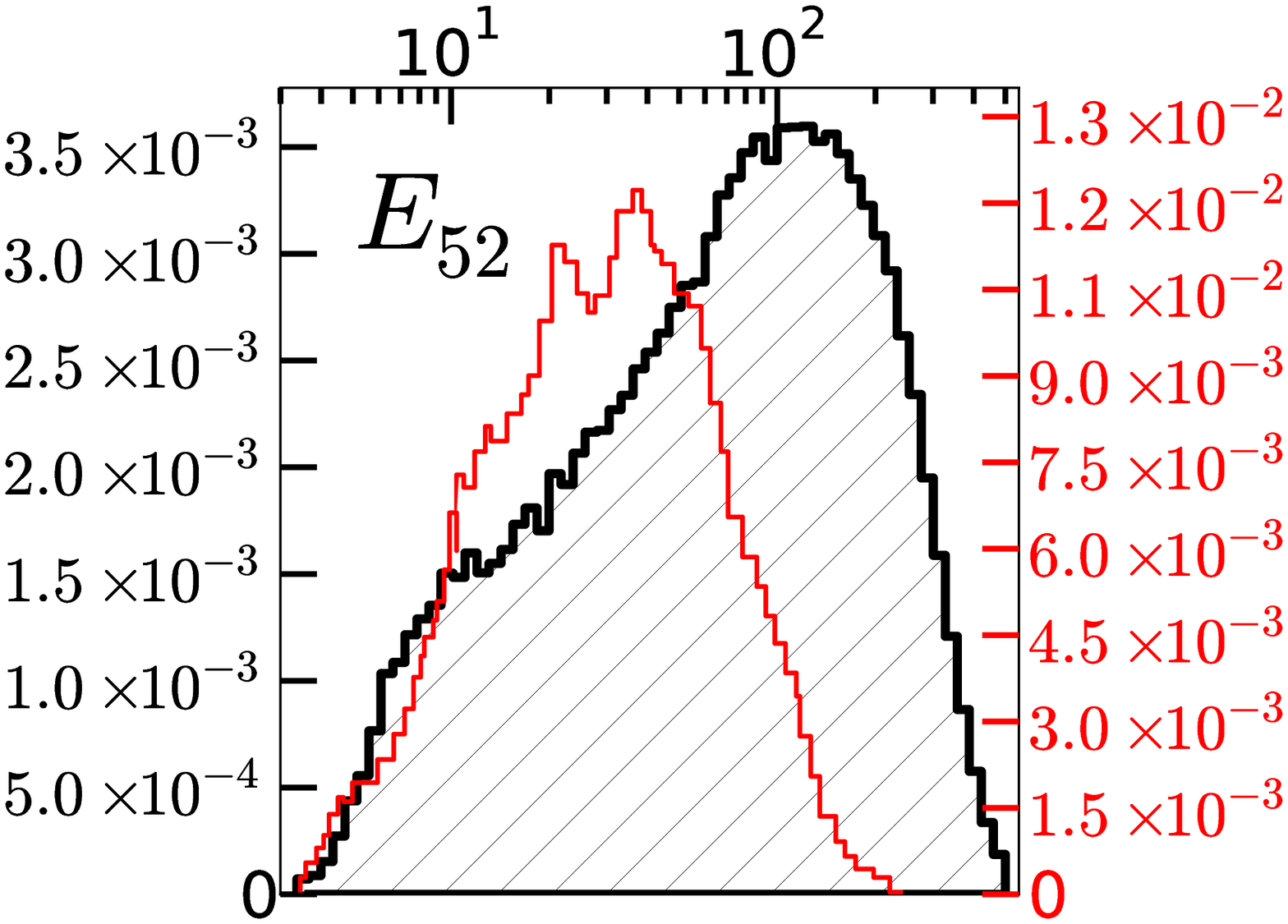} &
 \includegraphics[width=0.31\columnwidth]{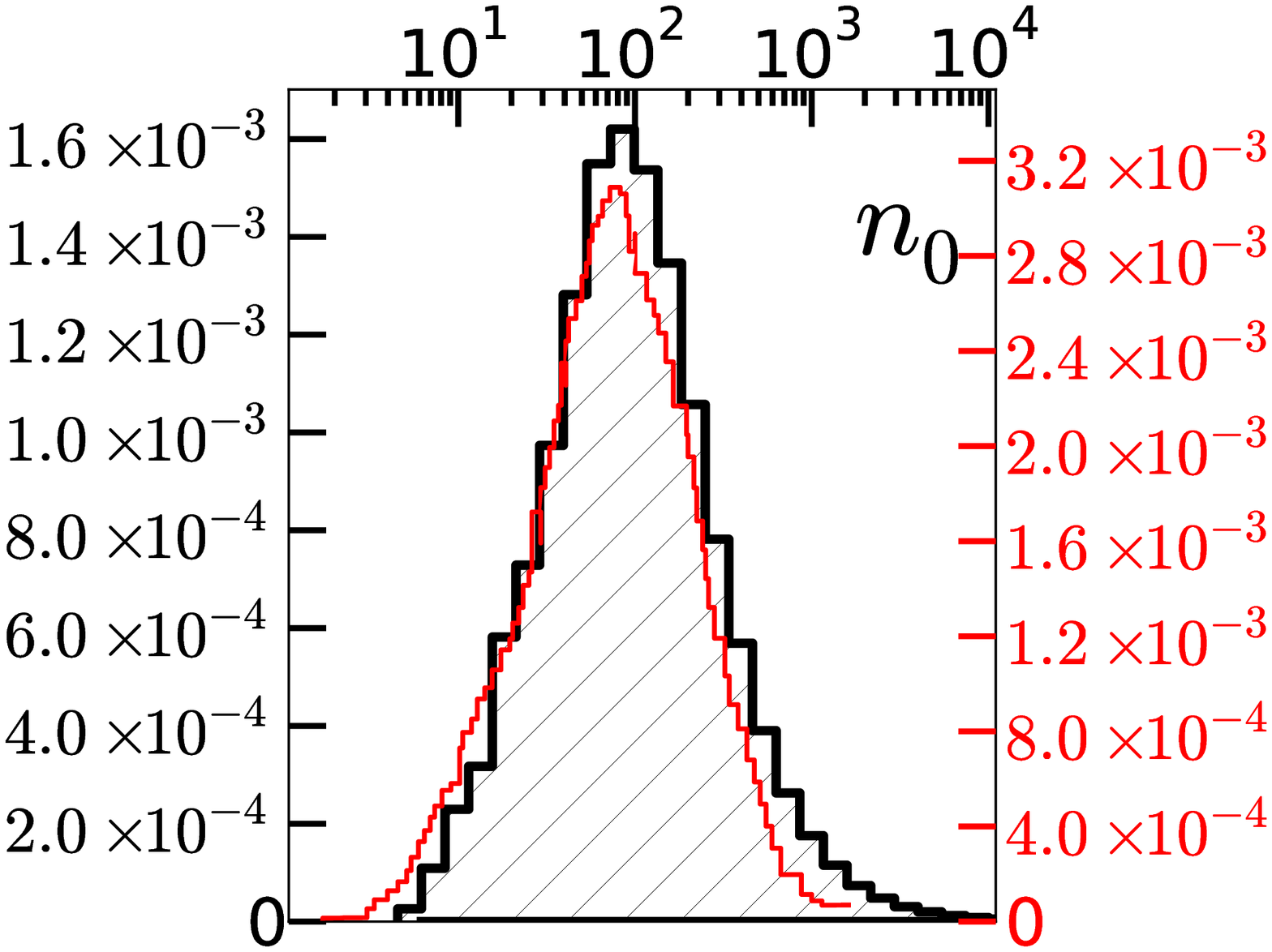} \\
 \includegraphics[width=0.31\columnwidth]{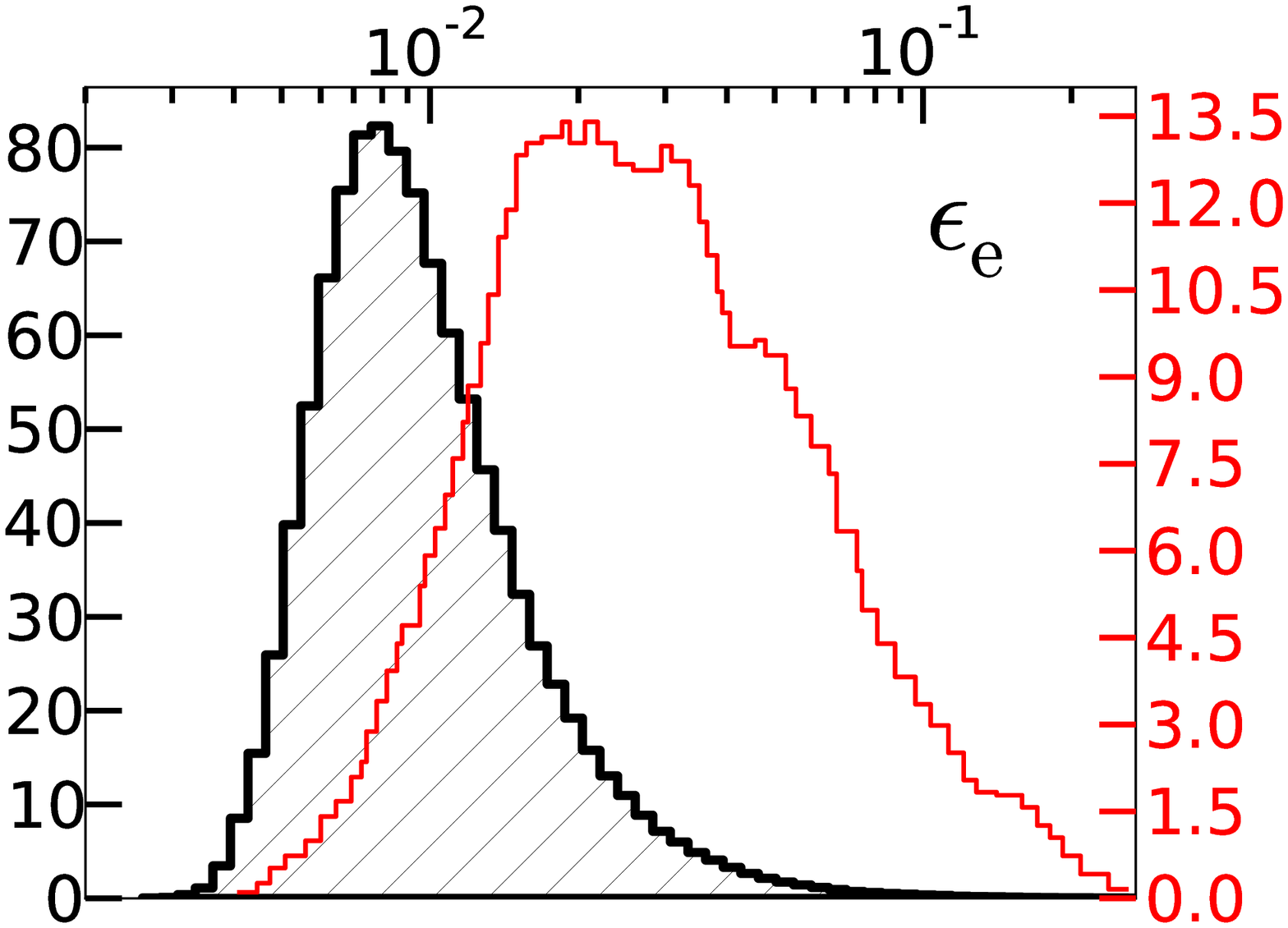} &
 \includegraphics[width=0.31\columnwidth]{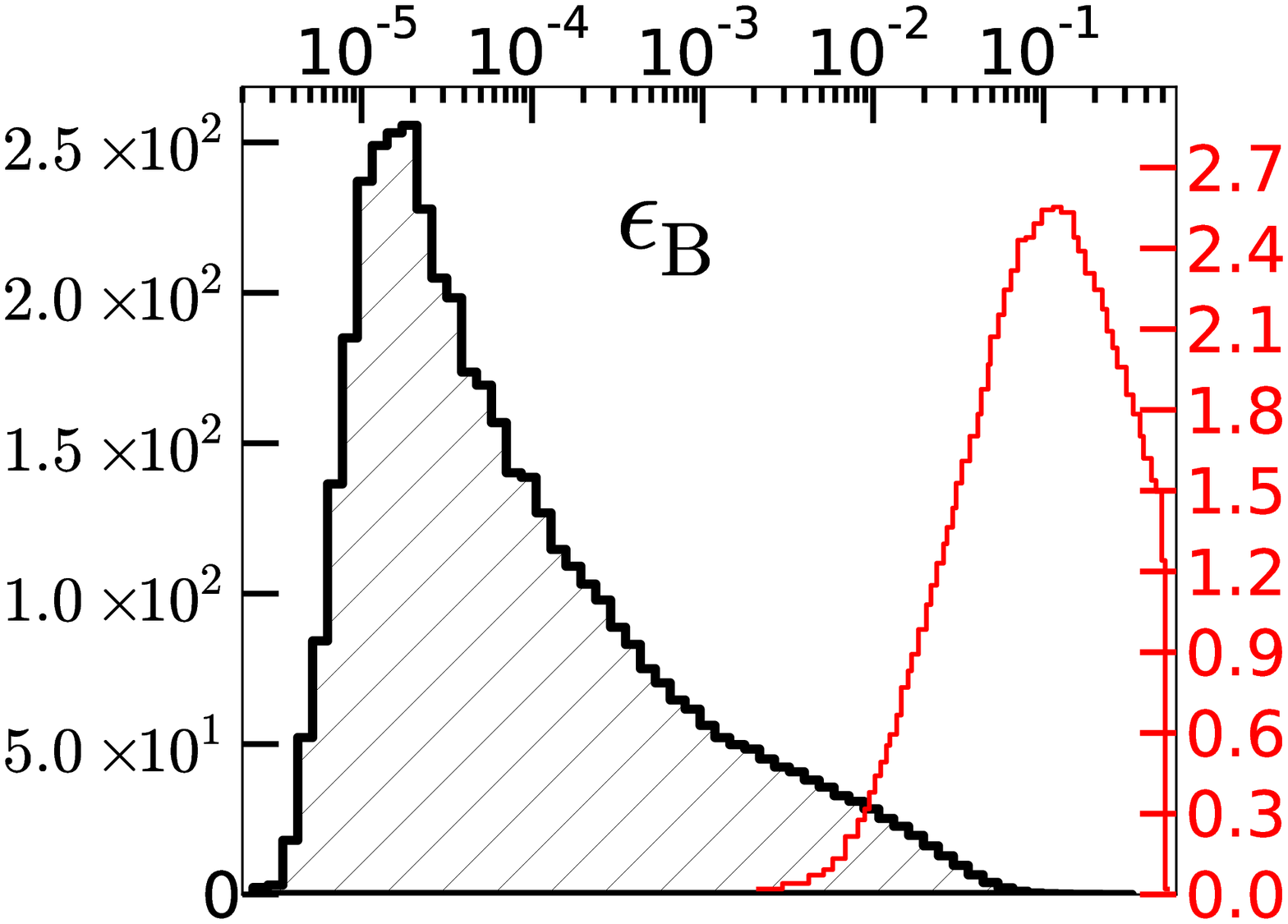} &
 \includegraphics[width=0.31\columnwidth]{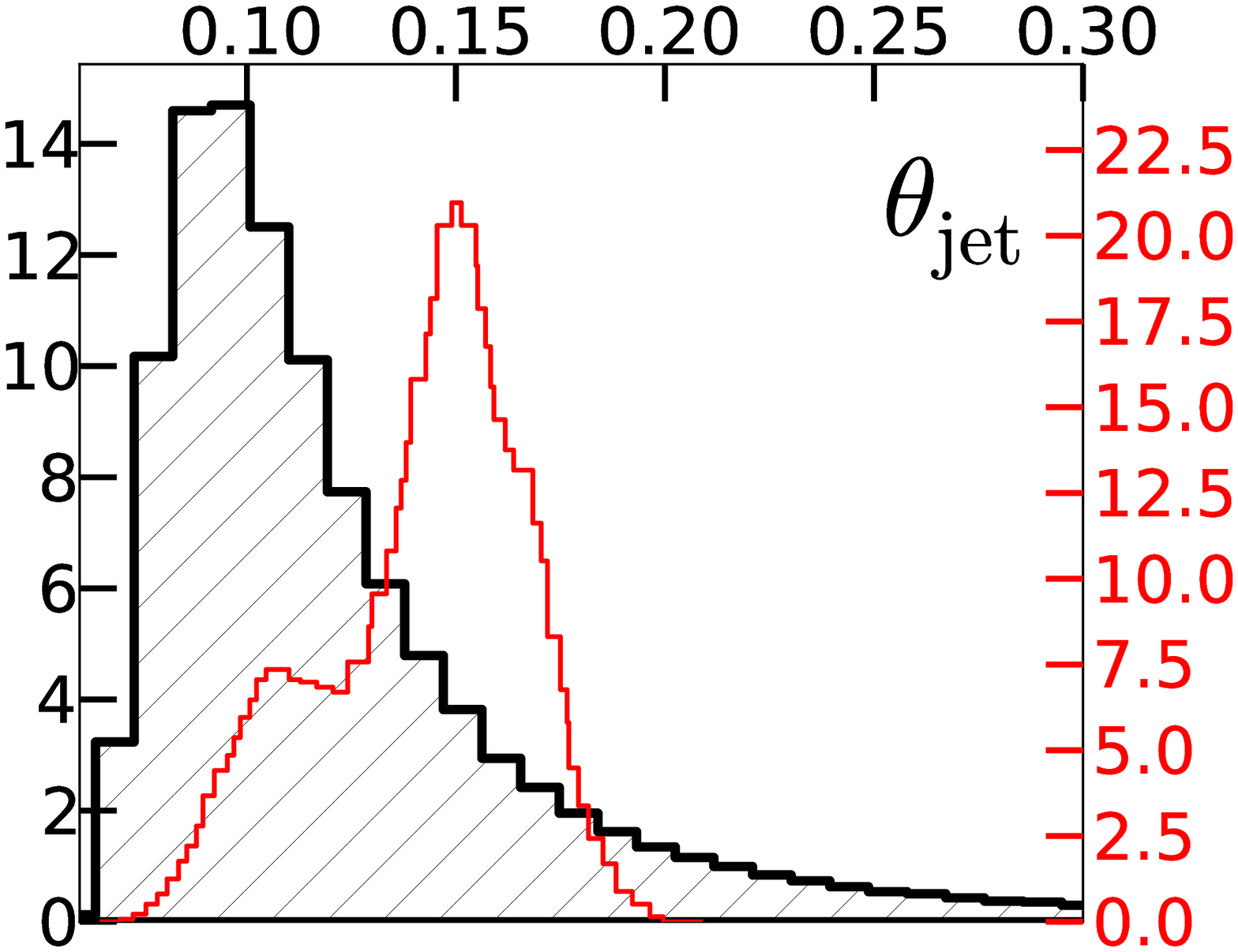} \\
\end{tabular}
\caption{Posterior probability density functions for the physical parameters of GRB~050904
(black curves and hatched regions; for details, see
Section \ref{text:modelling}), compared with the results of \citet{gfm07} (red curves). The 
extinction (\AV, not shown), is essentially unconstrained by the data, with the posterior density 
being very similar to the input (Jeffreys) prior. Note that these are \textit{density} functions, 
normalized such that the integral $\int_{-\infty}^{\infty}f(x)\,\mathrm{d}x=1$. Therefore the mode 
of one of these distributions may be different from the median value of the parameter, as the 
latter is computed using the corresponding probability \textit{mass} function. We have assumed that 
the `posterior distributions' presented in \citet{gfm07} also refer to density functions, and have 
normalized them to integrate to $1$.
} \label{fig:050904_MC_compISM}
\end{figure}

We find strong correlations between all four physical parameters (\epse, \epsb, \dens, and 
$E_{\rm K,iso}$; Figure \ref{fig:050904_MC_compISM}). Detailed investigation using the analytical 
expressions for the spectra in terms of the spectral break frequencies given in GS02 reveals the 
cause to be multiple levels of degeneracy. For instance, the characteristic synchrotron frequency 
is not well constrained, since it is located below the frequencies covered by our radio 
observations at all times. At the same time, $\nu_{\rm a}$ and the flux density at this frequency, 
$F_{\nu,\rm a}$, are not independently constrained, since this frequency lies below both the NIR 
and 
the X-rays. It is possible to change the two together in a way that leaves the NIR and X-ray light 
curves unchanged, without violating the radio limits. This latter degeneracy is the primary source 
of the observed correlations. We note that this degeneracy could have been broken with simultaneous 
detections in the radio and NIR. 

\subsection{GRB~090423}
\label{text:090423}
GRB~090423 was discovered with \Swift/BAT on 2009 April 23 at 7:55:19\,UT \citep{gcn9198}. The 
burst duration was $T_{90} = 10.3\pm1.1$\,s \citep{gcn9204}, with a fluence of $F_{\gamma} = (5.9 
\pm 0.4) \times10^{-7}$\,erg\,cm$^{-2}$ (15--150\,keV). The afterglow was detected by \Swift/XRT 
and ground-based near-infrared (NIR) follow-up observations, and the redshift, $z=8.26$, was 
confirmed by NIR spectroscopy \citep{sdvc+09,tfl+09}. The burst was also observed with the Spitzer 
Space Telescope \citep{gcn9582}, the Combined Array for Research in Millimeter-wave Astronomy 
\citep[CARMA;][]{cff+10}, the Plateau de Bure Interferometer \citep[PdBI;][]{gcn9273,duplm+12}, the
IRAM 30m telescope \citep{gcn9322}, the Westerbrock Synthesis Radio Telescope 
\citep[WSRT;][]{gcn9503}, and the VLA \citep{cff+10}.

We analyzed XRT data for this burst using methods similar to GRB~050904 and
GRB~120521C. We assume 
$N_{\rm H, MW} = 2.89\times10^{20}$\,cm${}^{-2}$ \citep{kbh+05}.
The best-fit
neutral hydrogen column density intrinsic to the host is 
$N_{\rm H, int} = \left(8.1^{+8.6}_{-6.5}\right)\times10^{22}$\,cm${}^{-2}$. 
In our temporally-resolved spectral analysis, we find that the X-ray photon index is consistent 
with $\Gamma = 2.03\pm 0.09$ ($68\%$ confidence interval) for all XRT data following 260\,s after 
the GRB trigger. We use this value of the photon index to convert the observed $0.3$--$10$\,keV
light curve to a flux density at 1.5\,keV (to facilitate comparison with \citealt{cff+10}). We 
compile all available photometry, together with our XRT analysis, in Table 
\ref{tab:090423:data:all}.

There are 134\,ks of unpublished X-ray data in the \Chandra\ archive for this GRB
(PI: Garmire), taken between 16 and 42~d after the burst and distributed
across five epochs. We downloaded and analyzed all available data from the
\Chandra\ archive. The GRB is marginally detected in three  of the five
epochs. We stacked observations taken close in time (epochs 1 and 2; epochs 3,
4, and 5) and restricted the energy range to $0.3$--$2$\,keV to increase the
signal-to-noise ratio. The GRB is marginally detected in both stacks. We report
the results of photometry using $1\arcsec.5$ apertures in Table \ref{tab:090423:data:Chandra}. We 
convert the measured count rates into flux densities at 1.5\,keV using the XRT photon index of 
$\Gamma=2.03$.

The 8.46\,GHz radio light curve peaks at a similar flux density as does the NIR light curve,
which strongly argues against a wind-like medium and suggests a constant-density
environment. We also note that the millimeter observations reported in \citet{duplm+12} are
inconsistent with the forward-shock synchrotron model, since the flux densities at $97$\,GHz are 
much higher than in any other waveband, whereas the ISM model for GRB afterglows predicts that 
light curves at each frequency would reach the same peak flux density prior to the jet break. The 
millimeter data are shown in Figure \ref{fig:090423_mm} for completeness, but have not been
included in the analysis. This was also noted by \cite{cff+10}, who suggested that the millimeter 
data and the first radio detection at 2.2\,d possibly included emission from a reverse shock. We 
investigate this possibility further in Appendix \ref{text:090423_RS}.

Our best-fit model requires that the afterglow be in the slow cooling phase with the spectral
ordering $\nu_{\rm a} < \nu_{\rm m} < \nu_{\rm c}$ and a peak flux density of 
$F_{\nu,\rm{max}}\approx 142\,\mu$Jy. At 1~day, the characteristic synchrotron frequency is 
$\nu_{\rm m} \approx 7.7\times10^{12}$\,Hz, while the cooling break is in the X-rays, at 
$4.5\times10^{17}$\,Hz ($1.8$\,keV). However, the data do not constrain \nua. In the ISM model \nua\ 
remains fixed before the jet break and falls as $t^{-0.2}$ after the jet break. Hence the only 
observational constraint on \nua\ is that it is located below the radio band at all times. The model 
shown in Figure \ref{fig:090423_mm} is therefore only one of a family of models that match the data 
and have $\nu_{\rm a} \lesssim 8$\,GHz.
Using the values of \numax, \nuc\ and $F_{\nu,\rm{max}}$ from our best-fit model and
the functional dependence of the microphysical parameters, \epse, \epsb, \dens, and \E\ on the
measured quantities \nua\, \numax, \nuc, and $F_{\nu,\rm{max}}$, we derive the following
constraints: 
$\epsilon_{\rm e}\approx 0.13{\nu_{\rm a, 8}}^{\nicefrac{5}{6}}$,
$\epsilon_{\rm B}\approx 4.0\times10^{-4}{\nu_{\rm a, 8}}^{\nicefrac{-5}{2}}$,
$n_{\rm 0}\approx7.5\times10^{-2}{\nu_{\rm a, 8}}^{\nicefrac{25}{6}}\,\rm{cm}^{-3}$, and
$E_{\rm K, iso, 52}\approx 72{\nu_{\rm a, 8}}^{\nicefrac{-5}{6}}$, 
where ${\nu_{\rm a, 8}}$ is the self-absorption frequency in units of $10^{8}$\,Hz. Imposing
the theoretical restriction, $\epsilon_{\rm e} < \nicefrac{1}{3}$, we can further restrict the 
self-absorption frequency to $\nu_{\rm a} \lesssim 3.1\times10^{8}\,$Hz. This allows us to
place an \textit{upper bound} on the circumburst density, $n_{\rm 0} \lesssim 8.3\,{\rm
cm}^{-3}$, and \textit{lower bounds} on the isotropic equivalent energy, $E_{\rm K, iso, 52} \gtrsim
28$ and $\epsilon_{\rm B} \gtrsim 2.4\times10^{-5}$. Similarly, imposing $\epsilon_{\rm B} <
\nicefrac{1}{3}$, we can place \textit{lower bounds} on the self-absorption frequency,
$\nu_{\rm a} \gtrsim 6.8\times10^{6}\,$Hz and the circumburst density, $n_{\rm 0} \gtrsim 
1.0\times10^{-6}\, {\rm cm}^{-3}$, and \textit{upper bounds} on the isotropic equivalent energy, 
$E_{\rm K, iso, 52} \lesssim 6.8\times10^{2}$ and $\epsilon_{\rm e} \gtrsim 1.4\times10^{-2}$. 

To further explore the degeneracies in the physical parameters of the explosion, we carried out an 
MCMC analysis with $p$ fixed at our best-fit value of $2.56$ (Figures \ref{fig:090423_mcmcgrid} and 
\ref{fig:090423_hists}). Our measured correlations between \E, \dens, \epse, and \epsb\ are 
consistent with the expected analytic relations. We find a small amount of extinction within the 
host galaxy ($A_{\rm V} = 0.15\pm0.02$\,mag), which is consistent with the low value of 
extinction ($A_{\rm V}\lesssim 0.1$\,mag) inferred by other authors based on the X-ray and NIR 
observations alone \citep{tfl+09,zwt+11}. 

A previous analysis of GRB~090423 claimed no jet break to $\approx 45$~d \citep{cff+10}. However, 
our model requires a jet break at $t_{\rm jet}\approx15$~d, driven by the late-time \Spitzer\ 
$3.6$\,$\mu$m detection, as well as the radio non-detection at 62~d. In particular, \numax\ 
passes through the radio band at 58 days while the radio light curve peaks at about 20~d, the 
signature of a jet break. In our model, the afterglow is optically thin at 8.46\,GHz at all times. 
Following the jet break, the $\nu^{1/3}$ part of the synchrotron spectrum transitions from 
$t^{\nicefrac{1}{2}}$ to $t^{\nicefrac{-1}{3}}$, followed by a transition to $t^{-p}$ when \numax\ 
crosses the radio band at 58~d, matching the observations. While the late-time \Chandra\ data 
do not show an obvious break, the model with $t_{\rm jet}\approx15$\,d is consistent with 
the full X-ray light curve including the \Chandra\ photometry and is required by the full model. 
From our MCMC analysis, we find $\theta_{\rm jet} = 1.5^{+0.7}_{-0.3}$\,degrees (68\% credible 
region). We list the best-fit parameters in Table \ref{tab:bestfit} and the results of the MCMC 
analysis in Table \ref{tab:mcmc}.

Using the distribution of jet opening angles from our MCMC analysis and the isotropic-equivalent
$\gamma$-ray energy, $E_{\gamma,\rm iso}=(1.03\pm0.3)\times10^{53}$\,erg \citep{gcn9251}, we
compute a beaming-corrected $\gamma$-ray energy of 
$E_{\gamma}=(3.2^{+2.7}_{-1.7})\times10^{49}$\,erg.
The deduced value of the afterglow kinetic energy from the MCMC analysis is $E_{\rm K,iso} =
(3.4^{+1.1}_{-1.4})\times10^{54}$\,erg, corresponding to a beaming-corrected energy of $E_{\rm
K}=(1.1^{+0.4}_{-0.2})\times10^{51}$\,erg. Together, these results imply a low radiative 
efficiency, 
$\eta\equiv \frac{E_{\gamma}}{E_{\rm K}+E_{\gamma}}\sim0.03$. However, we note that the value of 
$E_{\rm K}$ is sensitive to the upper cutoff of the prior on the $E_{\rm K,iso}$ and is affected by 
the strong correlation between $E_{\rm K,iso}$ and the other parameters due to the weak constraint 
on \nua. In particular, lower values of the kinetic energy are allowed (with the constraint, 
$E_{\rm K,iso,52}\gtrsim 30$, corresponding to $E_{\rm K}\gtrsim3\times10^{50}$ and $\eta \sim 0.4$ 
for our best-fit value of $\theta_{\rm jet}=2.5^{\circ}$). Hence our estimate of $\eta\sim0.03$ 
should be considered a lower bound.

\section{The Physical Properties of High-Redshift GRBs}
\label{text:comparisons}
Having performed afterglow modeling of the three existing GRBs at $z\gtrsim6$ with radio through 
X-ray data to determine the properties of the explosion and environment, we now 
turn to the question of how these events compare with each other, and with GRBs at lower redshifts.
We compile measurements of $E_{\gamma}$, $\theta_{\rm jet}$, $E_{\rm K}$, and $n_{0}$ (or $A_*$) 
for lower-redshift ($z\lesssim1$) events from the 
literature \citep{pk02,yhsf03,fb05,gngf07,cfh+10,cfh+11}. 
Where only a lower limit (or no information) is available for the jet opening angle, we use 
$E_{\gamma, \rm iso}$ as an upper bound on $E_{\gamma}$. This combined comparison sample includes 
GRBs from the pre-\Swift\ era, as well as \Swift\ and \textit{Fermi} events.
 
All three $z\gtrsim6$ GRBs presented here are well-fit by a constant density ISM
model. In the case of GRBs~090423 and 120521C, the synchrotron self-absorption frequency is not
directly observed and hence the best-fit model is only representative of a family of solutions. 
Despite this uncertainty, we are able to bound \nua\ using constraints on the microphysical 
parameters \epse, \epsb~$ < \nicefrac{1}{3}$. We find $1.7\times10^{8}\,{\rm Hz}<\nu_{\rm 
a}<2.7\times10^{9}\,{\rm Hz}$ for GRB~120521C, and $7.9\times10^{6}\,{\rm Hz}<\nu_{\rm 
a}<3.2\times10^{8}\,{\rm Hz}$ for GRB~090423. The corresponding constraints on the physical 
parameters for these two GRBs are $2.8\times10^{-4}\lesssim n_{\rm 0}<27$\,cm$^{-3}$, $2.9\lesssim 
E_{\rm K, iso, 52}\lesssim29$, $3.4\times10^{-2}\lesssim\epsilon_{\rm e}<\nicefrac{1}{3}$, and
$3.5\times10^{-4}\lesssim\epsilon_{\rm b}<\nicefrac{1}{3}$ for 
GRB~120521C, and $1.7\times10^{-6}<n_{\rm 0}<8.2$, $24\lesssim E_{\rm K, iso, 
52}\lesssim5.1\times10^{2}$, $1.5\times10^{-2}\lesssim \epsilon_{\rm e}<\nicefrac{1}{3}$, and
$3.3\times10^{-5}<\epsilon_{\rm B}<\nicefrac{1}{3}$ for GRB~090423. Together with the 
high density of $n_{\rm 0}\sim600\,{\rm cm}^{-3}$ for GRB~050904, these three high-redshift GRBs 
span the lowest to the highest densities inferred from GRB afterglow modeling (Figure 
\ref{fig:comp_EK}).

The light curves of all three high-redshift events display the signature of a jet break. Using the 
jet break time, we constrain the opening angle of the jet in each case and find $\theta_{\rm 
jet}\sim1.5^{\circ}$--$6^{\circ}$. The median\footnote{The uncertainty on the median is computed 
using Greenwood's formula for the variance of the Kaplan-Meier estimate of the cumulative 
distribution function. This method accounts for both upper and lower limits, which exist in the 
data.} jet opening angle of the low-redshift sample is $\theta_{\rm jet}=7.4^{+11}_{-6.6}$ (95\% 
confidence interval, Figure \ref{fig:comp_Egamma}). Whereas this interval formally includes the 
measurements of $\theta_{\rm jet}$ for the high-redshift sample, we note that the observed values 
of $\theta_{\rm jet}$ for the high-redshift sample are all below the best estimate for the median 
of the comparison sample, suggesting that higher-redshift events may be more strongly collimated 
than their lower-redshift counterparts. If this difference is verified with future events, it would 
indicate that previous studies may have underestimated the beaming correction and therefore the 
rate of $z\gtrsim6$ GRBs.

We use the calculated values of $\theta_{\rm jet}$ to compute the beaming-corrected $\gamma$-ray 
and kinetic energies of the high-redshift GRBs and find that both $E_{\gamma}$ and $E_{\rm K}$ span 
the range of $3\times10^{49}$\,erg to $\sim10^{52}$\,erg. We confirm previous reports that 
GRB~050904 is one of the most energetic GRBs ever observed \citep{gfm07}. GRB 120521C falls in the 
lower half of the distribution of $E_{\gamma}$ and $E_{\rm K}$, whereas GRB~090423 lies at the lower 
end of the distribution of $E_{\gamma}$ and near the median of the distribution of $E_{\rm K}$. The 
median values of these parameters for the low-redshift sample are $E_{\gamma} = 
\left(8.1^{+11}_{-4.3}\right)\times10^{50}\,$erg and $E_{\rm 
K}=\left(3.8^{+17}_{-2.6}\right)\times10^{50}\,$erg. The values of $E_{\gamma}$ and $E_{\rm K}$ for 
the three high-redshift GRBs span the observed distributions and present no evidence for a 
substantial difference from the low-redshift sample. The inferred $\gamma$-ray efficiencies ($\eta 
\sim 0.5$) are also similar to the efficiencies of lower-redshift events.

From this comparison, we conclude that the existing sample of $z\gtrsim6$ GRBs displays the same 
wide range of circumburst densities and beaming-corrected energies as their lower redshift 
counterparts (Figure \ref{fig:comp_Egamma}). On the other hand, the $z\gtrsim6$ events seem to have 
smaller jet opening angles than the median of the distribution at lower redshifts, suggesting that 
there might be some evolution in jet collimation with redshift. 

\section{Conclusions}
\label{text:conclusions}
We present X-ray, optical/NIR, and radio observations of GRB~120521C and use broad-band modeling to 
deduce a redshift of $z=6.01^{+0.05}_{-0.09}$, consistent with $z\sim5.93^{+0.11}_{-0.14}$ derived 
from optical/NIR SED-fitting and $z\sim6.15$ estimated from a low signal-to-noise spectrum. 
This is only the third GRB at $z\gtrsim6$ for which detailed multi-wavelength observations allow us 
to extract the properties of the explosion. The data suggest a constant-density circumburst 
environment with $\log{(n_{\rm 0})}=-2.7^{+1.4}_{-1.0}$, a jet-opening angle of $\theta_{\rm 
jet}=3.0^{+2.3}_{-1.1}$\,deg, beaming-corrected kinetic and $\gamma$-ray energies of 
$E_{\rm K} =\left(3.1^{+1.9}_{-0.9}\right)\times10^{50}$\,erg and 
$E_{\gamma}=\left(2.6^{+4.4}_{-2.0}\right)\times10^{50}$\,erg, 
and negligible extinction, $A_{\rm V} \lesssim 0.05$\,mag. We also re-fit the other 
two GRBs at $z\gtrsim6$ with radio detections and compare the properties of the high-redshift sample 
with those of their lower-redshift counterparts. We find that GRBs at $z\gtrsim6$ exhibit a wide 
range of explosion energies, circumburst densities, and shock microphysical parameters. The energies 
and circumburst densities of these high-redshift events are comparable to those of their 
counterparts at $z\sim1$, and overall, they display no evidence for an evolution in the 
progenitor population compared to $z\sim1$ events. 

We note that GRBs at $z\gtrsim6$ may have systematically smaller jet opening angles, with a mean of 
$\theta_{\rm jet} =3.6\pm0.7$\,deg, which would increase the inferred GRB rate at these 
redshifts by a factor of $\approx$ 4.
We caution that our results are based on a small sample of three events at $z\gtrsim6$. The primary 
reason for the small sample size is the historically low detection rate of GRB afterglows at radio 
frequencies. Like previous authors, we note that the lack of early-time radio data makes it 
difficult to determine the synchrotron self-absorption frequency, which in turn results in 
parameter degeneracies, giving rise to uncertainties in these parameters of several orders of 
magnitude. Rapid-response radio observations are therefore essential for studying the properties of 
GRBs, both at low and high redshifts. The recent refurbishment and expansion of the Very Large 
Array has resulted in an improvement in sensitivity by an order of magnitude, while the Atacama 
Large Millimeter Array promises to be an excellent facility for the study of GRBs owing to its 
excellent sensitivity. Detailed studies of high-redshift candidate afterglows with these facilities 
(e.g. the recent $z=5.913$ GRB 130606A; \citealt{gcn14817}) will augment this sample and help bring 
the study of GRBs in the reionization era into the mainstream.

\acknowledgements
The National Radio Astronomy Observatory is a facility of the National Science Foundation operated 
under cooperative agreement by Associated Universities, Inc. Radio data for GRB 120521C were 
obtained under VLA project codes 12A-394 and 12A-480. Some of the data presented here were obtained 
at the Gemini-North Observatory, which is operated by the Association of Universities for Research 
in Astronomy (AURA) under a cooperative agreement with the NSF on behalf of the Gemini-North 
partnership. The William Herschel Telescope is operated on the island of La Palma by the Isaac 
Newton Group in the Spanish Observatorio del Roque de los Muchachos of the Instituto de Astrofisica 
de Canarias. This work made use of data supplied by the UK Swift Science Data Centre at the 
University of Leicester. We thank Ofer Yaron and S.~R.~Kulkarni for the Keck observations. The 
Berger GRB group at Harvard is supported by the National Science Foundation under Grant AST-1107973. 
TL acknowledges support by NRAO. KW acknowledges support by STFC. 

\appendix
\section{Weighting}
\label{text:weights}
The behavior of the various spectral power law segments of a synchrotron source as outlined in 
GS02 is strictly valid only when the various spectral break frequencies are located far apart. 
However, the break frequencies evolve as a function of time and can cross, leading to transitions 
from one spectral shape to another. Since the normalizations of the light curves in GS02 was 
calculated in the asymptotic limit, spectral transitions (that occur when break 
frequencies approach each other and cross) lead to artificial discontinuities in the model light 
curves. We smooth over these glitches by adding together weighted combinations of all spectra that 
are accessible with the specified physical parameters (see \S 5 of GS02). For instance, in the ISM 
model with \dens$E_{\rm K,iso,52}^{4/7}$\epsb$^{9/7}$ < 18, we expect the afterglow to evolve in 
the order spectrum 5 $\rightarrow$ 1 $\rightarrow$ 2. Consequently in this example, we add together 
a combination of spectra 5, 1, and 2 with time-varying weights such that the appropriate spectrum 
presents the dominant contribution in the corresponding asymptotic limit, whereas at a spectral 
transition (defined next), the two spectra on either side of the transition contribute equally.

For a transition from spectrum A to spectrum B, we define the transition time,
$t_{\rm AB}$ as the geometric mean of the time when spectrum A ceases to be
valid and the time when spectrum B first becomes valid. In the above
example with the spectra evolving in the order 5 $\rightarrow$ 1
$\rightarrow$ 2, there are two transition times, denoted as $t_{51}$ and
$t_{12}$, respectively.

Next, we construct weighting functions for each spectrum as follows. If a
spectrum is valid in the range $(-\infty,t_{\rm AB}]$ (such as spectrum 5 in
the example above), the weighting function ($w_{\rm L}$ for `left') is unity at
early times, and falls as a power law near $t_{\rm AB}$, being equal to
$\nicefrac{1}{2}$ at $t_{\rm AB}$:
\begin{equation}
 w_{\rm L}(t,t_{\rm AB}) = \frac{1}{1+\left(t/t_{\rm AB}\right)^{\eta}},
\end{equation}
where $\eta$ is an ad-hoc parameter that controls the smoothness of the transition.

Similarly, if a spectrum is valid in the range $[t_{\rm AB},\infty)$ (such as
spectrum 2 in the example above), the weighting function ($w_{\rm R}$ for
'right') rises as a power law at early times, is equal to $\nicefrac{1}{2}$ at
$t_{\rm AB}$, and asymptotes to unity as $t\rightarrow\infty$:
\begin{equation}
 w_{\rm R} (t,t_{\rm AB})= \frac{1}{1+\left(t/t_0\right)^{-\eta}}.
\end{equation}

Finally, for a spectrum that is bracketed by two transition times, $[t_1,t_2]$
(such as spectrum 1 above; note that this can be true of more that one
spectrum), we define a weighting function, $w_{\rm M}$ (for `mid'):
\begin{equation}
 w_{\rm M} (t,t_1,t_2)= w_{\rm R}(t, t_1) + w_{\rm L}(t, t_2) - 1.
\end{equation}

The compound spectrum at any instant, $F_{\nu}(\nu,t)$ is then computed by
adding together weighted contributions from all spectra allowed under the given
set of physical parameters. For instance, in the above example, 
\begin{equation}
 F_{\nu}(\nu,t) = \frac{w_{\rm L}(t,t_{51}) F_{\nu}^{(5)}(\nu,t) +
w_{M}(t,t_{51},t_{12}) F_{\nu}^{(1)}(\nu,t) + w_{\rm R}(t,t_{12})
F_{\nu}^{(2)}(\nu,t)}
{w_{\rm L}(t,t_{51}) + w_{M}(t,t_{51},t_{12}) + w_{\rm R}(t,t_{12})}
\end{equation}

Since these weighting functions are designed to evaluate to unity far away from a spectral 
transition and fall as a power law near transitions, the above expression evaluates to the correct 
spectral shapes in all asymptotic limits. The weighting functions for two adjoining spectra at the 
transition time are both equal to one half, so both neighboring spectra contribute equally at a
spectral transition; this results in smooth light curves at all frequencies even across spectral 
transitions. Finally, we note that the index $\eta$ is an arbitrary choice; we find that $\eta = 2$ 
(corresponding to weighting by hyperbolic tangent functions in log-space) works well and yields 
smooth light curves near transitions, without significantly disturbing the spectrum away from 
transitions.

\section{Scintillation}
\label{app:scintillation}
Radio emission from a GRB afterglow traversing the Milky Way is susceptible to scintillation -- 
scattering by inhomogeneities in the electron density distribution of the interstellar medium (ISM) 
along the line of sight. The phenomenon is often modeled as being produced at a scattering screen 
located between the source and the observer. The screen produces a speckle pattern on
the detection plane, resulting in a modulation of the flux as the observer moves through the 
speckles. The effect of scintillation decreases above a transition frequency, characteristic of the 
general direction of the line of sight through the Galaxy (typically around 10~GHz).

The spectrum of the electron density inhomogeneities in the ISM is
well-characterized by the Kolmogorov spectrum \citep{ars95},
\begin{equation}
 \Phi_{{\rm N}_{\rm e}}(\vec{q}) = C_{\rm N}^2 q^{-11/3},
\end{equation}
where $\vec{q}$ is the wave-vector and $C_{\rm N}^2$ is a normalization constant
that varies from place to place with the Galaxy. The scattering measure is defined as the integral 
of $C_{\rm N}^2$ from the observer to the scattering screen,
\begin{equation}
 SM = \int_0^{d_{\rm scr}}\!C_{\rm N}^2(x)\,\mathrm{d}(x).
\end{equation}
\citet{cl02} used pulsar observations to build a model of the electron density
distribution in the Galaxy. We use their model, NE2001
\footnote{\footnotesize\url{http://www.astro.cornell.edu/~cordes/NE2001/}}, to
determine the scattering measure and transition frequency along the line of
sight to the GRB. We then compute the distance to the scattering screen using
the formula \citep{cl02},
\begin{equation}
 d_{\rm scr} = 2\pi\left(\frac{n_0}{318.0\,{\rm GHz}}\right)^{3.4} SM^{-1.2}.
\end{equation}
The strength of the scattering can be quantified by a parameter, $U$,
defined as
\begin{equation}
 U^{5/3} = \xi = 7.9\times10^3 SM^{0.6} {d_{\rm scr}}^{0.5}
                 \left(\frac{\nu}{1\,{\rm GHz}}\right)^{-1.7},
\end{equation}
where $\nu$ is the observing frequency, with $U \ll 1$ and $U \gg 1$
corresponding to the weak and strong scattering regimes, respectively
\citep{gn06, wal98,wal01}.

Having calculated $U$, we follow the prescription of \citet{gn06} to compute the
modulation index, $m$, computing the source size from the formula in Appendix A
of \citet{gs02}. The expected scatter in the observed flux density due to
scintillation is then given by
\begin{equation}
 \Delta F _{\rm scint} = m F_{\rm model},
\end{equation}
where $F_{\rm model}$ is the predicted flux density from the afterglow
synchrotron model. We add this uncertainty in quadrature to the flux
density uncertainty in each data point prior to performing likelihood analyses.

\newcommand{\numrs}{$\nu_{\rm m, RS}$}
\newcommand{\nuars}{$\nu_{\rm a, RS}$}
\newcommand{\numfs}{$\nu_{\rm m, FS}$}
\newcommand{\nuafs}{$\nu_{\rm a, FS}$}
\newcommand{\fnumrs}{$F_{\nu, \rm m, RS}$}
\newcommand{\fnuars}{$F_{\nu, \rm a, RS}$}
\newcommand{\fnumfs}{$F_{\nu, \rm m, FS}$}
\newcommand{\fnuafs}{$F_{\nu, \rm a, FS}$}
  
\section{A Possible Reverse Shock in GRB~120521C}
\label{text:120521C_RS}
GRB~120521C exhibits excess radio emission at 21.8\,GHz at 1.15\,d compared to the best-fit forward 
shock (FS) model (Figure \ref{fig:120521C_multimodel_ISM}). In Section \ref{text:results} we 
suggested that this may be due to contribution from a reverse shock (RS). Here we discuss a 
self-consistent RS + FS model that accounts for this excess emission. We do not search all 
possible RS models exhaustively, since the excess emission is observed in only a single data point, 
but list a plausible model that accounts for the observations.

We begin with a general discussion of the radio light curve of reverse shocks in an ISM 
environment. In the standard afterglow model, the reverse shock produces a synchrotron spectrum 
with a characteristic synchrotron frequency ($\nu_{\rm m, RS}$), cooling frequency ($\nu_{\rm c, 
RS}$),  self-absorption frequency ($\nu_{\rm a, RS}$), and overall flux normalization ($F_{\nu, \rm 
m, RS}$). At the time the reverse shock traverses the ejecta, the deceleration time 
($t_{\rm dec}$), these parameters are linked to those of the forward shock by the relations, 
$\nu_{\rm m, RS}(t_{\rm dec}) = \nu_{\rm m, FS}(t_{\rm dec})/\Gamma^2$,
$\nu_{\rm c, RS}(t_{\rm dec}) = \nu_{\rm c, FS}(t_{\rm dec})$, and
$F_{\nu, \rm m, RS}(t_{\rm dec}) = \Gamma F_{\nu, \rm m, FS}(t_{\rm dec})$, where
$\Gamma$ is the initial Lorentz factor of the ejecta. We use the simplest model to explain the data 
for GRB~120521C and assume that the ejecta are in the slow cooling regime ($\nu_{\rm c, 
RS}>\nu_{\rm m, RS}$) after $t_{\rm dec}$, 
although it is possible that the opposite is true in the initial afterglow phase. At low 
frequencies 
and early times the reverse shock emission is expected to be self-absorbed 
\citep[e.g.][]{sp99a,bsfk03,mkm+10} and the light curve therefore depends upon the relative 
ordering 
of \numrs, \nuars, and the observing frequency.

We note that the 21.8\,GHz radio detection for GRB~120521C at $1.15$\,d, with an excess flux 
density of $80\,\mu$Jy compared to the FS model, is preceded by a deeper non-detection at the same 
frequency at 0.15\,d. Subtracting the FS contribution to the 21.8\,GHz flux density at 0.15\,d, 
we find an upper limit to the RS contribution at 0.15\,d of $\lesssim34\,\mu$Jy. The light curve at 
21.8\,GHz is thus clearly rising between 0.15 and 1.15\,d and falling thereafter, implying that it 
reached a peak some time between 0.15 and 1.15\,d and indicating that the putative RS component is 
self-absorbed at this frequency at 0.15\,d. Regardless of the ordering of \nuars\ and \numrs, 
a peak in the 21.8\,GHz light curve must correspond to the passage of \nuars through this 
frequency, since this is the only way to explain a late-time ($t>t_{\rm dec}$) turn-over in a 
RS light curve. If we assume that \numrs~$>$~\nuars, then \numrs\ must pass through 21.8\,GHz even 
later than the apparent peak of the 21.8\,GHz light curve at $\approx 1$\,d. Our ISM model 
indicates \numfs~$=5.5\times10^{11}$\,Hz at 1\,d, implying $\Gamma(t_{\rm dec}) = \sqrt{\nu_{\rm m, 
FS}/\nu_{\rm m, RS}}\lesssim5$, which is too low. 

We therefore look for a self-consistent RS solution with \numrs~$<$~\nuars\ at 0.15\,d. In this 
scenario, the light curve rises as $t^{5/4}$ prior to the passage of \nuars, and 
then declines as $t^{-\frac{3p+1}{4}}\sim t^{-1.88}$ (using $p=2.17$, the median value estimated for 
the FS). From the upper limit at $0.15$\,d we can determine the earliest time at which \nuars\ can 
pass through $21.8$\,GHz. We find \nuars~$=21.8$\,GHz at $\gtrsim0.66$\,d and 
\fnuars~$\lesssim0.2\,\mu$Jy. This method does not allow us to precisely locate \numrs, with the 
only constraint that it passes through 21.8\,GHz at $\lesssim0.66$\,d. 
If we additionally assume that $t_{\rm dec} \sim T_{\rm 90} \approx 27$\,s, we find a solution that 
satisfies the relations at the deceleration time with $\Gamma\sim70$ and 
\numrs~$\sim2\times10^{8}$\,Hz at $0.66$\,d. We show this combined RS+FS model in Figure 
\ref{fig:120521C_multimodel_ISM_RS} and note that this model obeys the NIR limits at $\approx 
0.21$\,d.

To summarize, there exists a combined RS + FS model that explains the excess flux density at 
21.8\,GHz at 1.15\,d. Assuming that the deceleration time is of the order of $T_{90}$, we arrive at 
an initial Lorentz factor of $\sim70$ for this GRB, of the correct order of magnitude for GRBs 
\citep{sp99a,bsfk03,sr03}.

\section{A Possible Reverse Shock in GRB~090423}
\label{text:090423_RS}
\begin{figure}[ht]
\centering
\includegraphics[width=\columnwidth]{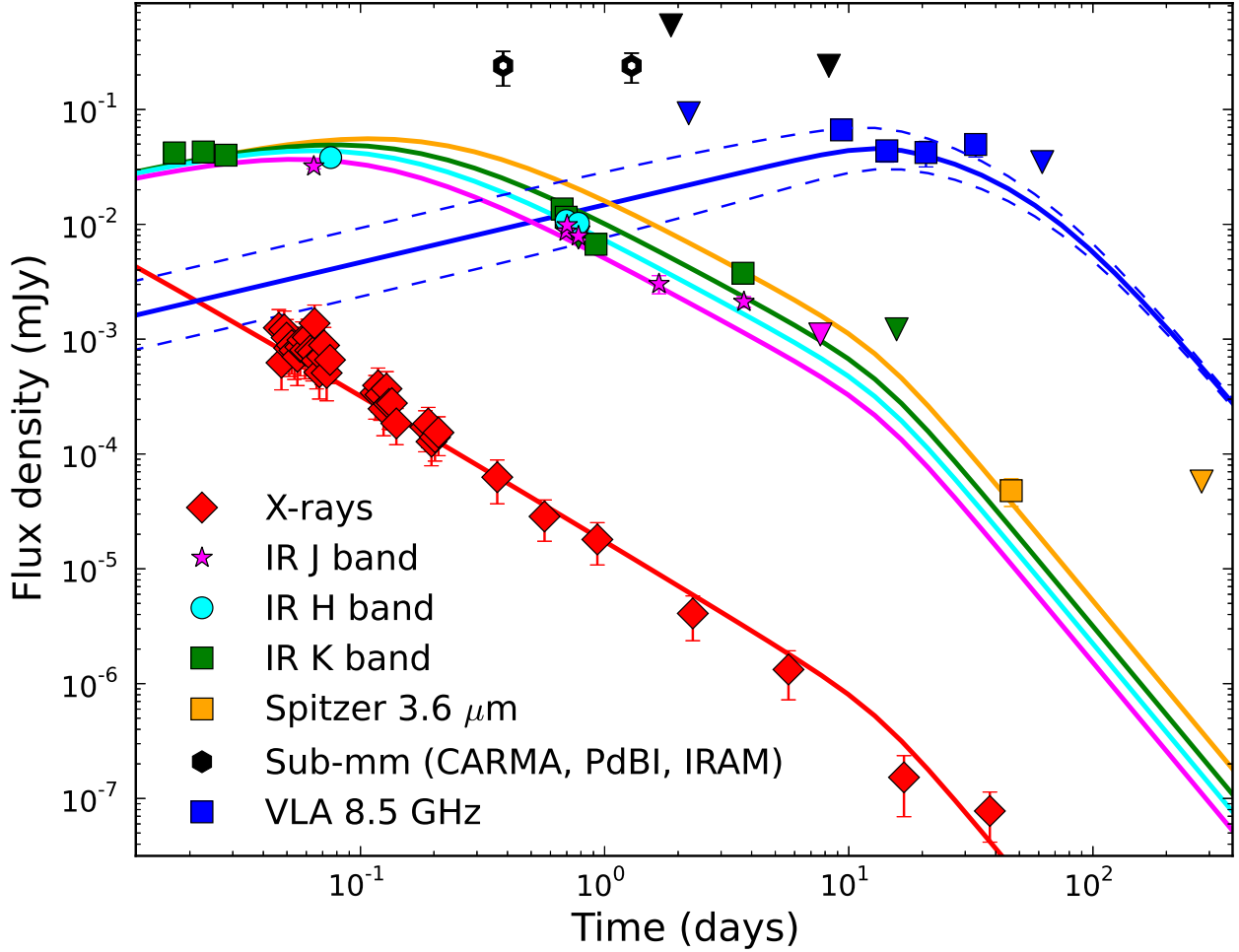}
\caption[]{Multi-wavelength modeling of GRB~090423 for a forward shock model with a 
homogeneous (ISM) environment \citep{gs02}. Triangles indicate upper limits and the dashed
lines show the point wise estimate of the 1$\sigma$ variation due
to scintillation. The X-ray points after 10 days are two separate stacks of
five \Chandra/ACIS observations. The millimeter data (CARMA, PdBI, IRAM) are
shown here for completeness, but are not included in the fit since the high
flux levels reported in these observations are not consistent with the peak
flux density observed in the NIR and radio bands.
This model corresponds to the parameters listed in Table \ref{tab:bestfit} and represents 
a family of models with identical light curves and $\nu_{\rm a}< 8.46$\,GHz. The full range of
model parameters allowed by the data are explored in Figure \ref{fig:090423_mcmcgrid}.
\label{fig:090423_mm}}
\end{figure}

\begin{figure}
\begin{tabular}{ccc}
\centering
 \includegraphics[width=0.31\columnwidth]{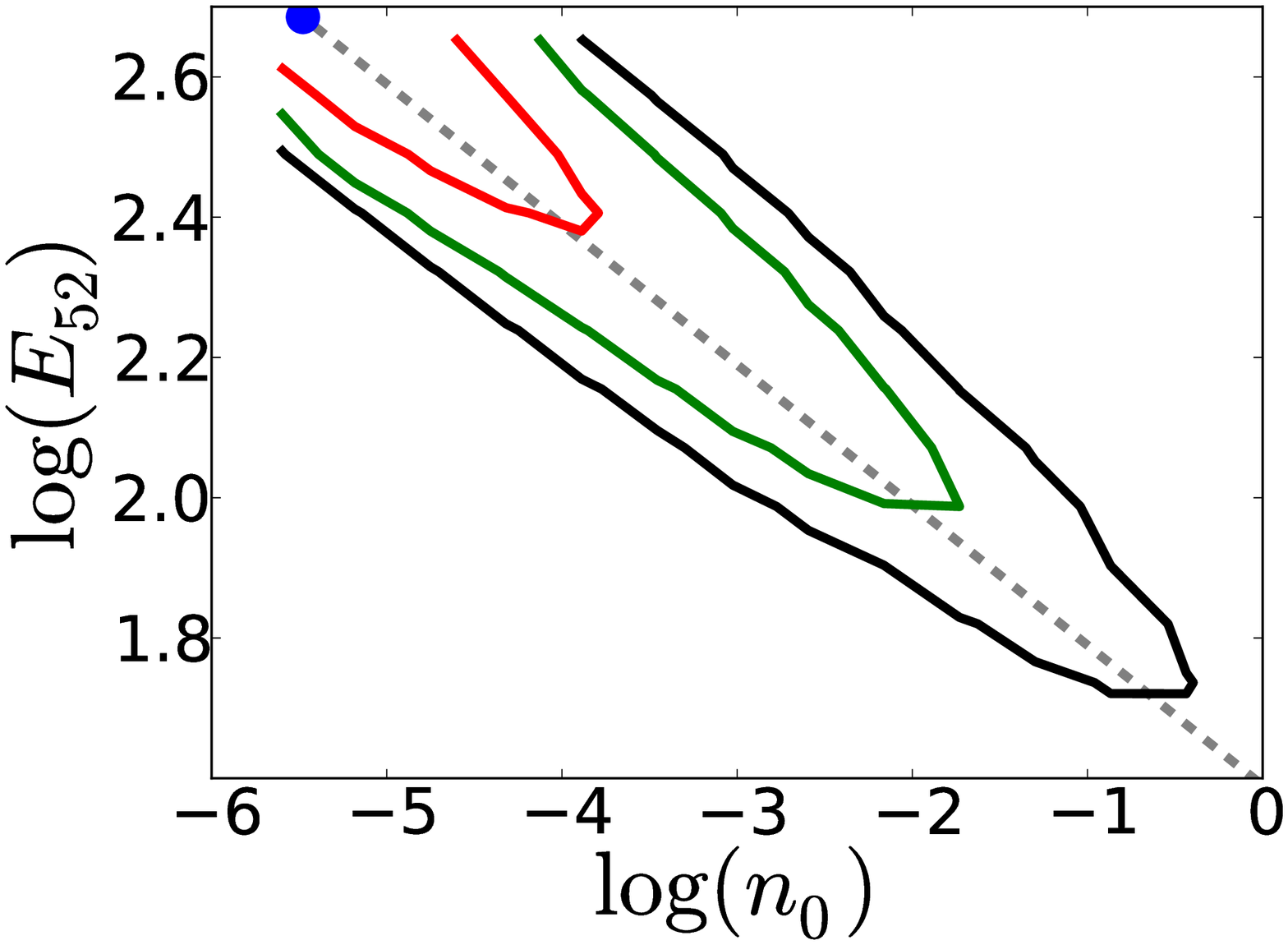} &
 \includegraphics[width=0.31\columnwidth]{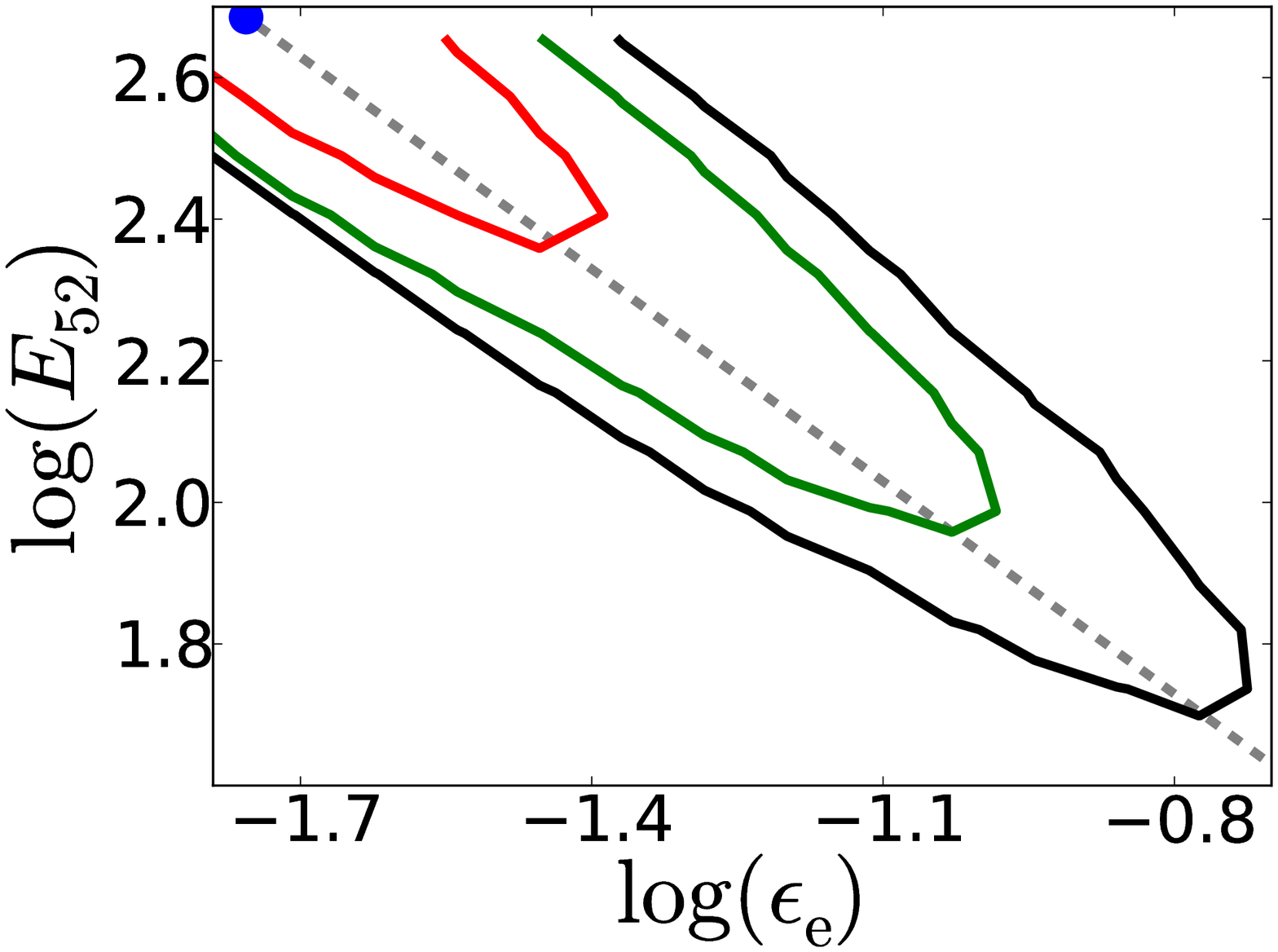} &
 \includegraphics[width=0.31\columnwidth]{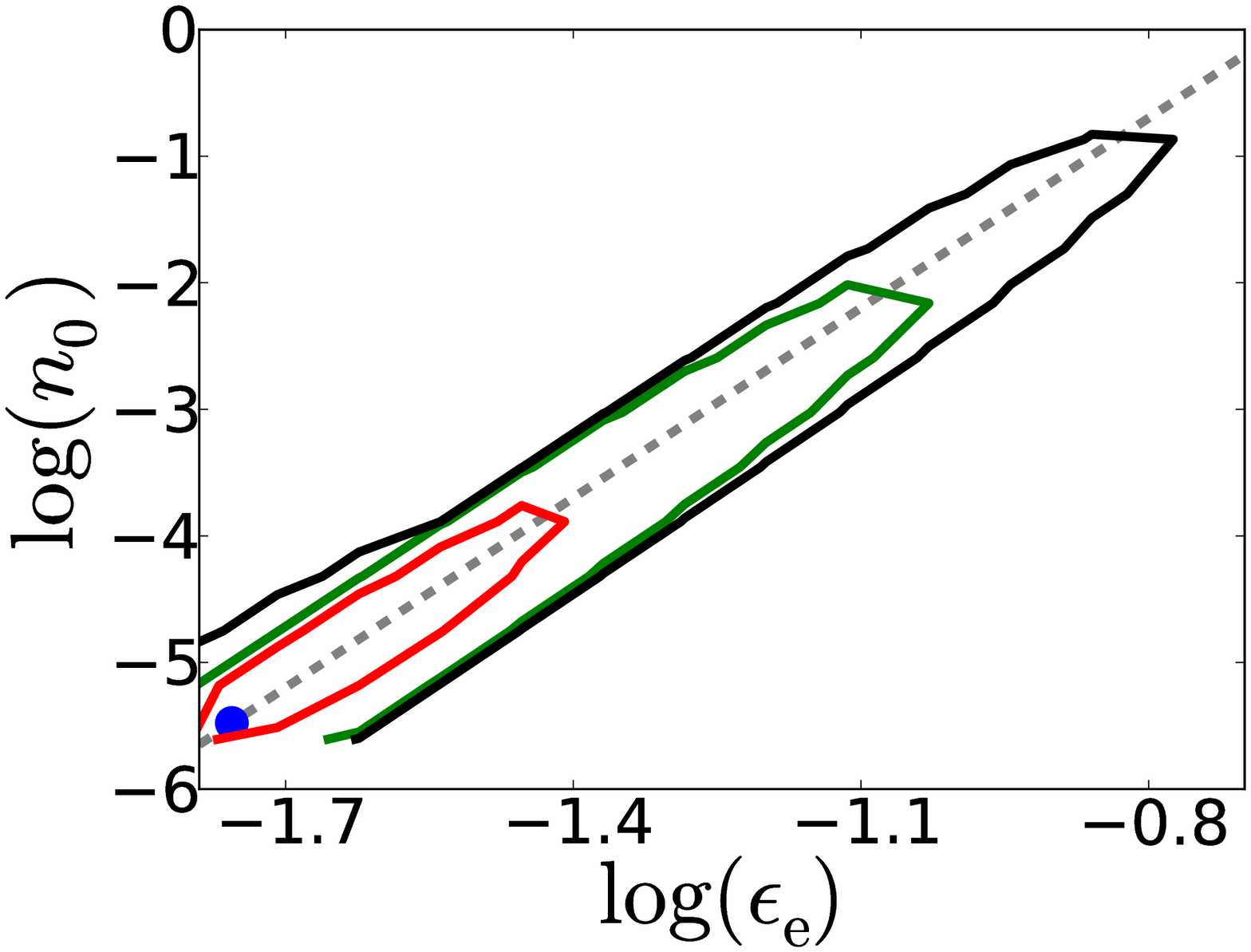} \\
 \includegraphics[width=0.31\columnwidth]{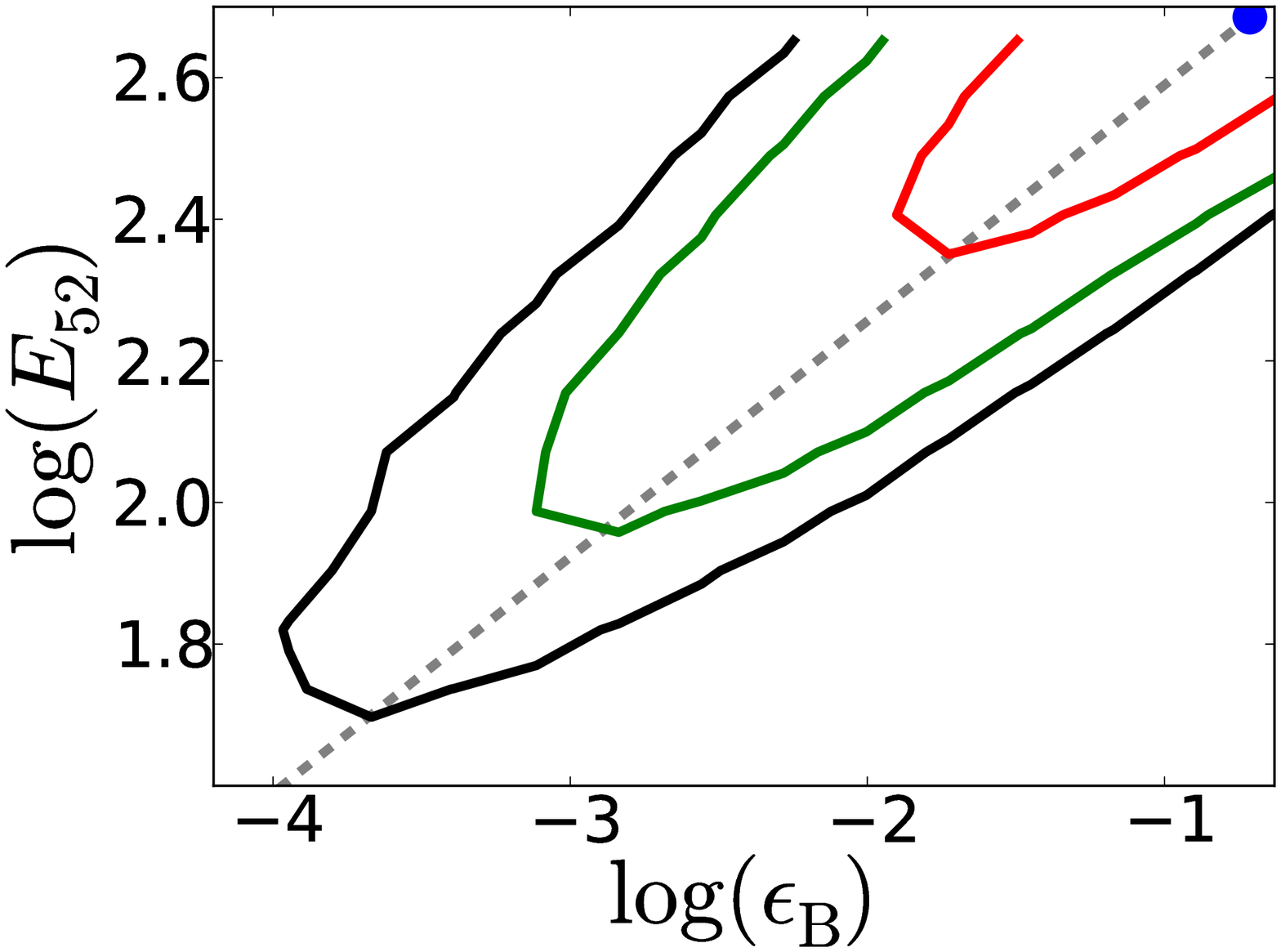} &
 \includegraphics[width=0.31\columnwidth]{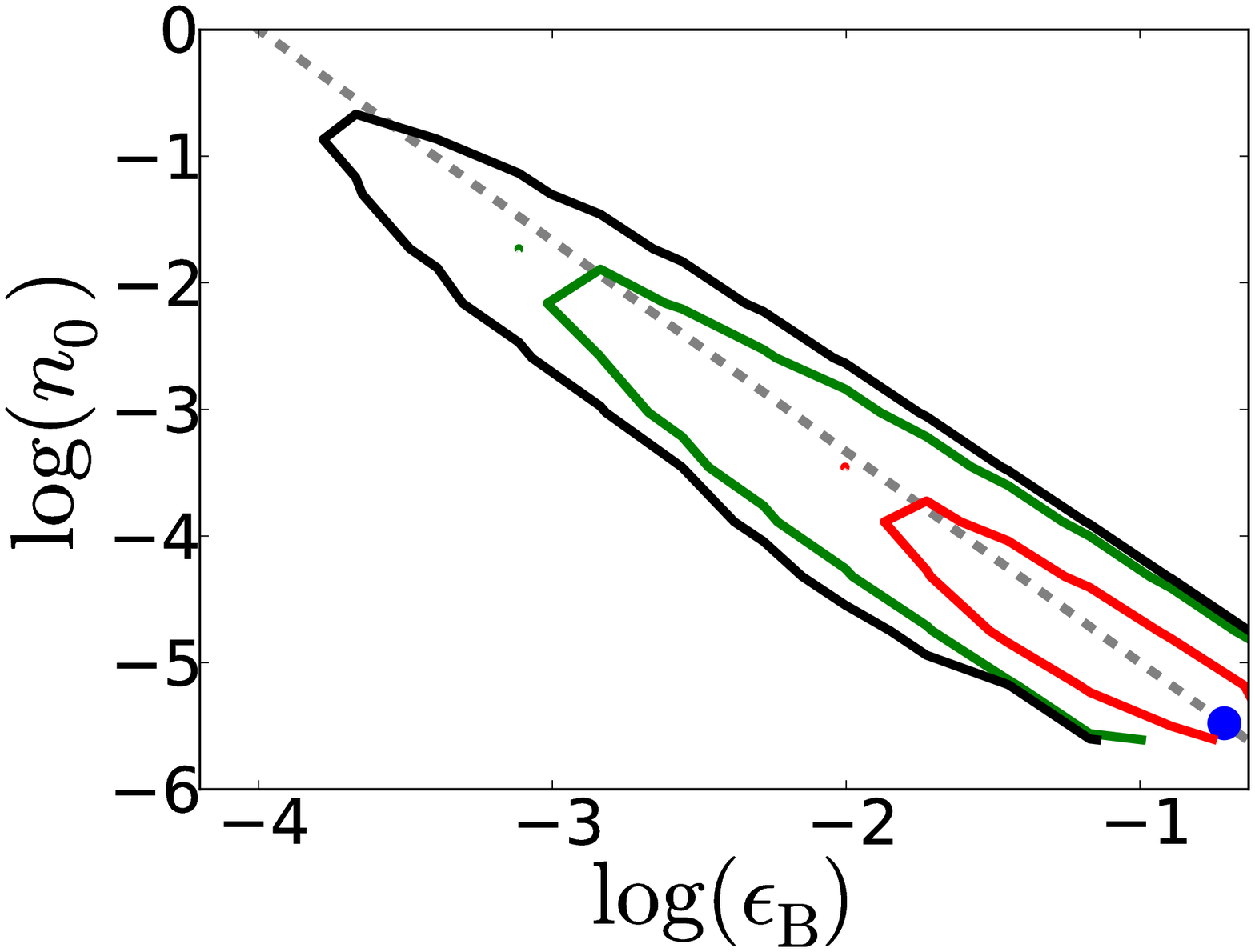} &
 \includegraphics[width=0.31\columnwidth]{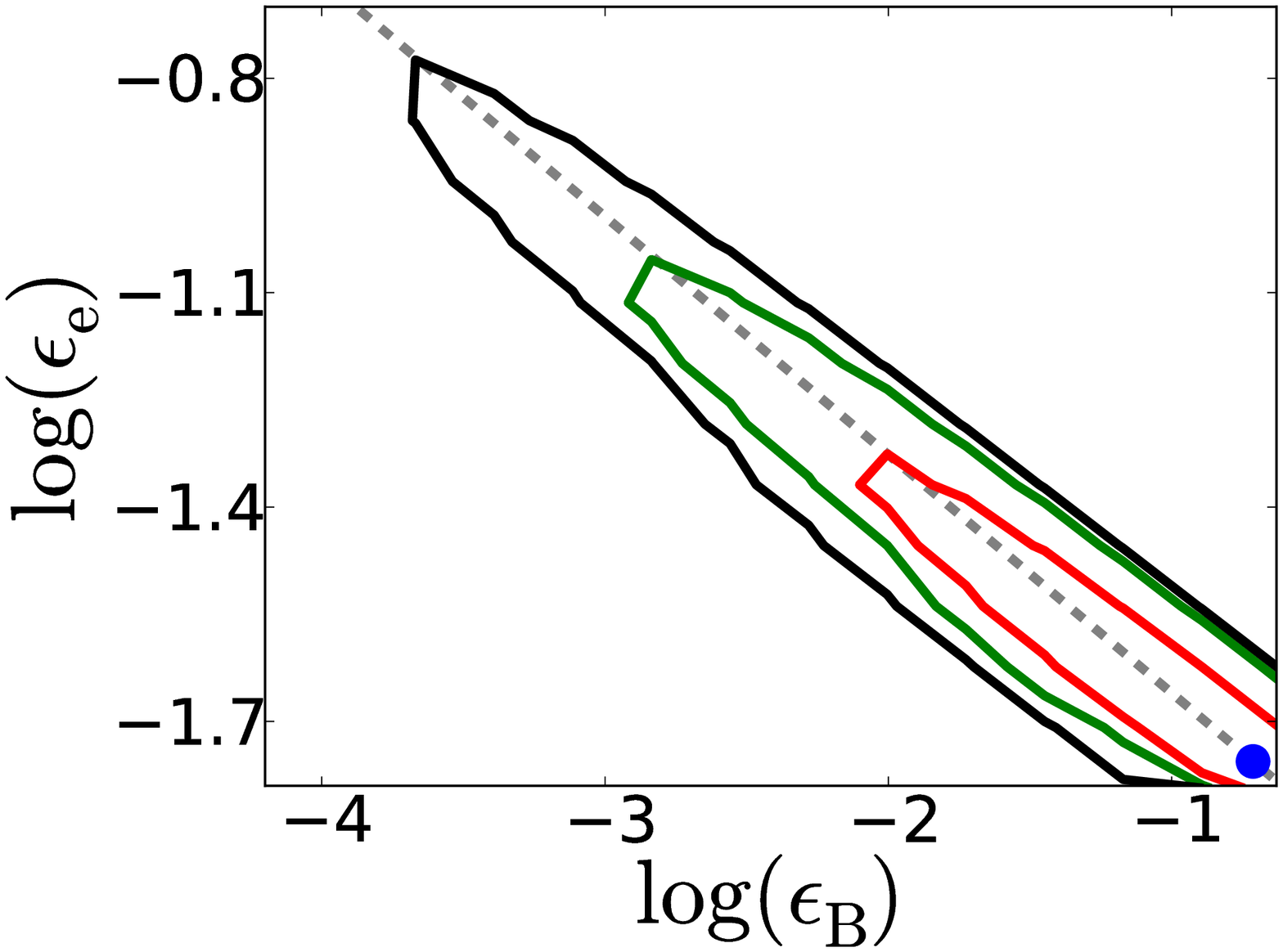} \\
\end{tabular}
\caption{1$\sigma$ (red), 2$\sigma$ (green), and 3$\sigma$ (black) contours for correlations
between the physical parameters, \E, \dens, \epse, and \epsb\ for GRB~0904023, for $p=2.56$ from
Monte Carlo simulations. We have restricted $E_{\rm K, iso, 52} < 500$, $\epsilon_{\rm e} <
\nicefrac{1}{3}$, and $\epsilon_{\rm B} < \nicefrac{1}{3}$. The dashed grey lines indicate the 
expected relations between these parameters when \nua\ is not fully constrained:
$E_{\rm K, iso, 52}\propto n_{0}^{-1/5}$, $E_{\rm K, iso, 52}\propto \epsilon_{\rm e}^{-1}$,
$n_0\propto \epsilon_{\rm e}^{5}$,
$E_{\rm K, iso, 52}\propto \epsilon_{\rm B}^{1/3}$,
$n_{0}\propto \epsilon_{\rm B}^{-5/3}$,
$\epsilon_{\rm e}\propto\epsilon_{\rm B}^{-1/3}$,
normalized to pass through the highest-likelihood point (blue dot).
The contours lie parallel to these lines, indicating that the primary source of uncertainty in
the physical parameters comes from the poor observational constraint on \nua.
See the on line version of this Figure for additional plots of correlations between these
parameters and $t_{\rm jet}$ and $A_{\rm V}$. \label{fig:090423_mcmcgrid}}
\end{figure}

\begin{figure}
\begin{tabular}{ccc}
\centering
 \includegraphics[width=0.31\columnwidth]{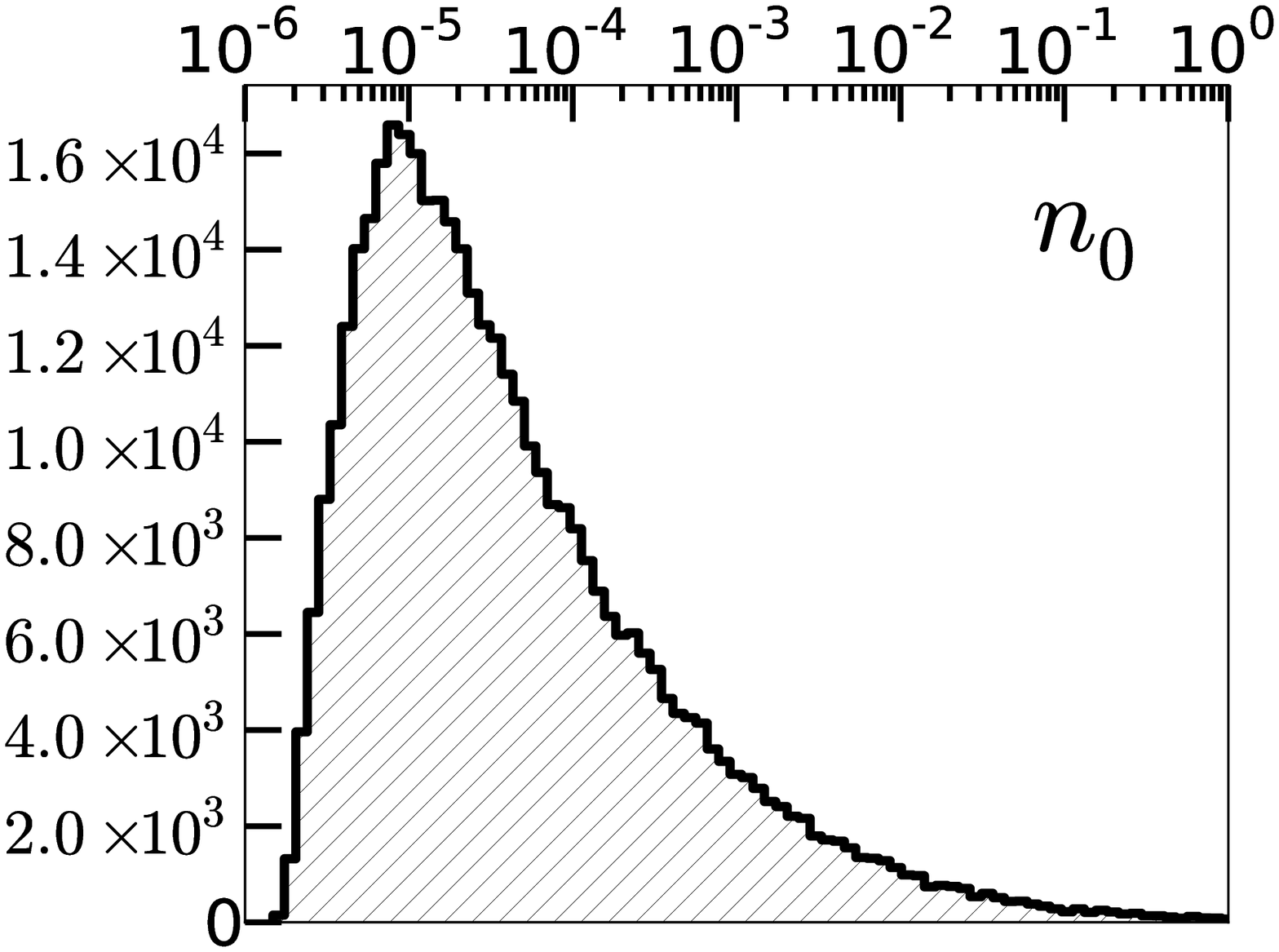} &
 \includegraphics[width=0.31\columnwidth]{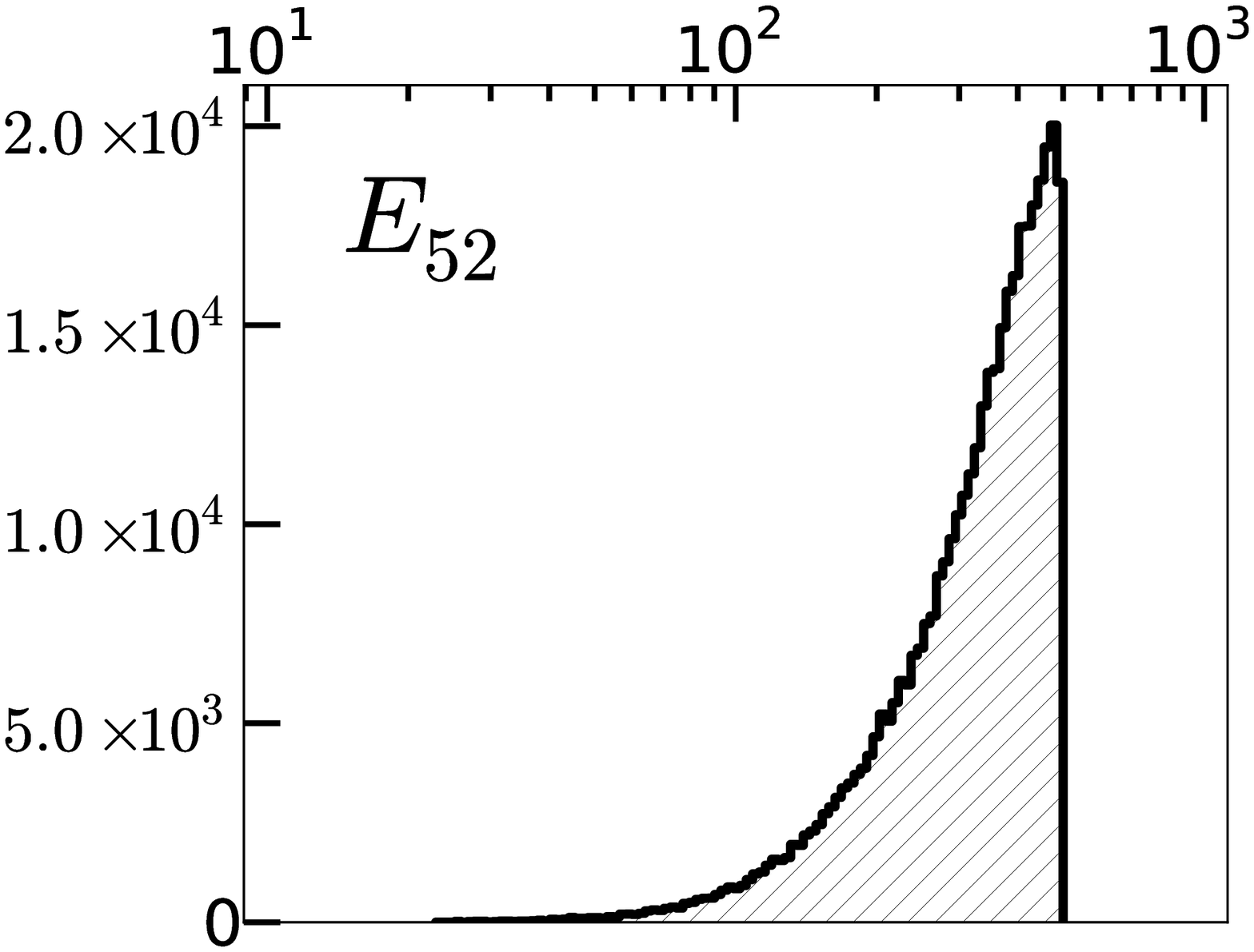} &
 \includegraphics[width=0.31\columnwidth]{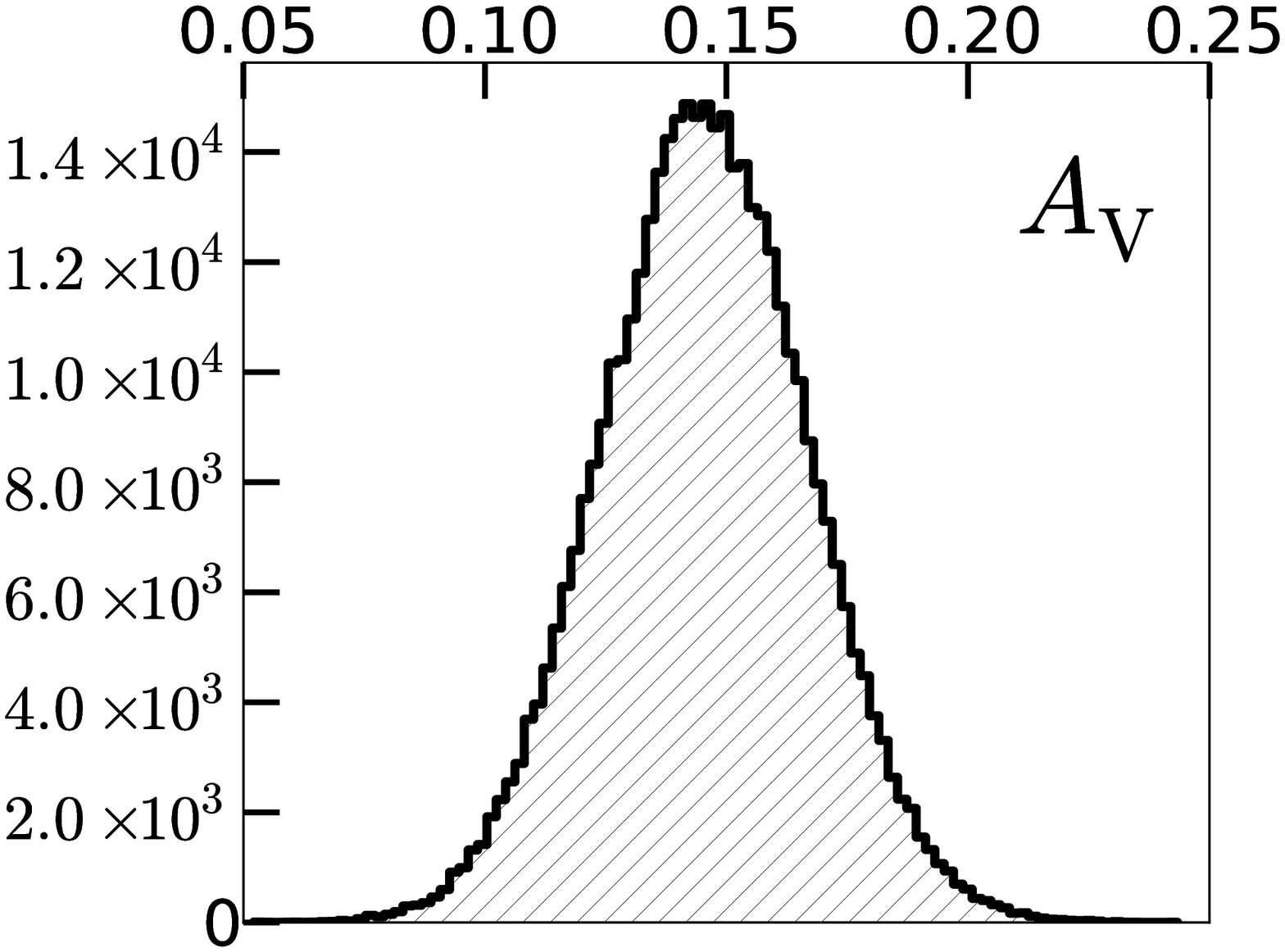} \\
 \includegraphics[width=0.31\columnwidth]{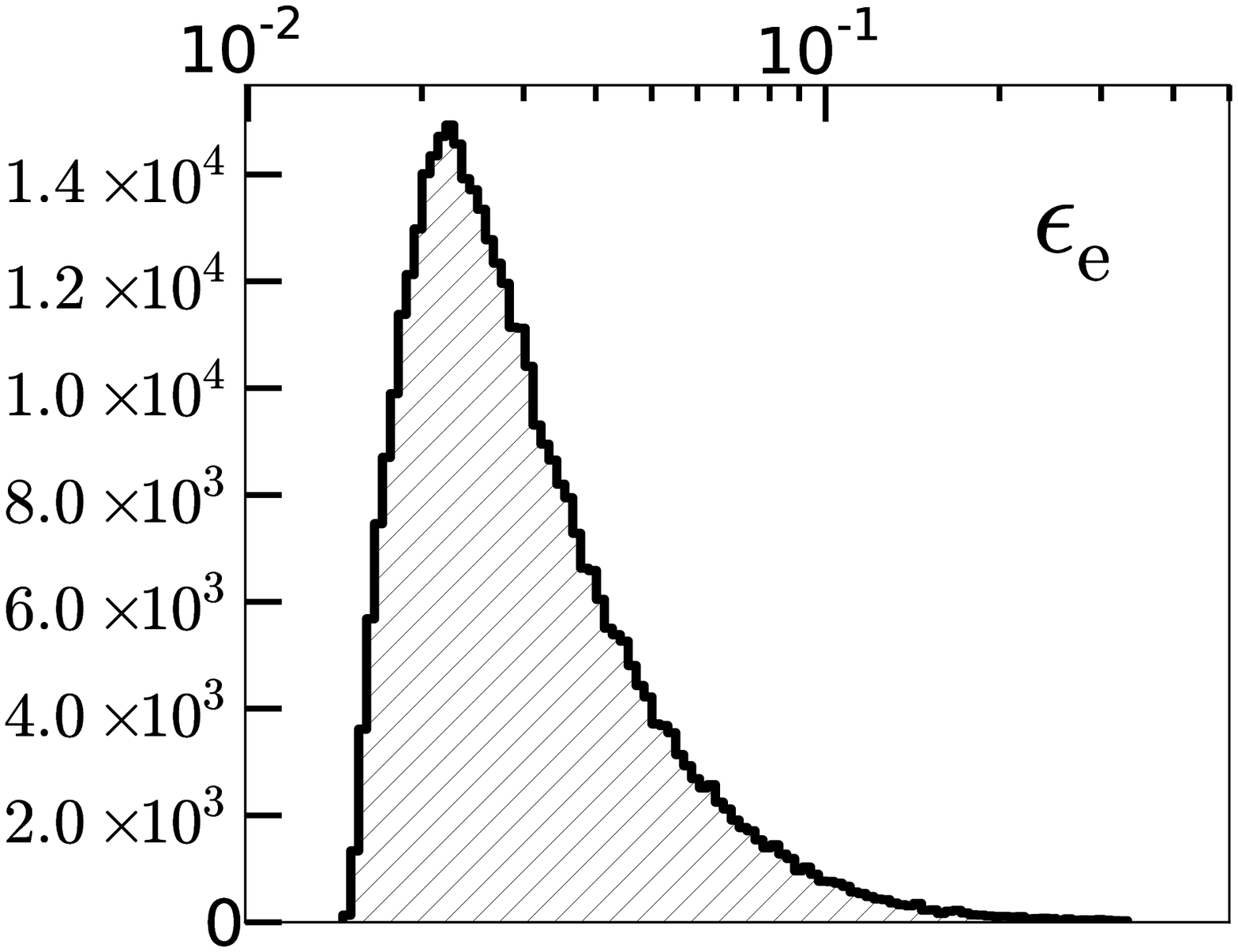} &
 \includegraphics[width=0.31\columnwidth]{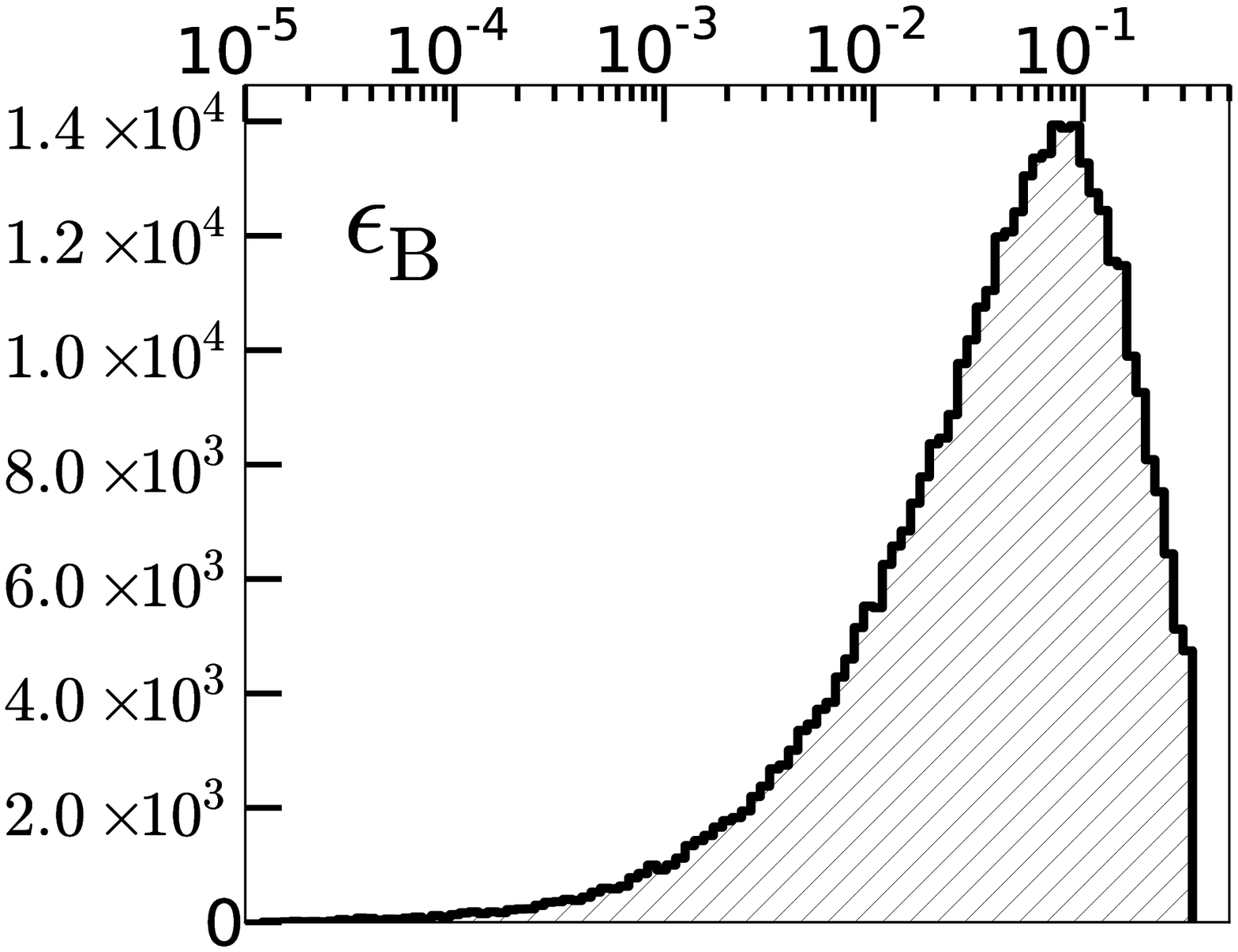} &
 \includegraphics[width=0.31\columnwidth]{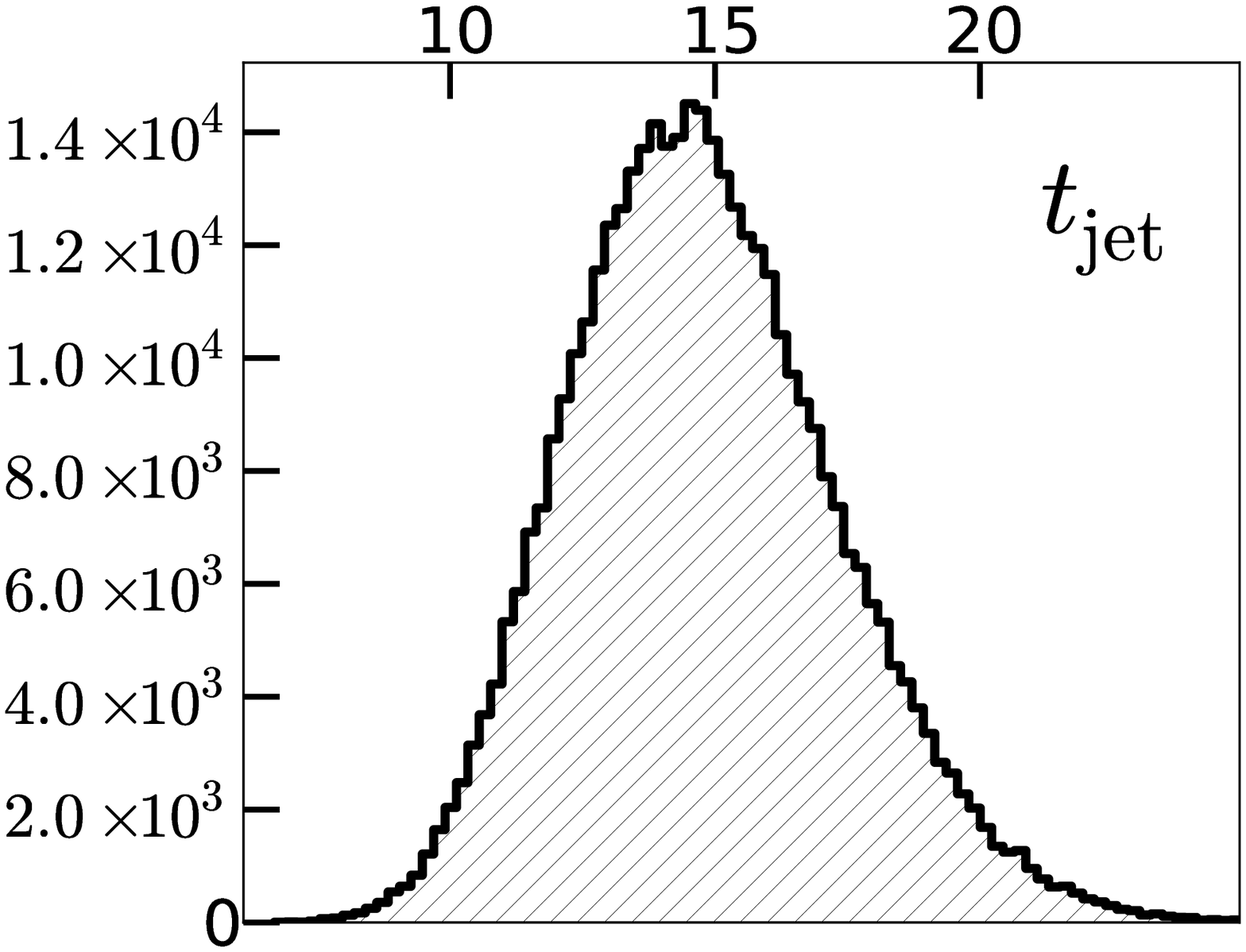} \\
\end{tabular}
\caption{Posterior probability density functions for the physical parameters for GRB~0904023 from
MCMC simulations $(p=2.56)$. We have restricted $E_{\rm K, iso, 52} < 500$, $\epsilon_{\rm e} <
\nicefrac{1}{3}$, and $\epsilon_{\rm B} < \nicefrac{1}{3}$.
\label{fig:090423_hists}}
\end{figure}

\begin{figure}
 \begin{tabular}{cc}
  \centering
  \includegraphics[width=0.45\columnwidth]{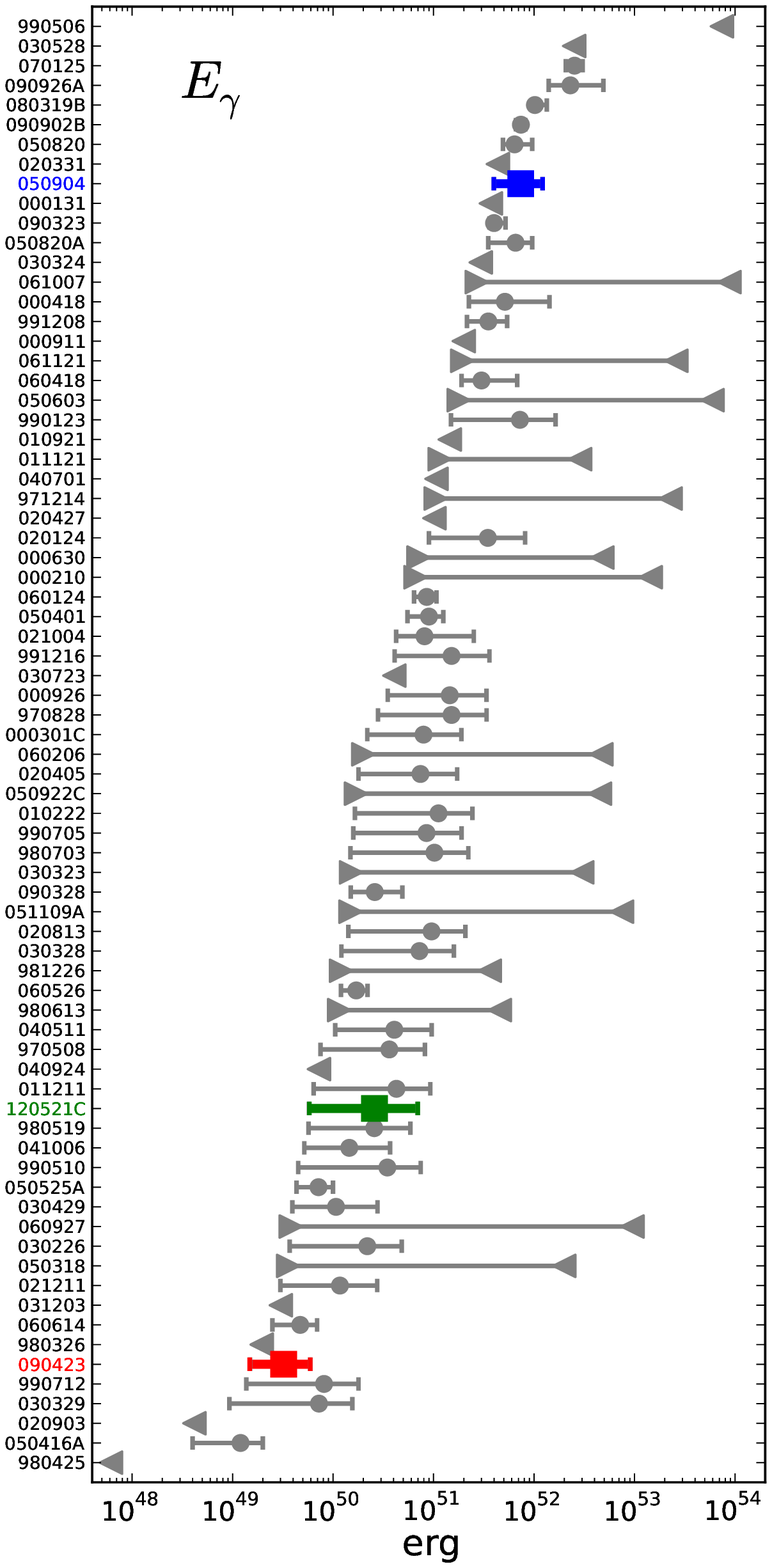} &
  \includegraphics[width=0.45\columnwidth]{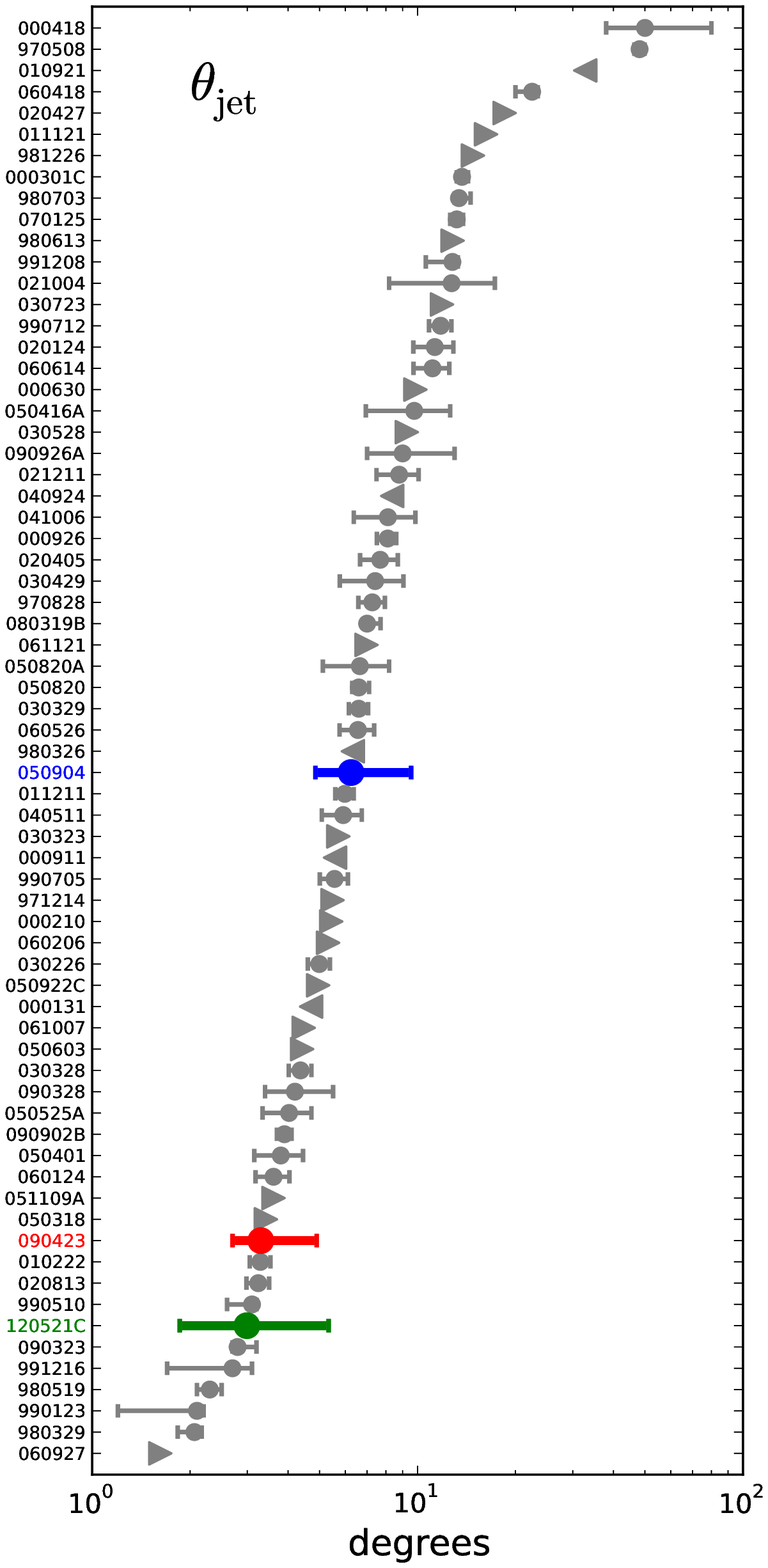}\\
 \end{tabular}
 \caption{Beaming-corrected $\gamma$-ray energy (left) and jet opening angle (right)
for the $z\gtrsim6$ GRBs 050904 (blue), 090423 (red), and 120521C (green), together with a 
comparison sample of lower-redshift long GRBs \citep[grey;][]{fb05, gngf07, cfh+10, cfh+11}. The 
isotropic-equivalent $\gamma$-ray energy for GRB~050904 is taken from \citet{agf+08}, and for 
GRB~090423 from \citet{sdvc+09}. The three GRBs at $z\gtrsim6$ do not appear distinct from the 
comparison sample in $E_{\gamma}$, but appear to all reside at lower values of $\theta_{\rm jet}$ 
than the median for lower-redshift GRBs.}
 \label{fig:comp_Egamma}
\end{figure}

\begin{figure}
  \centering
  \begin{tabular}{cc}
   \centering
   \includegraphics[width=0.45\columnwidth]{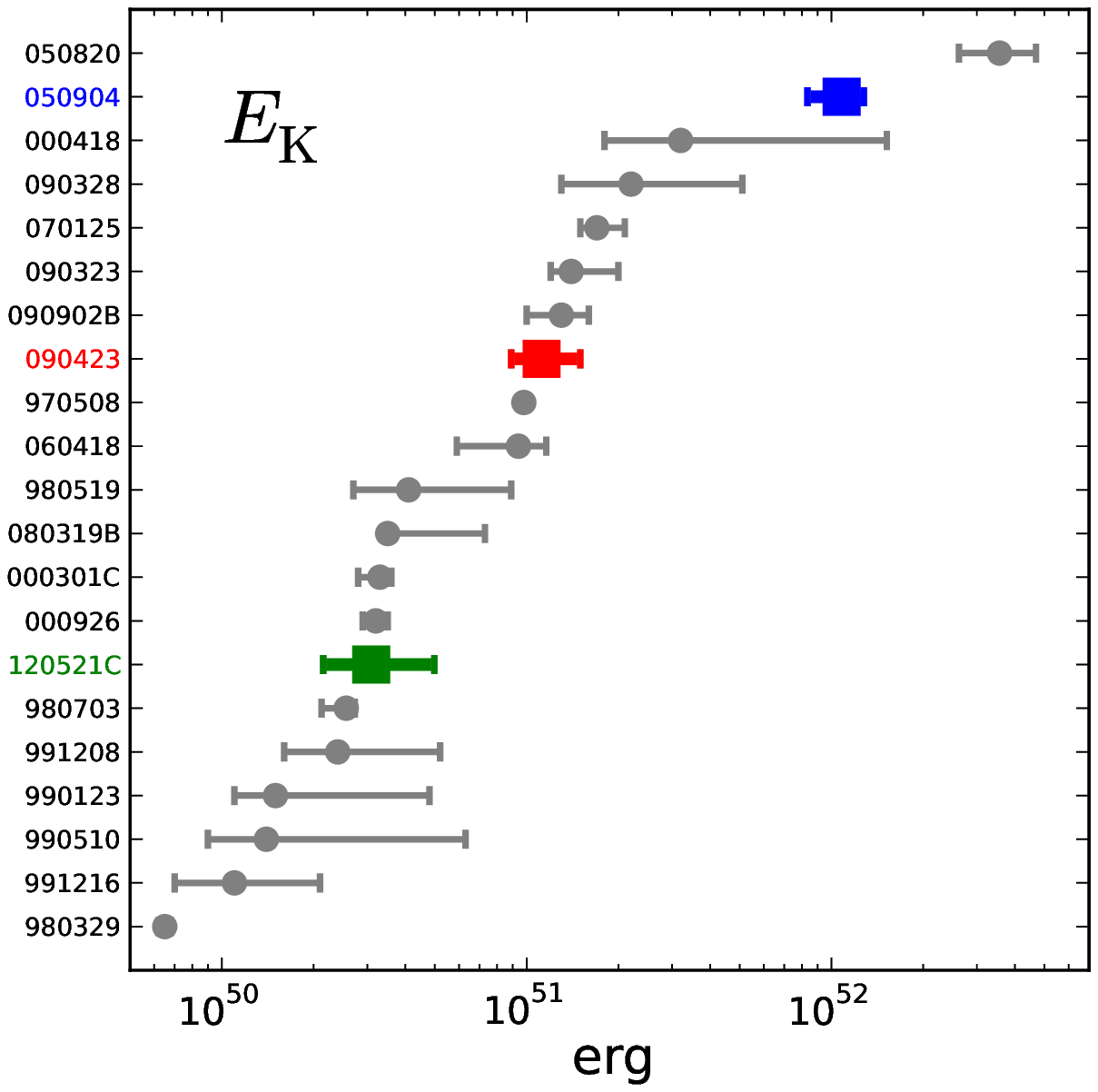} &
   \includegraphics[width=0.45\columnwidth]{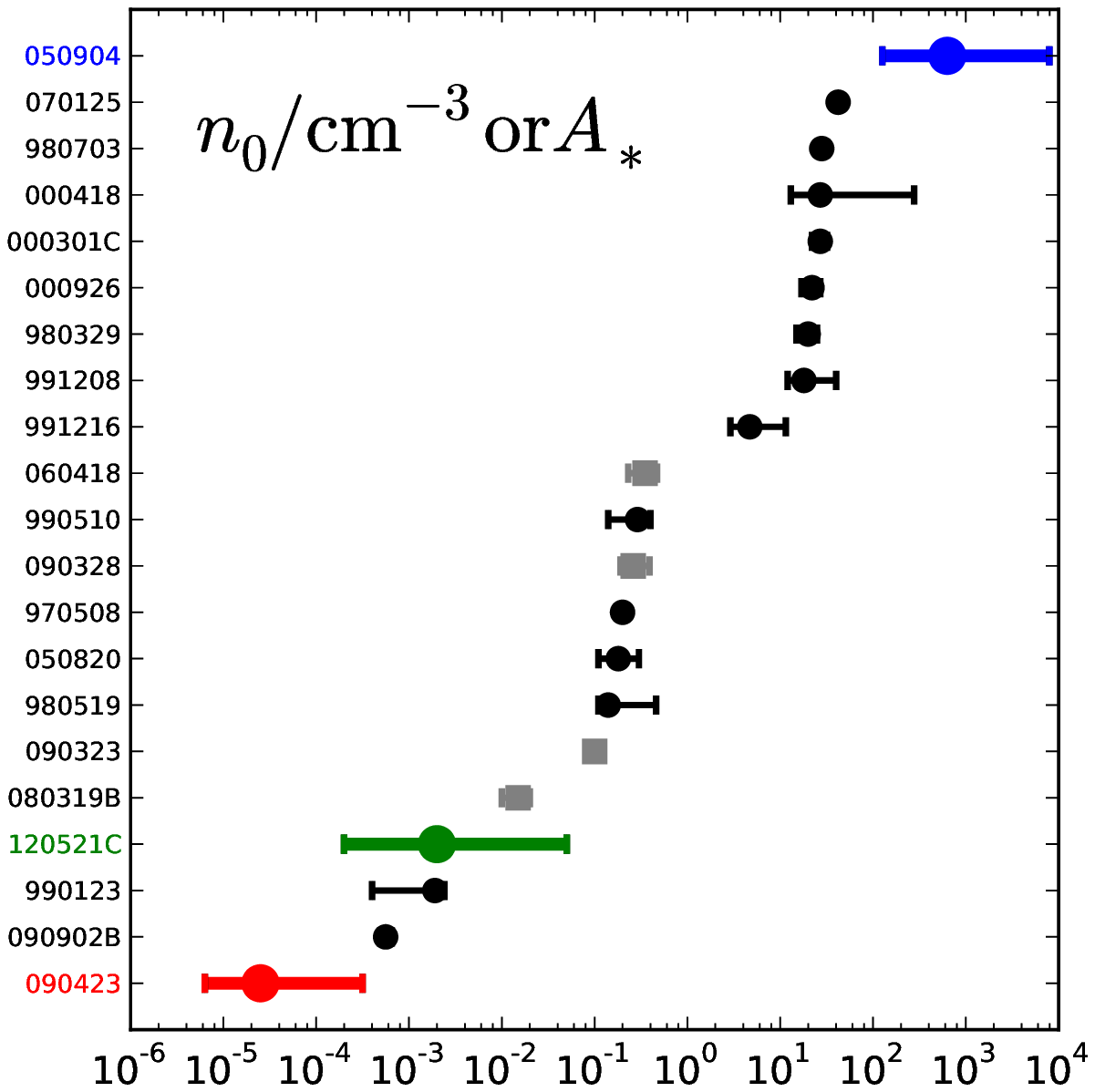}\\
  \end{tabular}
 \caption{Beaming-corrected kinetic energy (left) and circumburst density (right) for both ISM 
(black circles) and wind-like environments (grey squares). The three $z\gtrsim6$ GRBs,
050904 (blue), 090423 (red), and 120521C (green), do not appear distinct from the low 
redshift comparison sample \citep[grey and black;][]{pk02, yhsf03, ccf+08, cfh+10, cfh+11}.}
 \label{fig:comp_EK}
\end{figure}

\begin{figure}
  \centering    
   \includegraphics[width=\columnwidth]{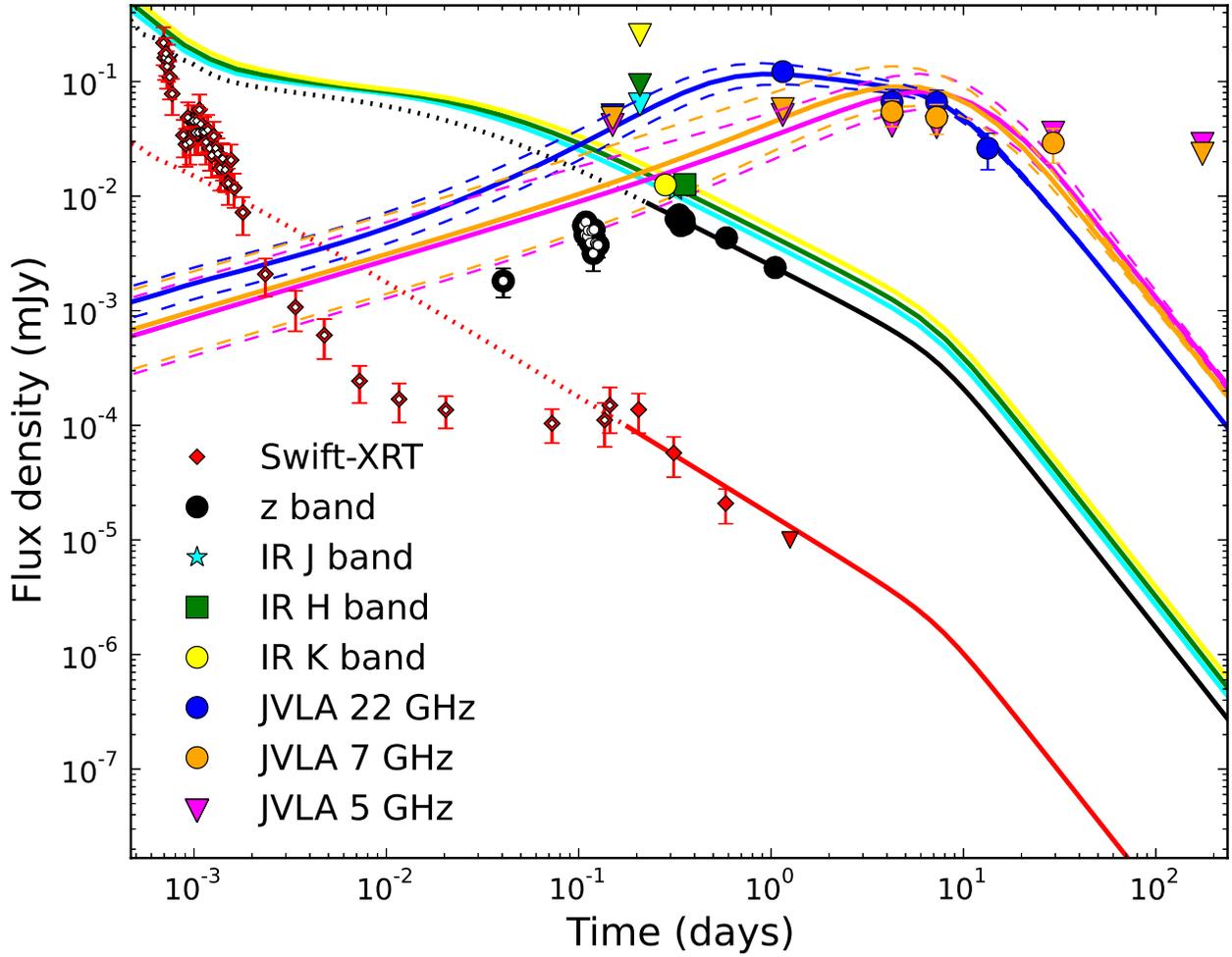}
  \caption{Same as Figure \ref{fig:120521C_multimodel_ISM}, with an additional reverse shock 
component to account for the high flux density of the first 21.8\,GHz detection at 1.15\,d. See 
Appendix \ref{text:120521C_RS} for details.}
 \label{fig:120521C_multimodel_ISM_RS}
\end{figure}

The millimeter detections at a flux level of $240\,\mu$Jy at $\approx 0.4$ and $1.3$\,d for 
GRB~090423 are much brighter than expected from the forward shock alone. Based on our best-fit ISM 
model (Section \ref{text:090423} and Figure \ref{fig:090423_mm}), the expected contribution of the 
FS to the millimeter flux density is $20\,\mu$Jy and $35\,\mu$Jy respectively, corresponding to an 
excess flux density of $220\,\mu$Jy and $205\,\mu$Jy. We now consider the hypothesis that this 
excess is due to reverse shock emission and perform an analysis similar to that for GRB~120521C 
(Appendix \ref{text:120521C_RS}).

\begin{figure}
  \centering    
   \includegraphics[width=\columnwidth]{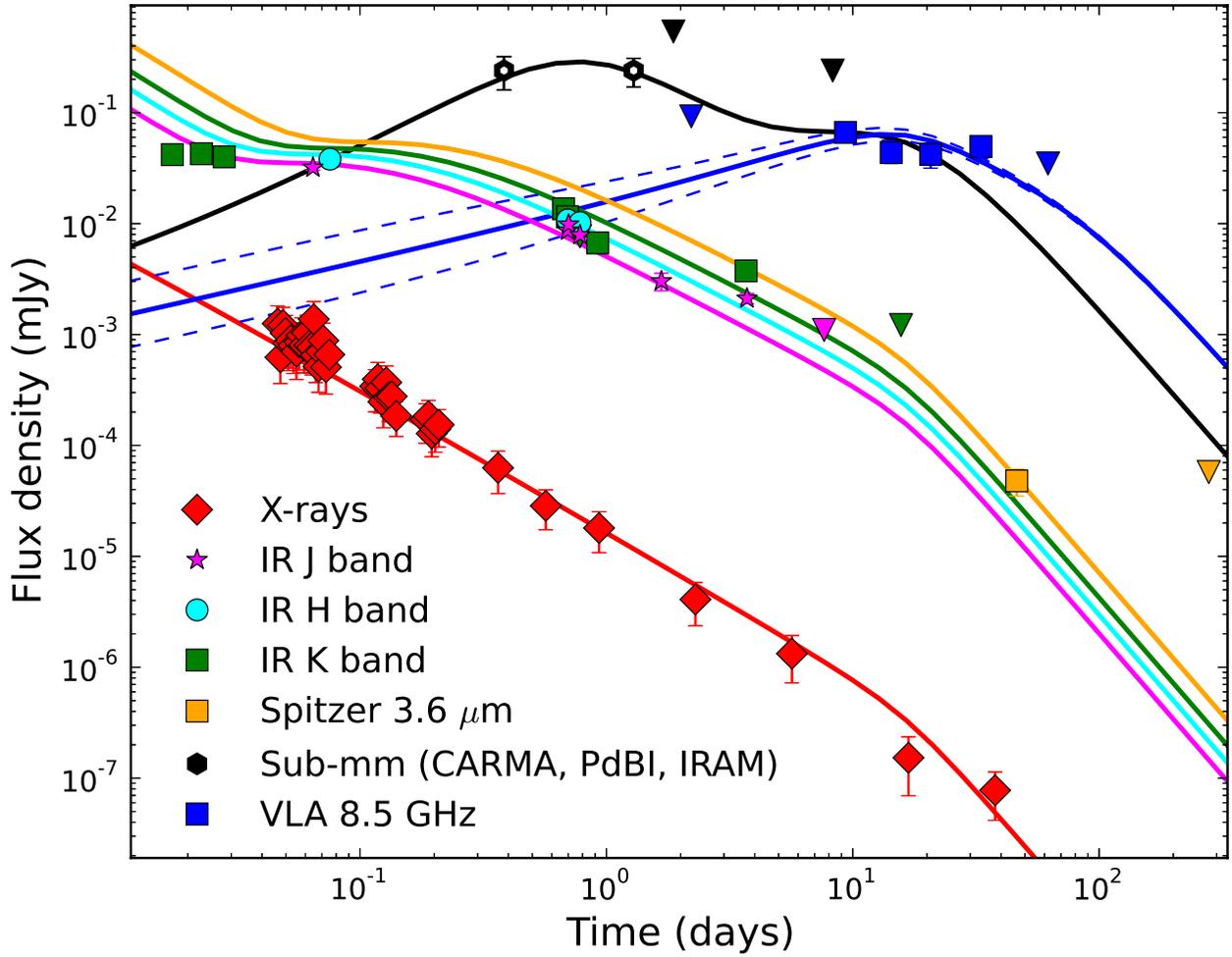}
  \caption{Same as Figure \ref{fig:090423_mm}, with an additional reverse shock 
component to account for the mm detections at 0.4 and 1.3\,d. The combined RS+FS model 
over-predicts the NIR $K$-band data between 0.01 and 0.05\,d. See Appendix \ref{text:090423_RS} for 
details.}
 \label{fig:090423_multimodel_RS}
\end{figure}

As in the case of GRB~120521C, we find that we must have \nuars~$<$~\numrs\ to avoid a low 
value of $\Gamma\sim5$. Given the millimeter data, a similar analysis to that of GRB~120521C 
indicates that the light curve must have peaked at $\approx0.8$\,d with a flux density of $\approx 
0.5\,$mJy (we use $p=2.56$ derived from the forward shock). The data do not directly constrain 
\numrs. If we assume that $t_{\rm dec} \sim T_{\rm 90} \approx 10$\,s we find a solution that 
satisfies the relations between the RS and FS at the deceleration time (see Appendix 
\ref{text:120521C_RS}) with $\Gamma \sim 500$ and \numrs~$=3.5\times10^7$\,Hz at 0.80\,d. In this 
case, the combined RS+FS model (Figure 
\ref{fig:090423_multimodel_RS}) over-predicts the NIR $K$-band observations around $0.02$\,d. 
However, we note that these observations take place at the same time as an X-ray plateau, which 
could result from energy injection. This would reduce the contribution of the FS to the NIR $K$-band 
light curve.

In summary, a combined RS + FS model with $\Gamma \sim 500$ can explain the significant excess flux 
density in the millimeter. The model over-predicts the NIR $K$-band observations at 0.01 to 0.05\,d, 
which could potentially be explained by a lower contribution from the FS than expected, due to 
energy injection over this period.

\bibliographystyle{apj}

\sisetup{round-mode = figures, round-precision = 3, scientific-notation = true}

\clearpage
\begin{table}
\caption{\Swift\ XRT Observations of GRB~120521C}
\label{tab:120521C:data:xrt}
\centering
\begin{tabular}{cccc}
 \toprule
 $\Delta t$ & Flux density  & Uncertainty & Detection?\\
 (days)  & (mJy)         & (mJy)       & ($1=$ Yes)\\
 \midrule
 \num{0.205} & \num{0.000137} & \num{5.19e-05} & \num{1} \\
 \num{0.312} & \num{5.73e-05} & \num{2.21e-05} & \num{1} \\
 \num{0.581} & \num{2.08e-05} & \num{6.94e-06} & \num{1} \\
 \num{1.25}  & \num{2.99e-05} & \num{9.98e-06} & \num{0} \\
 \bottomrule
\end{tabular}
\end{table}

\clearpage
\begin{ThreePartTable}
\begin{longtable}{lllccccc}
\caption[Optical and Near-Infrared Observations
of GRB~120521C]{\hbox {Optical and Near-Infrared Observations
of GRB~120521C}}\label{tab:120521C:data:IR}\\
\hline \hline \\[-2ex]
   \multicolumn{1}{c}{$\Delta t$} &
   \multicolumn{1}{c}{Telescope} &
   \multicolumn{1}{c}{Instrument} &
   \multicolumn{1}{c}{Band} &   
   \multicolumn{1}{c}{Frequency} &
   \multicolumn{1}{c}{Flux density\tablenotemark{a}} &
   \multicolumn{1}{c}{Uncertainty\tablenotemark{a}} &
   \multicolumn{1}{c}{Detection?} \\
   \multicolumn{1}{c}{(days)} &
   \multicolumn{1}{c}{} &
   \multicolumn{1}{c}{} &
   \multicolumn{1}{c}{} &
   \multicolumn{1}{c}{(Hz)} &
   \multicolumn{1}{c}{(mJy)} &
   \multicolumn{1}{c}{(mJy)} &
   \multicolumn{1}{c}{($1=$ Yes)}\\[0.5ex] \hline
   \\[-1.8ex]
\endfirsthead

%This is the header for the remaining page(s) of the table...
\multicolumn{8}{c}{{\tablename} \thetable{} -- Continued} \\[0.5ex]
  \hline \hline \\[-2ex]
   \multicolumn{1}{c}{$\Delta t$} &
   \multicolumn{1}{c}{Telescope} &
   \multicolumn{1}{c}{Instrument} &
   \multicolumn{1}{c}{Band} &
   \multicolumn{1}{c}{Frequency} &
   \multicolumn{1}{c}{Flux density\tablenotemark{a}} &
   \multicolumn{1}{c}{Uncertainty\tablenotemark{a}} &
   \multicolumn{1}{c}{Detection?} \\
   \multicolumn{1}{c}{(days)} &
   \multicolumn{1}{c}{} &
   \multicolumn{1}{c}{} &
   \multicolumn{1}{c}{} &
   \multicolumn{1}{c}{(Hz)} &
   \multicolumn{1}{c}{(mJy)} &
   \multicolumn{1}{c}{($1\sigma$, mJy)} &
   \multicolumn{1}{c}{($1=$ Yes)}\\[0.5ex] \hline
   \\[-1.8ex]
\endhead

%This is the footer for all pages except the last page of the table...
  \multicolumn{8}{l}{{Continued on Next Page\ldots}} \\
\footnotetext{Flux densities are not corrected for Galactic extinction.}
\endfoot

%This is the footer for the last page of the table...
  \\[-1.8ex] \hline \hline
\endlastfoot
\num{0.0316} & WHT & ACAM & $R$ & \num{4.81e+14} & \num{0.000585} & \num{0.000195} & \num{0} \\
\num{0.0372} & WHT & ACAM & $I$ & \num{3.93e+14} & \num{0.00109} & \num{0.000362} & \num{0} \\
\num{0.0379} & NOT &  & $R$ & \num{4.81e+14} & \num{0.000702} & \num{0.000234} & \num{0} \\
\num{0.0405} & WHT & ACAM &$z$& \num{3.46e+14} & \num{0.00146} & \num{0.000408} & \num{1} \\
\num{0.0433} & NOT &  & $I$ & \num{3.93e+14} & \num{0.00135} & \num{8.00e-05} & \num{0} \\
\num{0.106} & WHT & ACAM &$z$& \num{3.46e+14} & \num{0.00444} & \num{0.000555} & \num{1} \\
\num{0.108} & WHT & ACAM &$z$& \num{3.46e+14} & \num{0.00369} & \num{0.000669} & \num{1} \\
\num{0.109} & WHT & ACAM &$z$& \num{3.46e+14} & \num{0.00476} & \num{0.000615} & \num{1} \\
\num{0.111} & WHT & ACAM &$z$& \num{3.46e+14} & \num{0.0036} & \num{0.000625} & \num{1} \\
\num{0.112} & WHT & ACAM &$z$& \num{3.46e+14} & \num{0.00402} & \num{0.000651} & \num{1} \\
\num{0.115} & WHT & ACAM &$z$& \num{3.46e+14} & \num{0.00313} & \num{0.000717} & \num{1} \\
\num{0.117} & WHT & ACAM &$z$& \num{3.46e+14} & \num{0.00398} & \num{0.000653} & \num{1} \\
\num{0.119} & WHT & ACAM &$z$& \num{3.46e+14} & \num{0.00253} & \num{0.000748} & \num{1} \\
\num{0.12} & WHT & ACAM &$z$& \num{3.46e+14} & \num{0.00408} & \num{0.000635} & \num{1} \\
\num{0.122} & WHT & ACAM &$z$& \num{3.46e+14} & \num{0.0031} & \num{0.000725} & \num{1} \\
\num{0.124} & WHT & ACAM &$z$& \num{3.46e+14} & \num{0.00301} & \num{0.000649} & \num{1} \\
\num{0.126} & WHT & ACAM &$z$& \num{3.46e+14} & \num{0.00300} & \num{0.00068} & \num{1} \\
\num{0.208} & PAIRITEL &  &$K$& \num{1.37e+14} & \num{0.255} & \num{0.0848} & \num{0} \\
\num{0.208} & PAIRITEL &  &$H$& \num{1.84e+14} & \num{0.0932} & \num{0.031} & \num{0} \\
\num{0.208} & PAIRITEL &  &$J$& \num{2.38e+14} & \num{0.0633} & \num{0.0211} & \num{0} \\
\num{0.282} & UKIRT & WFCAM &$K$& \num{1.37e+14} & \num{0.0125} & \num{0.00134} & \num{1} \\
\num{0.318} & UKIRT & WFCAM &$J$& \num{2.38e+14} & \num{0.0112} & \num{0.00108} & \num{1} \\
\num{0.321} & Gemini-North& GMOS &$z$& \num{3.46e+14} & \num{0.00632} & \num{0.000316} & \num{1} \\
\num{0.324} & Gemini-North& GMOS &$z$& \num{3.46e+14} & \num{0.00664} & \num{0.000332} & \num{1} \\
\num{0.326} & Gemini-North& GMOS &$z$& \num{3.46e+14} & \num{0.00659} & \num{0.000329} & \num{1} \\
\num{0.329} & Gemini-North& GMOS &$z$& \num{3.46e+14} & \num{0.00601} & \num{0.000301} & \num{1} \\
\num{0.332} & Gemini-North& GMOS &$z$& \num{3.46e+14} & \num{0.00686} & \num{0.000343} & \num{1} \\
\num{0.334} & Gemini-North& GMOS &$z$& \num{3.46e+14} & \num{0.00627} & \num{0.000313} & \num{1} \\
\num{0.336} & Gemini-North& GMOS &$z$& \num{3.46e+14} & \num{0.00623} & \num{0.000311} & \num{1} \\
\num{0.339} & Gemini-North& GMOS &$z$& \num{3.46e+14} & \num{0.00553} & \num{0.000277} & \num{1} \\
\num{0.341} & Gemini-North& GMOS &$z$& \num{3.46e+14} & \num{0.00647} & \num{0.000323} & \num{1} \\
\num{0.344} & Gemini-North& GMOS &$z$& \num{3.46e+14} & \num{0.00604} & \num{0.000302} & \num{1} \\
\num{0.347} & Gemini-North& GMOS &$z$& \num{3.46e+14} & \num{0.00593} & \num{0.000296} & \num{1} \\
\num{0.349} & Gemini-North& GMOS &$z$& \num{3.46e+14} & \num{0.00594} & \num{0.000297} & \num{1} \\
\num{0.352} & Gemini-North& GMOS &$z$& \num{3.46e+14} & \num{0.00619} & \num{0.00031} & \num{1} \\
\num{0.354} & Gemini-North& GMOS &$z$& \num{3.46e+14} & \num{0.00569} & \num{0.000284} & \num{1} \\
\num{0.356} & UKIRT & WFCAM &$H$& \num{1.84e+14} & \num{0.0126} & \num{0.00135} & \num{1} \\
\num{0.514} & WHT & ACAM & $g$ & \num{6.29e+14} & \num{0.000114} & \num{3.8e-05} & \num{0} \\
\num{0.516} & Keck & LRIS & $I$ & \num{3.93e+14} & \num{0.000453} & \num{0.000151} & \num{0} \\
\num{0.579} & Gemini-North & GMOS & $I$ & \num{3.93e+14} & \num{0.000495} & \num{0.000165} & 
\num{0} \\
\num{0.586} & Gemini-North& GMOS &$z$& \num{3.46e+14} & \num{0.00433} & \num{0.000374} & \num{1} \\
\num{1.05} & WHT & ACAM &$z$& \num{3.46e+14} & \num{0.00191} & \num{0.000108} & \num{1} \\

\end{longtable}
\begin{tablenotes}
\item[a] Not corrected for Galactic extinction
\end{tablenotes}
\end{ThreePartTable}

\begin{table}
\begin{ThreePartTable}
\caption{\Swift\ UVOT Observations of GRB~120521C}
\label{tab:120521C:data:uvot}
\centering
\begin{tabular}{cccc}
 \toprule
 $\Delta t$ & Filter  & Frequency  & $3\sigma$ Flux Upper Limit \tablenotemark{a}\\
 (days)  &         & (Hz)       & (mJy)\\
 \midrule
  \num{1.5907e-02} & B        & \num{6.9250e+14} & \num{2.8387e-02} \\
  \num{1.6055e-02} & UVM2     & \num{1.3450e+15} & \num{1.3113e-02} \\
  \num{1.4277e-02} & U        & \num{8.5630e+14} & \num{9.5506e-03} \\
  \num{1.6770e-02} & V        & \num{5.5500e+14} & \num{5.4590e-02} \\
  \num{1.7337e-02} & UVW1     & \num{1.1570e+15} & \num{9.5866e-03} \\
  \num{1.6487e-02} & UVW2     & \num{1.4750e+15} & \num{8.7097e-03} \\
  \num{1.3334e-02} & WHITE    & \num{8.6400e+14} & \num{3.7121e-03} \\
  \num{1.0321e-01} & B        & \num{6.9250e+14} & \num{1.3541e-02} \\
  \num{7.4747e-02} & UVM2     & \num{1.3450e+15} & \num{1.0309e-02} \\
  \num{1.4464e-01} & U        & \num{8.5630e+14} & \num{8.6864e-03} \\
  \num{2.0687e-01} & V        & \num{5.5500e+14} & \num{4.6748e-02} \\
  \num{1.4259e-01} & UVW1     & \num{1.1570e+15} & \num{3.8169e-03} \\
  \num{2.0528e-01} & UVW2     & \num{1.4750e+15} & \num{1.9738e-03} \\
  \num{1.0784e-01} & WHITE    & \num{8.6400e+14} & \num{2.9149e-03} \\
  \num{5.7824e-01} & WHITE    & \num{8.6400e+14} & \num{8.9244e-04} \\
  \num{1.5257e+00} & UVM2     & \num{1.3450e+15} & \num{1.5728e-03} \\
 \bottomrule
\end{tabular}
\begin{tablenotes}
\item[a] Not corrected for Galactic extinction
\end{tablenotes}
\end{ThreePartTable}
\end{table}

\clearpage
\begin{table}
\caption{VLA Observations of GRB~120521C}
\label{tab:120521C:data:VLA}
\centering
\begin{tabular}{ccccccc}
\toprule
$\Delta t$ & VLA & Frequency & Integration time &
Integrated Flux & Uncertainty & Detection? \\
(days) & Configuration & (GHz) & (min) & density ($\mu$Jy) & ($\mu$Jy) & (1 = Yes) \\
\midrule
0.15 & CnB &  4.9 & 15.28 & $41.7$ & 13.9 & 0\\
     &     &  6.7 & 15.28 & $48.0$ & 16.0 & 0\\
     &     & 21.8 & 15.07 & $50.7$ & 16.9 & 0\\

1.15 & CnB &  4.9 & 10.12 & $51.0$ & 17.0 & 0\\
     &     &  6.7 & 10.12 & $57.3$ & 19.1 & 0\\
     &     & 21.8 & 15.07 & $112$   & 18.5 & 1\\

4.25 & B   &  4.9 & 15.27 & $41.1$ & 13.7 & 0\\
     &     &  6.7 & 15.27 & $54.5$  & 14.3 & 1\\
     &     & 21.8 & 14.52 & $66.5$  & 18.6 & 1\\

7.25 & B   &  4.9 & 15.12 & $39.9$ & 13.3 & 0\\
     &     &  6.7 & 15.12 & $48.8$  & 14.2 & 1\\
     &     & 21.8 & 12.97 & $65.8$  & 18.3 & 1\\

12.27 & B   & 21.8 & 32.95 & $30.6$ & 10.2 & 0\\
14.27 & B   & 21.8 & 32.68 & $38.4$ & 12.8 & 0\\
13.27\tablenotemark{a} & B   & 21.8 & --    & $26.2$  & 9.2  & 1\\

29.25 & B   &  4.9 & 24.87 & $35.7$ &  11.9 & 0\\
      &     &  6.7 & 24.87 & $29.1$ &   9.7 & 1\\

174.66& A   &  4.9 & 46.43 & $28.5$ &   9.5 & 0\\
      & A   &  6.7 & 46.43 & $23.4$ &   7.8 & 0\\
\bottomrule

\end{tabular}
\tablenotetext{a}{Weighted sum of data at 12.27 and 14.27~d.}
\end{table}

\clearpage
\begin{table}
\caption{Parameters from optical/NIR SED modeling of GRB~120521C}
\centering
\label{tab:120521C_sedfit}
\begin{tabular}{lcc}
\toprule
Parameter  & Best-fit & $68\%$ Credible Regions \\
\midrule
$z$       &  6.03 &  5.93$^{+0.11}_{-0.14}$\\
$\beta$   &$-0.34$& $-0.16^{+0.34}_{-0.25}$\\
$A_V$     &  0    &  0.11$^{+0.22}_{-0.10}$\\
\bottomrule
\end{tabular}
\end{table}

\clearpage
\begin{landscape}
\begin{table}
\caption{Best fit forward shock parameters}
\centering
\label{tab:bestfit}
\begin{tabular}{lcccc}
\toprule
Parameter   & \multicolumn{2}{c}{120521C\,\tablenotemark{\S}} &090423\,\tablenotemark{\S} &050904\\
            & ISM   & wind & & \\
\midrule
$z$         & 6.04 & 5.70 & 8.23 (fixed) & 6.29 (fixed)\\
$p$         & 2.12 & 2.03 & 2.56 (fixed) & 2.07        \\
$\epsilon_e$& $1.5\times10^{-1}\nu_{\rm a,8}^{\nicefrac{5}{6}}$  ($3.4\times10^{-2}$) & 0.26  &
              $1.3\times10^{-1}\nu_{\rm a,8}^{\nicefrac{5}{6}}$  ($1.6\times10^{-2}$)
            & $9.1\times10^{-3}$\\
$\epsilon_b$& $4.0\times10^{-3}\nu_{\rm a,8}^{-\nicefrac{5}{2}}$  ($3.2\times10^{-1}$)
            & $2.7\times10^{-3}$ &
              $4.0\times10^{-4}\nu_{\rm a,8}^{-\nicefrac{5}{2}}$  ($2.7\times10^{-1}$)
            & $2.0\times10^{-2}$\\
$n_0$       & $4.4\times10^{-1}\nu_{\rm a,8}^{\nicefrac{25}{6}}$ ($3.1\times10^{-4}$) & ...
            & $7.5\times10^{-2}\nu_{\rm a,8}^{\nicefrac{25}{6}}$ ($2.4\times10^{-6}$)
            & $3.2\times10^2$\\
$A_{*}$     & ...          & 0.81 & ...   & ...  \\
$E_{\rm K,iso,52}$ (erg)  & $6.7\nu_{\rm a,8}^{-\nicefrac{5}{6}}$ ($2.9\times10^{1}$) & 1.8
                          & $7.2\times10^{1}\nu_{\rm a,8}^{-\nicefrac{5}{6}}$ ($4.8\times10^{2}$)
                          & $2.4\times10^2$\\
$t_{\rm jet}$ (d)  & 7.4  & $\gtrsim8$\,\tablenotemark{\ast} & 16.7 & 1.5   \\
$\theta_{\rm jet}$ (deg)  & 2.3  & $\gtrsim 10$ &  2.5 & 5.4 \\
$A_V$  (mag)  & $\lesssim0.05$   & $\lesssim0.05$  & 0.17 & $\lesssim0.05$ \\
\midrule
$E_{\gamma,\rm iso}$ (erg) & \multicolumn{2}{c}{$(1.9\pm0.8)\times10^{53}$}
                           & $(1.0\pm0.3)\times10^{53}$\,\tablenotemark{\dag}
                           & $(1.24\pm0.13)\times10^{54}$\,\tablenotemark{\ddag}\\
$E_{\gamma}$         (erg) & $(1.5\pm0.6)\times10^{50}$& $\gtrsim 2.9\times10^{51}$
                           & $(9.5\pm2.9)\times10^{49}$  & $(5.5\pm0.6)\times10^{51}$\\
$E_{\rm K}$          (erg) & $5.4\times10^{49}\nu_{\rm a,8}^{-\nicefrac{5}{6}}$
                           & $\gtrsim 2.7\times10^{50}$
                           & $6.9\times10^{50}\nu_{\rm a,8}^{-\nicefrac{5}{6}}$
                           & $1.1\times10^{52}$\\
$E_{\rm tot}$        (erg) & $1.8\times10^{50}$\,\tablenotemark{\dag\dag}&$\gtrsim3.2\times10^{51}$ 
                           & $5.3\times10^{51}$\,\tablenotemark{\ddag\ddag}& $1.7\times10^{52}$\\
$\eta_{\rm rad}=\frac{E_{\gamma}}{E_{\rm tot}}$ & 0.83 & 0.91 & 0.02 & 
0.32\\
\bottomrule
\end{tabular}
\tablenotetext{\S}{The best-fit values of the physical parameters, \epse, \epsb, \dens, \E\ for 
GRBs 120521C and 090423 have been scaled to $\nu_{\rm a,8}=\nu_{\rm a}/10^8\,{\rm Hz}$. The values 
of these parameters corresponding to the highest likelihood model are given in parentheses and
correspond to $\nu_{\rm a} = 1.75\times10^8$\,Hz and $\nu_{\rm a} = 8.6\times10^6$\,Hz for
GRB~120521C and GRB~090423, respectively.}
\tablenotetext{\ast}{The lower end of the $90\%$ credible interval from MCMC simulations (see 
Table \ref{tab:mcmc}). The jet break time is not well constrained in the wind model for 
GRB~120521C.}
\tablenotetext{\dag}{\citet{gcn9251}}
\tablenotetext{\ddag}{\citet{agf+08}}
\tablenotetext{\dag\dag}{Assuming \nua$ = 1.75\times10^8$\,Hz, the best-fit value}
\tablenotetext{\ddag\ddag}{Assuming \nua$ = 8.6\times10^6$\,Hz, the best-fit value}
\end{table}
\end{landscape}

\clearpage
\begin{table}
\caption{Summary statistics from MCMC analyses}
\centering
\label{tab:mcmc}
\begin{tabular}{lcccc}
\toprule
Parameter   & \multicolumn{2}{c}{120521C} & 090423 & 050904 \\
            & ISM   & wind & & \\
\midrule
$z$         & $6.01^{+0.05}_{-0.09}$ & $5.71^{+0.04}_{-0.03}$ & 8.23 (fixed) & 6.29 (fixed)\\
$p$         & $2.17^{+0.09}_{-0.07}$ & $2.05^{+0.04}_{-0.02}$ & 2.56 (fixed) & $2.07\pm0.02$\\
$\epsilon_e$& $4.5^{+6.7}_{-2.4}\times10^{-2}$   & $0.20^{+0.09}_{-0.9}$
            & $2.7^{+2.0}_{-0.7}\times10^{-2}$   & $1.2^{+1.5}_{-0.5}\times10^{-2}$ \\
$\epsilon_b$& $0.7^{+1.5}_{-0.6}\times10^{-2}$   & $2.4^{+6.9}_{-1.7}\times10^{-3}$
            & $4.8^{+9.5}_{-3.9}\times10^{-2}$   & $1.3^{+2.2}_{-1.1}\times10^{-2}$\\
$\log{n_0}$ & $-2.7^{+1.4}_{-1.0}$   & ...        & $-4.6^{+1.1}_{-0.6}$ & $2.8^{+1.1}_{-0.7}$\\
$A_{*}$     & ...          & $0.79^{+0.65}_{-0.44}$  & ...                   & ...  \\
$E_{\rm K,iso,52}$ (erg)  & $2.2^{+3.7}_{-1.4}\times10^1$       & $1.9^{+1.4}_{-0.9}$
            & $3.4^{+1.1}_{-1.4}\times10^{2}$   &  $1.7^{+1.2}_{-1.0}\times10^{2}$\\
$t_{\rm jet}$ (d)  & $6.8^{+3.8}_{-2.4}$         & $\gtrsim6$\,\tablenotemark{\ast}
                            & $14.6^{+2.7}_{-2.3}$ & $1.5^{+0.2}_{-0.1}$ \\
$\theta_{\rm jet}$ (deg) & $3.0^{+2.3}_{-1.1}$     & $\gtrsim 9$\,\tablenotemark{\ast}
             &  $1.5^{+0.7}_{-0.3}$ & $6.2^{+3.3}_{-1.4}$ \\
$A_V$  (mag)  & $<0.05$    & $<0.05$  & $0.15\pm0.02$ & $<0.05$ \\
\midrule
$E_{\gamma,\rm iso}$(erg) & \multicolumn{2}{c}{$(1.9\pm0.8)\times10^{53}$}
                          & $(1.0\pm0.3)\times10^{53}$\,\tablenotemark{\dag}
                          & $(1.24\pm0.13)\times10^{54}$\,\tablenotemark{\ddag}\\
$E_{\gamma}$        (erg) & $2.6^{+4.4}_{-2.0}\times10^{50}$ & $\gtrsim2.1\times10^{51}$
                          & $3.2^{+2.7}_{-1.7}\times10^{49}$ & $7.4^{+4.8}_{-3.4}\times10^{51}$ \\
$E_{\rm K}$         (erg) & $3.1^{+1.9}_{-0.9}\times10^{50}$
                          & $\gtrsim 5.2\times10^{49}$\,\tablenotemark{\ast}
                          & $1.1^{+0.4}_{-0.2}\times10^{51}$ & $1.1^{+0.2}_{-0.2}\times10^{52}$\\
$E_{\rm tot} = E_{\gamma}+E_{\rm K}$  (erg)
                          & $6\times10^{50}$ & $2\times10^{51}$ 
                          & $1\times10^{51}$ & $2\times10^{52}$\\
$\eta_{\rm rad}=\frac{E_{\gamma}}{E_{\rm tot}}$ & 0.5 & 0.1\,\tablenotemark{\dag\dag} 
                                               & 0.03 & 0.4\\
\bottomrule
\end{tabular}
\tablenotetext{\ast}{The lower end of the $90\%$ credible interval. The jet break time is not well
constrained in the wind model for GRB~120521C.}
\tablenotetext{\dag}{\citet{gcn9251}}
\tablenotetext{\ddag}{\citet{agf+08}}
\tablenotetext{\dag\dag}{Using isotropic-equivalent energies}
\end{table}

\clearpage
\begin{ThreePartTable}
\begin{longtable}{c c c c c c c c}
\caption{Multi-wavelength Observations of GRB~050904}
\label{tab:050904:data:all}\\

%This is the header for the first page of the table...
\hline \hline \\[-2ex]
   \multicolumn{1}{c}{$\Delta t$} &
   \multicolumn{1}{c}{Telescope} &
   \multicolumn{1}{c}{Instrument} &
   \multicolumn{1}{c}{Band} &
   \multicolumn{1}{c}{Frequency} &
   \multicolumn{1}{c}{Flux density\tablenotemark{a}} &
   \multicolumn{1}{c}{Uncertainty\tablenotemark{a}} &
   \multicolumn{1}{c}{Detection} \\
   \multicolumn{1}{c}{(days)} &
   \multicolumn{1}{c}{} &
   \multicolumn{1}{c}{} &
   \multicolumn{1}{c}{} &
   \multicolumn{1}{c}{(GHz)} &
   \multicolumn{1}{c}{(mJy)} &
   \multicolumn{1}{c}{($1\sigma$, mJy)} &
   \multicolumn{1}{c}{(1 = Yes)} \\[0.5ex] \hline
   \\[-1.8ex]
\endfirsthead

%This is the header for the remaining page(s) of the table...
\multicolumn{7}{c}{{\tablename} \thetable{} -- Continued} \\[0.5ex]
  \hline \hline \\[-2ex]
   \multicolumn{1}{c}{$\Delta t$} &
   \multicolumn{1}{c}{Telescope} &
   \multicolumn{1}{c}{Instrument} &
   \multicolumn{1}{c}{Band} &
   \multicolumn{1}{c}{Frequency} &
   \multicolumn{1}{c}{Flux density\tablenotemark{a}} &
   \multicolumn{1}{c}{Uncertainty\tablenotemark{a}} &
   \multicolumn{1}{c}{Detection} \\
   \multicolumn{1}{c}{(days)} &
   \multicolumn{1}{c}{} &
   \multicolumn{1}{c}{} &
   \multicolumn{1}{c}{} &
   \multicolumn{1}{c}{(Hz)} &
   \multicolumn{1}{c}{(mJy)} &
   \multicolumn{1}{c}{($1\sigma$, mJy)} &
   \multicolumn{1}{c}{(1 = Yes)} \\[0.5ex] \hline
   \\[-1.8ex]
\endhead

%This is the footer for all pages except the last page of the table...
  \multicolumn{7}{l}{{Continued on Next Page\ldots}} \\
\endfoot

%This is the footer for the last page of the table...
  \\[-1.8ex] \hline \hline
\endlastfoot
\num{0.0202} & Swift & XRT & 1keV & \num{2.42e+17} & \num{0.00148} & \num{0.000485} & \num{1} \\
\num{0.128} & SOAR &  & $J$ & \num{2.43e+14} & \num{0.184} & \num{0.00689} & \num{1} \\
\num{0.135} & SOAR &  & $J$ & \num{2.43e+14} & \num{0.185} & \num{0.00695} & \num{1} \\
\num{0.142} & SOAR &  & $J$ & \num{2.43e+14} & \num{0.146} & \num{0.00547} & \num{1} \\
\num{0.312} & SOAR &  & $J$ & \num{2.43e+14} & \num{0.0555} & \num{0.00822} & \num{1} \\
\num{0.324} & SOAR &  & $Ks$ & \num{1.37e+14} & \num{0.132} & \num{0.00882} & \num{1} \\
\num{0.408} & UKIRT & WFCAM & $H$ & \num{1.82e+14} & \num{0.0566} & \num{0.00322} & \num{1} \\
\num{0.411} & UKIRT & WFCAM & $J$ & \num{2.43e+14} & \num{0.0398} & \num{0.00226} & \num{1} \\
\num{0.424} & UKIRT & WFCAM & $K$ & \num{1.37e+14} & \num{0.0755} & \num{0.00429} & \num{1} \\
\num{0.44} & IRTF &  & $K$ & \num{1.37e+14} & \num{0.0646} & \num{0.00181} & \num{1} \\
\num{0.487} & UKIRT & WFCAM & $J$ & \num{2.43e+14} & \num{0.0322} & \num{0.00214} & \num{1} \\
\num{0.505} & VLA &  & X & \num{8.46e+09} & \num{0.174} & \num{0.058} & \num{0} \\
\num{0.609} & Swift & XRT & 1\,keV & \num{2.42e+17} & \num{6.09e-05} & \num{2.37e-05} & \num{1} \\
\num{1.03} & TNG & NICS & $J$ & \num{2.43e+14} & \num{0.0234} & \num{0.00322} & \num{1} \\
\num{1.09} & VLT-UT1 & ISAAC & $J$ & \num{2.43e+14} & \num{0.0171} & \num{0.000642} & \num{1} \\
\num{1.1} & VLT-UT1 & ISAAC & $H$ & \num{1.82e+14} & \num{0.0236} & \num{0.00157} & \num{1} \\
\num{1.12} & SOAR &  & $Y$ & \num{2.91e+14} & \num{0.014} & \num{0.00379} & \num{1} \\
\num{1.12} & VLT-UT1 & ISAAC & $Ks$ & \num{1.37e+14} & \num{0.0324} & \num{0.00216} & \num{1} \\
\num{1.17} & SOAR &  & $J$ & \num{2.43e+14} & \num{0.0139} & \num{0.00236} & \num{1} \\
\num{1.39} & VLA &  & $X$ & \num{8.46e+09} & \num{0.075} & \num{0.025} & \num{0} \\
\num{1.91} & Swift & XRT & 1\,keV & \num{2.42e+17} & \num{5.37e-06} & \num{2.33e-06} & \num{1} \\
\num{2.09} & VLT-UT1 & ISAAC & $J$ & \num{2.43e+14} & \num{0.00797} & \num{0.00053} & \num{1} \\
\num{2.12} & VLT-UT1 & ISAAC & $H$ & \num{1.82e+14} & \num{0.0106} & \num{0.000706} & \num{1} \\
\num{2.15} & VLT-UT1 & ISAAC & $K$s & \num{1.37e+14} & \num{0.0144} & \num{0.000958} & \num{1} \\
\num{2.22} & SOAR &  & $J$ & \num{2.43e+14} & \num{0.00929} & \num{0.00219} & \num{1} \\
\num{2.27} & SOAR &  & $Y$ & \num{2.91e+14} & \num{0.00835} & \num{0.00307} & \num{1} \\
\num{3.1} & VLT-UT1 & ISAAC & $J$ & \num{2.43e+14} & \num{0.00345} & \num{0.000263} & \num{1} \\
\num{4.16} & VLT-UT1 & ISAAC & $J$ & \num{2.43e+14} & \num{0.00266} & \num{0.000204} & \num{1} \\
\num{5.32} & VLT-UT1 & ISAAC & $J$ & \num{2.43e+14} & \num{0.00166} & \num{0.000318} & \num{1} \\
\num{5.41} & VLA &  & X & \num{8.46e+09} & \num{0.075} & \num{0.025} & \num{0} \\
\num{6.22} & VLA &  & X & \num{8.46e+09} & \num{0.072} & \num{0.024} & \num{0} \\
\num{6.54} & Swift & XRT & 1\,keV & \num{2.42e+17} & \num{1.9e-06} & \num{8.59e-07} & \num{0} \\
\num{7.18} & VLT-UT1 & ISAAC & $J$ & \num{2.43e+14} & \num{0.000834} & \num{0.00167} & \num{0} \\
\num{20.1} & VLA &  & X & \num{8.46e+09} & \num{0.111} & \num{0.037} & \num{0} \\
\num{23.2} & HST & NICMOS & F160W & \num{1.82e+14} & \num{0.00013} & \num{2.5e-05} & \num{1} \\
\num{29.1} & VLA &  & X & \num{8.46e+09} & \num{0.09} & \num{0.03} & \num{0} \\
\num{33.4} & VLA &  & X & \num{8.46e+09} & \num{0.105} & \num{0.035} & \num{0} \\
\num{34.2} & VLA &  & X & \num{8.46e+09} & \num{0.069} & \num{0.023} & \num{0} \\
\num{35} & VLA &  & X & \num{8.46e+09} & \num{0.116} & \num{0.018} & \num{1} \\
\num{37.5} & VLA &  & X & \num{8.46e+09} & \num{0.067} & \num{0.017} & \num{1} \\
\num{44} & VLA &  & X & \num{8.46e+09} & \num{0.081} & \num{0.027} & \num{0} \\

\end{longtable}
\begin{tablenotes}
\small
\item[a] Not corrected for Galactic extinction
\item NIR observations are from \citet{hnr+06}, \citet{gfm07}, and \cite{bcc+07}. Radio 
observations are from \citet{fck+06}. We report the \Swift\ photometry included in our model (Figure 
\ref{fig:050904_mm}). We do not use the $Riz$ photometry in our model fitting (see Section 
\ref{text:050904}) and do not list them here.
\end{tablenotes}
\end{ThreePartTable}

\clearpage
\begin{ThreePartTable}
\begin{longtable}{c c c c c c c c}
\caption{Multi-wavelength Observations of GRB~090423}
\label{tab:090423:data:all}\\

%This is the header for the first page of the table...
\hline \hline \\[-2ex]
   \multicolumn{1}{c}{$\Delta t$} &
   \multicolumn{1}{c}{Telescope} &
   \multicolumn{1}{c}{Instrument} &
   \multicolumn{1}{c}{Band} &
   \multicolumn{1}{c}{Frequency} &
   \multicolumn{1}{c}{Flux density\tablenotemark{a}} &
   \multicolumn{1}{c}{Uncertainty\tablenotemark{a}} &
   \multicolumn{1}{c}{Detection} \\
   \multicolumn{1}{c}{(days)} &
   \multicolumn{1}{c}{} &
   \multicolumn{1}{c}{} &
   \multicolumn{1}{c}{} &
   \multicolumn{1}{c}{(GHz)} &
   \multicolumn{1}{c}{(mJy)} &
   \multicolumn{1}{c}{($1\sigma$, mJy)} &
   \multicolumn{1}{c}{(1 = Yes)} \\[0.5ex] \hline
   \\[-1.8ex]
\endfirsthead

%This is the header for the remaining page(s) of the table...
\multicolumn{7}{c}{{\tablename} \thetable{} -- Continued} \\[0.5ex]
  \hline \hline \\[-2ex]
   \multicolumn{1}{c}{$\Delta t$} &
   \multicolumn{1}{c}{Telescope} &
   \multicolumn{1}{c}{Instrument} &
   \multicolumn{1}{c}{Band} &
   \multicolumn{1}{c}{Frequency} &
   \multicolumn{1}{c}{Flux density\tablenotemark{a}} &
   \multicolumn{1}{c}{Uncertainty\tablenotemark{a}} &
   \multicolumn{1}{c}{Detection} \\
   \multicolumn{1}{c}{(days)} &
   \multicolumn{1}{c}{} &
   \multicolumn{1}{c}{} &
   \multicolumn{1}{c}{} &
   \multicolumn{1}{c}{(Hz)} &
   \multicolumn{1}{c}{(mJy)} &
   \multicolumn{1}{c}{($1\sigma$, mJy)} &
   \multicolumn{1}{c}{(1 = Yes)} \\[0.5ex] \hline
   \\[-1.8ex]
\endhead

%This is the footer for all pages except the last page of the table...
  \multicolumn{7}{l}{{Continued on Next Page\ldots}} \\
\endfoot

%This is the footer for the last page of the table...
  \\[-1.8ex] \hline \hline
\endlastfoot

\num{0.0173} & UKIRT & WFCAM &$K$& \num{1.37e+14} & \num{0.0419} & \num{0.00209} & \num{1} \\
\num{0.0227} & UKIRT & WFCAM &$K$& \num{1.37e+14} & \num{0.0427} & \num{0.00213} & \num{1} \\
\num{0.0281} & UKIRT & WFCAM &$K$& \num{1.37e+14} & \num{0.0400} & \num{0.00200} & \num{1} \\
\num{0.0463} & Swift & XRT &1.5\,kev& \num{3.63e+17} & \num{0.00126} & \num{0.000555} & \num{1} \\
\num{0.0475} & Swift & XRT &1.5\,kev& \num{3.63e+17} & \num{0.000624} & \num{0.000262} & \num{1} \\
\num{0.0487} & Swift & XRT &1.5\,kev& \num{3.63e+17} & \num{0.00121} & \num{0.000542} & \num{1} \\
\num{0.0498} & Swift & XRT &1.5\,kev& \num{3.63e+17} & \num{0.00103} & \num{0.000452} & \num{1} \\
\num{0.051} & Swift & XRT &1.5\,kev& \num{3.63e+17} & \num{0.000837} & \num{0.000363} & \num{1} \\
\num{0.0523} & Swift & XRT &1.5\,kev& \num{3.63e+17} & \num{0.000886} & \num{0.000389} & \num{1} \\
\num{0.0536} & Swift & XRT &1.5\,kev& \num{3.63e+17} & \num{0.000797} & \num{0.000351} & \num{1} \\
\num{0.0552} & Swift & XRT &1.5\,kev& \num{3.63e+17} & \num{0.000702} & \num{0.000308} & \num{1} \\
\num{0.0566} & Swift & XRT &1.5\,kev& \num{3.63e+17} & \num{0.000854} & \num{0.000375} & \num{1} \\
\num{0.0579} & Swift & XRT &1.5\,kev& \num{3.63e+17} & \num{0.000982} & \num{0.000432} & \num{1} \\
\num{0.0593} & Swift & XRT &1.5\,kev& \num{3.63e+17} & \num{0.000802} & \num{0.00035} & \num{1} \\
\num{0.0608} & Swift & XRT &1.5\,kev& \num{3.63e+17} & \num{0.00104} & \num{0.000453} & \num{1} \\
\num{0.0622} & Swift & XRT &1.5\,kev& \num{3.63e+17} & \num{0.000796} & \num{0.000346} & \num{1} \\
\num{0.0637} & Swift & XRT &1.5\,kev& \num{3.63e+17} & \num{0.000765} & \num{0.000332} & \num{1} \\
\num{0.0644} & Gemini-North & NIRI &$J$& \num{2.38e+14} & \num{0.032} & \num{0.0016} & \num{1} \\
\num{0.0649} & Swift & XRT &1.5\,kev& \num{3.63e+17} & \num{0.00137} & \num{0.000608} & \num{1} \\
\num{0.0661} & Swift & XRT &1.5\,kev& \num{3.63e+17} & \num{0.000649} & \num{0.00028} & \num{1} \\
\num{0.0678} & Swift & XRT &1.5\,kev& \num{3.63e+17} & \num{0.000515} & \num{0.000214} & \num{1} \\
\num{0.0695} & Swift & XRT &1.5\,kev& \num{3.63e+17} & \num{0.000704} & \num{0.000302} & \num{1} \\
\num{0.0709} & Swift & XRT &1.5\,kev& \num{3.63e+17} & \num{0.000883} & \num{0.000383} & \num{1} \\
\num{0.0727} & Swift & XRT &1.5\,kev& \num{3.63e+17} & \num{0.000508} & \num{0.000217} & \num{1} \\
\num{0.075} & Swift & XRT &1.5\,kev& \num{3.63e+17} & \num{0.000659} & \num{0.000248} & \num{1} \\
\num{0.0755} & Gemini-North & NIRI &$H$& \num{1.84e+14} & \num{0.0381} & \num{0.00191} & \num{1} \\
\num{0.115} & Swift & XRT &1.5\,kev& \num{3.63e+17} & \num{0.000342} & \num{0.000141} & \num{1} \\
\num{0.118} & Swift & XRT &1.5\,kev& \num{3.63e+17} & \num{0.000395} & \num{0.000165} & \num{1} \\
\num{0.121} & Swift & XRT &1.5\,kev& \num{3.63e+17} & \num{0.000328} & \num{0.000132} & \num{1} \\
\num{0.124} & Swift & XRT &1.5\,kev& \num{3.63e+17} & \num{0.000247} & \num{0.000103} & \num{1} \\
\num{0.128} & Swift & XRT &1.5\,kev& \num{3.63e+17} & \num{0.00037} & \num{0.000151} & \num{1} \\
\num{0.131} & Swift & XRT &1.5\,kev& \num{3.63e+17} & \num{0.000272} & \num{0.000108} & \num{1} \\
\num{0.134} & Swift & XRT &1.5\,kev& \num{3.63e+17} & \num{0.000277} & \num{0.000114} & \num{1} \\
\num{0.14} & Swift & XRT &1.5\,kev& \num{3.63e+17} & \num{0.000184} & \num{6.41e-05} & \num{1} \\
\num{0.185} & Swift & XRT &1.5\,kev& \num{3.63e+17} & \num{0.000171} & \num{6.67e-05} & \num{1} \\
\num{0.19} & Swift & XRT &1.5\,kev& \num{3.63e+17} & \num{0.000184} & \num{7.02e-05} & \num{1} \\
\num{0.195} & Swift & XRT &1.5\,kev& \num{3.63e+17} & \num{0.000128} & \num{4.91e-05} & \num{1} \\
\num{0.202} & Swift & XRT &1.5\,kev& \num{3.63e+17} & \num{0.00014} & \num{5.3e-05} & \num{1} \\
\num{0.209} & Swift & XRT &1.5\,kev& \num{3.63e+17} & \num{0.000154} & \num{5.71e-05} & \num{1} \\
\num{0.363} & Swift & XRT &1.5\,kev& \num{3.63e+17} & \num{6.28e-05} & \num{2.6e-05} & \num{1} \\
\num{0.384} & PdBI &  &  & \num{9.7e+10} & \num{0.24} & \num{0.0800} & \num{1} \\
\num{0.567} & Swift & XRT &1.5\,kev& \num{3.63e+17} & \num{2.86e-05} & \num{1.12e-05} & \num{1} \\
\num{0.67} & VLT-UT4 & HAWKI &$K$& \num{1.37e+14} & \num{0.0136} & \num{0.000681} & \num{1} \\
\num{0.696} & ESO2.2m & GROND &$K$& \num{1.37e+14} & \num{0.0115} & \num{0.00135} & \num{1} \\
\num{0.696} & ESO2.2m & GROND &$H$& \num{1.84e+14} & \num{0.0108} & \num{0.000825} & \num{1} \\
\num{0.696} & ESO2.2m & GROND &$J$& \num{2.38e+14} & \num{0.00865} & \num{0.000662} & \num{1} \\
\num{0.702} & VLT-UT4 & HAWKI &$J$& \num{2.38e+14} & \num{0.00984} & \num{0.000492} & \num{1} \\
\num{0.781} & ESO2.2m & GROND &$K$& \num{1.37e+14} & \num{0.00769} & \num{0.00256} & \num{0} \\
\num{0.781} & ESO2.2m & GROND &$H$& \num{1.84e+14} & \num{0.0102} & \num{0.00068} & \num{1} \\
\num{0.781} & ESO2.2m & GROND &$J$& \num{2.38e+14} & \num{0.00789} & \num{0.000526} & \num{1} \\
\num{0.922} & UKIRT & WFCAM &$K$& \num{1.37e+14} & \num{0.0067} & \num{0.000922} & \num{1} \\
\num{0.934} & Swift & XRT &1.5\,kev& \num{3.63e+17} & \num{1.8e-05} & \num{7.21e-06} & \num{1} \\
\num{1.29} & PdBI &  &  & \num{9.7e+10} & \num{0.24} & \num{0.0700} & \num{1} \\
\num{1.67} & VLT-UT1 & ISAAC &$J$& \num{2.38e+14} & \num{0.00303} & \num{0.000546} & \num{1} \\
\num{1.87} & CARMA &  &  & \num{9.7e+10} & \num{0.54} & \num{0.18} & \num{0} \\
\num{2.21} & VLA &  & X & \num{8.46e+09} & \num{0.0927} & \num{0.0309} & \num{0} \\
\num{2.3} & Swift & XRT &1.5\,kev& \num{3.63e+17} & \num{4.09e-06} & \num{1.72e-06} & \num{1} \\
\num{2.44} & IRAM30m &  &  & \num{2.5e+11} & \num{0.23} & \num{0.32} & \num{0} \\
\num{3.69} & VLT-UT4 & HAWKI &$K$& \num{1.37e+14} & \num{0.00375} & \num{0.000213} & \num{1} \\
\num{3.72} & VLT-UT4 & HAWKI &$J$& \num{2.38e+14} & \num{0.00211} & \num{0.000225} & \num{1} \\
\num{5.66} & Swift & XRT &1.5\,kev& \num{3.63e+17} & \num{1.33e-06} & \num{6.04e-07} & \num{1} \\
\num{7.65} & VLT-UT4 & HAWKI &$J$& \num{2.38e+14} & \num{0.0011} & \num{0.000366} & \num{0} \\
\num{8.29} & PdBI &  &  & \num{9.7e+10} & \num{0.24} & \num{0.0800} & \num{0} \\
\num{9.34} & VLA &  & X & \num{8.46e+09} & \num{0.0664} & \num{0.0114} & \num{1} \\
\num{14.3} & VLA &  & X & \num{8.46e+09} & \num{0.0437} & \num{0.0089} & \num{1} \\
\num{15.7} & VLT-UT4 & HAWKI &$K$& \num{1.37e+14} & \num{0.00122} & \num{0.000406} & \num{0} \\
\num{20.7} & VLA &  & X & \num{8.46e+09} & \num{0.0422} & \num{0.0106} & \num{1} \\
\num{29.4} & VLA &  & X & \num{4.9e+09} & \num{0.044} & \num{0.025} & \num{0} \\
\num{33.1} & VLA &  & X & \num{8.46e+09} & \num{0.0496} & \num{0.011} & \num{1} \\
\num{46.3} & Spitzer & IRAC & $3.6$\,$\mu$m & \num{8.4e+13} & \num{4.79e-05} & \num{1.3e-05} & 
\num{1} \\
\num{62} & VLA &  & X & \num{8.46e+09} & \num{0.0342} & \num{0.0116} & \num{0} \\
\num{279} & Spitzer & IRAC & $3.6$\,$\mu$m & \num{8.4e+13} & \num{5.75e-05} & \num{1.92e-05} & 
\num{0} \\

\end{longtable}
\begin{tablenotes}
\small
\item[a] Not corrected for Galactic extinction
\item All observations except the \Swift\ and \Spitzer\ data are collected from 
\citet{sdvc+09,tfl+09,cff+10,gcn9322,gcn9503}, and \citet{duplm+12}.
\end{tablenotes}
\end{ThreePartTable}

\clearpage
\begin{table}
\caption{\Chandra\ Observations of GRB~090423}
\label{tab:090423:data:Chandra}
\centering
\begin{tabular}{ccccc}
\toprule
 Epoch & $\Delta t$\tablenotemark{a} & Exposure time  & Count rate\tablenotemark{b} &
       1.5\,keV Flux\\
       & (days) & (ks) & ($10^{-4}$\,s${}^{-1}$) & density (mJy) \\
\midrule
1--2 & 16.8 \tablenotemark{c} & 31.9 & $1.1 \pm 0.6$   & $(1.5\pm0.8)\times10^{-7}$\\
3--5 & 37.8 \tablenotemark{c} &102.2 & $0.56 \pm 0.26$ & $(7.7\pm3.6)\times10^{-8}$\\
\bottomrule
\end{tabular}
\tablenotetext{1}{time to mid-exposure}
\tablenotetext{2}{$0.3$--$2$\,keV, $1.5$\arcsec (radius) aperture}
\tablenotetext{3}{mean time since GRB, weighted by exposure time of individual epochs}
\end{table}

\clearpage

\clearpage

\clearpage

\clearpage

\clearpage

\clearpage

\clearpage

\clearpage

\clearpage

\clearpage

\clearpage

\clearpage

\clearpage

\clearpage

\clearpage

\clearpage

\clearpage

\clearpage

\clearpage

\clearpage

\clearpage

\clearpage

\end{document}